\documentclass{aa}  

\usepackage{graphicx}
\usepackage{upgreek}
  
\usepackage{txfonts}
\usepackage{color}
\newcommand{\OI}{O\,{\sc i}}

\newcommand{\CII}{C\,{\sc ii}}
\newcommand{\HII}{H\,{\sc ii}}

\begin{document}

   \title{Molecular tracers of radiative feedback in Orion (OMC-1)\thanks{\textit{Herschel} is an ESA space observatory with science instruments provided by European-led Principal Investigator consortia and with important participation from NASA.}$^{,}$\thanks{Includes IRAM 30m observations. IRAM is supported by INSU/CNRS (France), MPG (Germany), and IGN (Spain).}
}

\subtitle{Widespread CH$^+$\,($J$\,=\,1--0), CO\,(10--9), HCN\,(6--5), and HCO$^+$\,(6--5) emission}

 \titlerunning{Stellar radiative feedback in OMC-1} 
\authorrunning{Goicoechea et al.}

   \author{Javier R. Goicoechea\inst{1}
          \and 
           Miriam G. Santa-Maria \inst{1}
          \and
          Emeric Bron\inst{1}
          \and
          David Teyssier\inst{2},\\
          Nuria Marcelino\inst{1}
          \and 
          Jos\'e Cernicharo\inst{1}
          \and   
          Sara Cuadrado\inst{1}
          }

   \institute{Instituto de F\'{\i}sica Fundamental
     (CSIC). Calle Serrano 121, E-28006, Madrid, Spain.
              \email{javier.r.goicoechea@csic.es}
         \and
    Telespazio Vega UK Ltd for ESA/ESAC. Urbanizaci\'on
Villafranca del Castillo, Villanueva de la
Ca\~{n}ada, E-28692 Madrid, Spain.          
             }
   \date{Received 9 October 2018 / Accepted 30 November 2018}

% \abstract{}{}{}{}{} 
% 5 {} token are mandatory
 
  \abstract
  % context heading (optional)
  % {Mapping an area of 85~arcmin$^2$}   
  % {Range of critical densities}
  % aims heading (mandatory)
  %{Bla}
  % methods heading (mandatory)
  % {}
  % results heading (mandatory)
  % {Bla}
  % conclusions heading (optional), leave it empty if necessary 
  %{Scalings}

\abstract{Young massive stars regulate the physical conditions, ionization,
and fate of their natal molecular cloud and surroundings.  It is important to
find tracers that help quantifying the  stellar feedback processes that take place  at  different spatial scales. 
We present \mbox{$\sim$85~arcmin$^2$} \mbox{($\sim$1.3\,pc$^2$)}
velocity-resolved maps of several submillimeter  molecular lines, taken with \mbox{\textit{Herschel}/HIFI},  toward the closest high-mass star-forming region, the  Orion molecular cloud~1 core \mbox{(OMC-1)}. The  observed rotational lines include probes of 
warm and dense molecular gas that are difficult, if not impossible, to detect from ground-based telescopes: \mbox{CH$^+$~($J$\,$=$\,1--0)}, \mbox{CO~($J$\,$=$\,10--9)}, \mbox{HCO$^+$~($J$\,$=$\,6--5)} and  
\mbox{HCN~($J$\,$=$\,6--5)}, and \mbox{CH~($N$,\,$J$\,$=$1,\,3/2--1,\,1/2)}. These lines trace an extended but thin layer
(\mbox{$A_{\rm V}\simeq$3--6\,mag} or \mbox{$\sim$10$^{16}$\,cm})  of molecular gas  at high \mbox{thermal pressure},
\mbox{$P_{\rm th}= n_{\rm H} \cdot T_{\rm k} \approx 10^7-10^9$ cm$^{-3}$\,K}, 
associated with the far ultraviolet (FUV) irradiated surface of \mbox{OMC-1}. 
The intense FUV radiation field, emerging from  massive stars in the \mbox{Trapezium}
cluster, heats, compresses and photoevaporates the cloud edge. It also  triggers the formation of specific reactive molecules such as CH$^+$.
 We find  that the \mbox{CH$^+$~($J$\,$=$\,1--0)} emission spatially correlates with the
 flux of FUV photons impinging the cloud: \mbox{$G_0$ from $\sim$10$^3$ to $\sim$10$^5$}. This correlation is supported by constant-pressure photodissociation region (PDR) models in the parameter space  \mbox{$P_{\rm th}/G_0\approx[5\cdot10^3-8\cdot10^4]$ cm$^{-3}$\,K} 
 where many observed PDRs seem to lie. The  \mbox{CH$^+$\,($J$\,$=$\,1--0)} emission spatially correlates with the  extended infrared emission from  \mbox{vibrationally excited  H$_2$\,($v$\,$\geq$\,1)}, and with that of [\CII]\,158\,$\upmu$m and \mbox{CO~$J$\,$=$\,10--9}, all emerging from \mbox{FUV-irradiated gas}. These  correlations link the presence of CH$^+$  to the availability  of C$^+$ ions and of \mbox{FUV-pumped} \mbox{H$_2$\,($v$\,$\geq$\,1)} molecules. We conclude that the parsec-scale CH$^+$ emission and narrow-line (\mbox{$\Delta$v\,$\simeq$\,3\,km\,s$^{-1}$}) \mbox{mid-$J$ CO}   emission arises from extended PDR gas and not from fast shocks. PDR line tracers are the smoking gun of the  stellar feedback from
young  massive stars. The PDR cloud surface component in \mbox{OMC-1}, with a  mass density of \mbox{120--240\,$M_{\odot}$\,pc$^{-2}$}, represents
$\sim$5\%~to $\sim$10\%\, of the total gas mass, however, it dominates the emitted line luminosity; the average \mbox{CO~$J$\,$=$\,10--9} surface luminosity in the mapped region being $\sim$35 times brighter than that of  \mbox{CO~$J$\,$=$\,2--1}. These results provide insights into the source of submillimeter CH$^+$ and \mbox{mid-$J$ CO} emission from distant star-forming galaxies.}

   \maketitle
%
%________________________________________________________________

\section{Introduction}  

Massive stars ($>$8\,$M_{\odot}$) dominate the injection of radiative energy
into the interstellar medium (ISM) through ultraviolet (UV) photons, and  of mechanical energy through \mbox{stellar winds}, supernova explosions, and merger encounters  \citep{Beuther07,Zinnecker07,Tan14,Krumholz14}. \mbox{Massive stars} are born inside dense giant molecular cloud (GMC) cores
\mbox{($n_{\rm H}$\,$\gtrsim$\,10$^5$\,cm$^{-3}$)}. Protostars of different masses develop inside these star-forming sites \mbox{\citep[e.g.,][]{McKee07}}. Their outflows shock the ambient cloud, heating and compressing the molecular gas around them to high temperatures and densities.  These young protostellar systems emit, on spatial scales of $\sim$0.05\,pc, high infrared (IR) luminosities, as well as  vibrationally and rotationally excited H$_2$, CO, and H$_2$O lines from shocked gas   \citep[e.g.,][and references therein]{vanDishoeck11}. At the scales of an entire GMC,
however, from several to hundred parsec most of the dust continuum  and gas line luminosity do not arise from individual protostars   but from the extended cloud component. 

Once a new O-type star or a massive star cluster is formed, the energy and momentum injected by photoionization, radiation pressure, and  stellar winds, ionize and erode the natal molecular cloud, creating  \HII~regions  and expanding bubbles \mbox{\citep[e.g.,][]{Krumholz14,Rahner17,Haid18}}. Photodisociation regions (PDRs)  develop at the interfaces between the hot ionized gas and the cold molecular gas, and more generally, at any slab of neutral gas (meaning that \mbox{hydrogen} atoms are not ionized) illuminated by stellar far-UV (FUV) photons
with energies in the range \mbox{$5\lesssim E< 13.6$\,eV} \mbox{\citep[e.g.,][]{Hollenbach97,Goico16}}. 
 
A strong FUV radiation field\footnote{$G_0$\,$=$\,1 equals to 1.6$\cdot$10$^{-3}$~erg\,cm$^{-2}$\,s$^{-1}$,  the  FUV flux (integrated from $\sim$912\,\AA~to $\sim$2400\,\AA) in the solar neighbourhood \citep{Habing68}.} 
($G_0$$>$10$^3$)  induces a plethora of dynamical effects \citep[e.g.,][]{Hill78,Hosokawa06,Wareing18,Bron18} and chemical
changes in the cloud \citep[e.g.,][]{Hoger95,Cuadrado15,Cuadrado16,Cuadrado17,Nagy17,Goico17}. This stellar feedback is not limited to the close vicinity of massive stars, but it can determine the gas physical conditions at scales of several parsec \citep[e.g.,][]{Stacey93, Herrmann97,Goi15} 
and drive the evolution of the natal cloud itself.
Finding observational tracers of the radiative and \mbox{mechanical} feedback from massive stars is relevant not only for local studies, but also in the more general framework of star formation in galaxies.  There it is not easy, at high red-shifts not even possible, to disentangle where the observed line emission is coming from: embedded star-forming sites, nuclear outflows, quiescent molecular clouds, diffuse halos, etc.

The brightest  FUV radiation line diagnostic is the  [\CII]\,158\,$\upmu$m fine-structure line of ionized carbon  C$^+$ \mbox{\citep[][]{Dalgarno72}}. With an ionization potential of 11.3\,eV,  C$^+$ can also be abundant in both the hot ionized and the cold atomic gas \mbox{\citep[e.g.,][]{Pineda13}}. Hence,  it is not always trivial to delimitate the origin
of the [\CII]\,158\,$\upmu$m line  and exploit its full diagnostic power \mbox{\citep[e.g.,][]{Pabst17}}. In consequence, it is important to have diverse observational probes of stellar feedback processes.

\subsection{Reactive ions as tracers of harsh interstellar conditions}

Reactive ions  such as CH$^+$, OH$^+$, or SH$^+$ are among the first molecules to form in initially atomic gas.
That is, \mbox{$x$(H atoms)\,$>$\,$x$(H$_2$ molecules)}, where $x$ refers to the species abundance with respect to H nuclei \citep[for a review see][]{Gerin16}. In \mbox{FUV-irradiated} neutral gas, carbon is mainly in the form of C$^+$ \citep[][]{Goldsmith12,Gerin15}. The formation of CH$^+$ depends on the flux of FUV photons, the gas temperature, and the abundance of vibrationally excited H$_{2}$\,($v$\,$\geq$\,1); either  \mbox{collisionally} excited or FUV-pumped \citep[][]{Sternberg_1995,Agundez10}.
 The main gas-phase reaction producing detectable quantities of CH$^+$ is:\\
\begin{equation}
\mbox{C$^+$~+~H$_2$\,($v$)~$\rightarrow$~CH$^+$~+~H},
\end{equation}\label{eq:chp}where $v$ refers to the specific vibrational level of H$_2$. This reaction has a very high endothermicity, \mbox{$\Delta E/k$\,$\simeq$\,4300~K}, if \mbox{$v$=0}. Thus, one would expect negligible CH$^+$ abundances in molecular clouds 
(where $T_{\rm k}$$\ll$\,$\Delta E/k$). \mbox{Reaction\,(1)}, however, becomes exothermic and fast for $v$\,$\geq$\,1 \citep{Hierl97,Zanchet13}, with the first H$_2$ vibrational level  lying  at \mbox{$E_{v=1}/k\,\simeq$\,5987~K}. Hence, compared to other molecules, 
CH$^+$ is expected to be abundant  only in high-temperature \mbox{FUV-irradiated} gas \mbox{\citep[e.g.,][]{Black98,Gerin16}}

Once formed, CH$^+$ is a very reactive ion; the timescale for its destruction in reactions with H$_2$ molecules or H atoms is comparable to, or shorter than, that for non-reactive collisions (elastic or inelastic). In dense  gas, 
\mbox{$n_{\rm H}$\,$\gtrsim$\,10$^5$\,cm$^{-3}$}, the lifetime of CH$^+$ is so short, a few hours, that the molecule may form and be destroyed without experiencing many non-reactive collisions with other species \citep{Black98, Nagy13, Godard13,Goico17}.

 CH$^+$ was one of the first  interstellar molecules
detected in the 1930s  \citep{Douglas41}. Due to the high \mbox{endothermicity}
of \mbox{reaction\,(1)} and limited  H$_2$\,($v$\,$\geq$\,1) column densities  in low-density gas, explaining the presence of CH$^+$ in diffuse clouds (typically with $G_0$$\approx$1, $n_{\rm H}$$\lesssim$100~cm$^{-3}$ and $T_{k}$$\lesssim$100~K) has been a long standing problem in \mbox{astrochemistry}. At such low densities, the \mbox{CH$^+$ $J$\,$=$\,1--0} line can only be detected
in \mbox{\textit{absorption}} against bright submillimeter (submm) continuum background  sources \citep{Falgarone10b,Godard12}. Low-velocity shocks  \citep{Elitzur78,Draine86,PdF86} and,
in particular, intermittent turbulence dissipation in a magnetized medium \citep{Falgarone95,Joulain98,Godard09,Godard14} are the most promising scenarios to explain the presence of CH$^+$ in diffuse gas. These theories 
are based in a \mbox{local} enhancement of the gas heating
 that raises the temperature to \mbox{$T_{\rm k}$\,$\gtrsim$\,1000~K} in  shear structures or vortices of only 
a few hundred~AU. At these high temperatures, and enhanced velocity drift between ions and
neutrals, \mbox{reaction\,(1)} becomes efficient.
Other, perhaps less succesful, scenarios include the turbulent mixing between the warm and the cold neutral phases of the ISM \citep{Lesaffre07} and the presence of extended warm H$_2$   \citep{Valdivia17}.

\begin{figure}[t]
\centering   
\includegraphics[scale=0.60, angle=0]{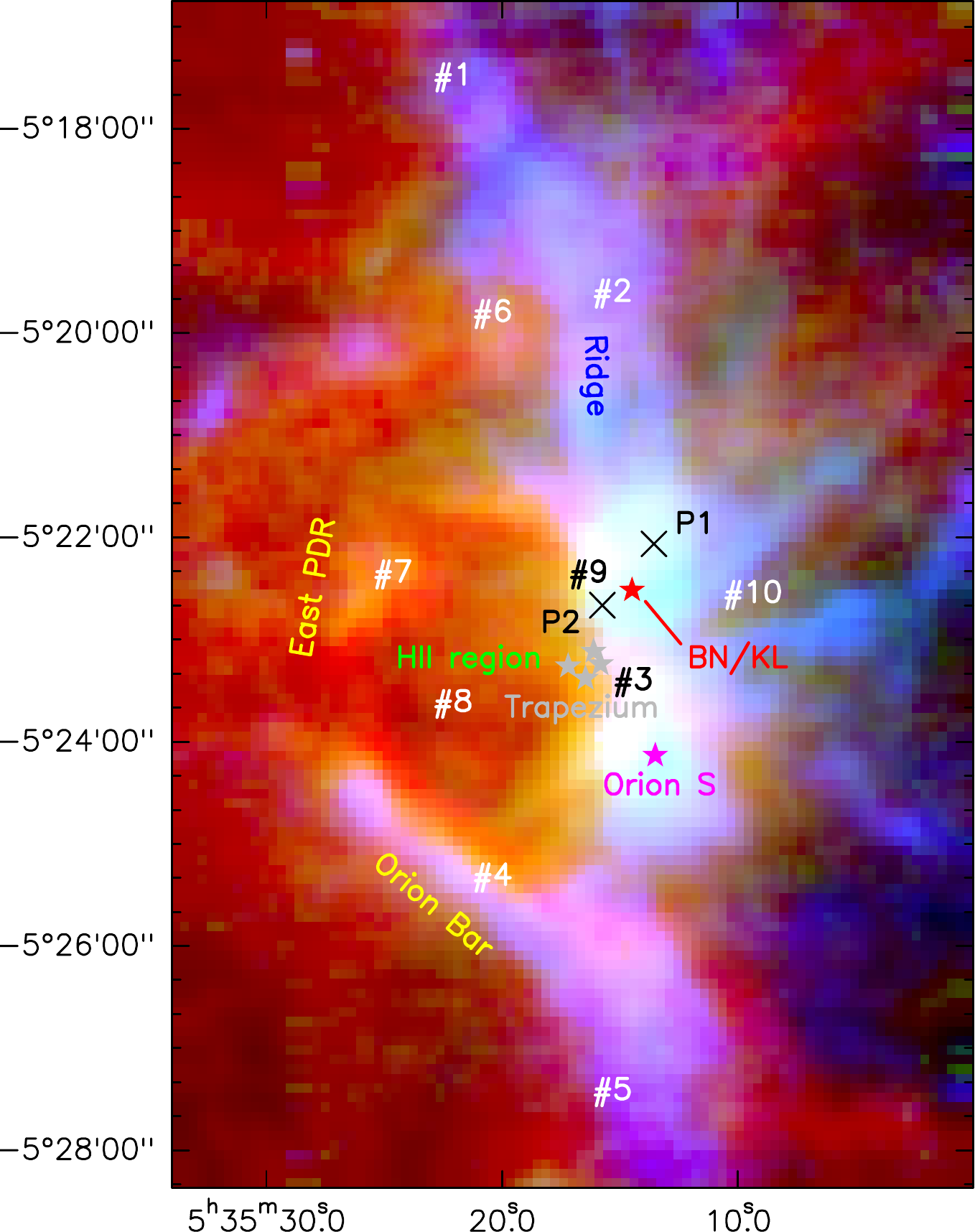}\\
\caption{RGB multitracer view of \mbox{OMC-1} at $\sim$12$''$ resolution. 
\textit{Red}:~[\CII]\,158\,$\upmu$m emission from FUV-illuminated surface
of the molecular cloud 
\citep[][]{Goi15}. \textit{Green}:~HCO$^+$ ($J$\,$=$\,3--2) from warm and dense  gas. \textit{Blue}: C$^{18}$O ($J$\,$=$\,2--1) from colder and more \mbox{FUV-shielded} gas. Main regions in \mbox{OMC-1} are labeled and representative positions discussed
in the text are indicated with numbers.}\label{fig:rgb}
\end{figure}

The detection of far-IR (FIR) and submm CH$^+$ rotational line \textit{emission} from strongly \mbox{FUV-irradiated}
\mbox{($G_0$\,$>$\,10$^3$)} and dense PDRs \mbox{($n_{\rm H}$\,$\gtrsim$\,10$^{5}$\,cm$^{-3}$)} 
\citep{Cernicharo97,Naylor10,Nagy13,Pilleri14,Joblin18}  suggested that the 
more intense FUV field and much higher densities compared to diffuse clouds enhances the abundance of H$_2$\,($v$\,$\geq$\,1) and also the  CH$^+$ production.
Indeed, PDR models using updated state-to-state chemical rates for \mbox{reaction\,(1)}
emphasised the role of \mbox{FUV-pumped H$_2$\,($v$\,$\geq$\,1)}  to explain the presence of CH$^+$ 
 in bright and dense PDRs  \citep{Agundez10,Faure17}. Later observations
showed that the \mbox{CH$^+$} emission can be relatively extended near young massive stars 	\mbox{\citep{Morris16}}.
In~addition, \textit{Herschel} also  detected  CH$^+$  broad line-emission from high- 
and low-mass protostars \citep{Falgarone10,Benz10,Benz16}. The preferred interpretation
is that CH$^+$ forms in irradiated magnetized shocks:
the FUV-irradiated walls of a protostellar outflow.
Last but not least, \mbox{CH$^+$~($J$\,$=$\,1--0)} emission has been recently detected by ALMA in star-forming galaxies at redshift of about two \citep{Falgarone17}.

 \begin{table*}[h]
 %\small
\caption{\label{table:freqs}Transition frequencies, upper energy levels, Einstein coefficients, and critical densities discussed in this work (sorted by frequencies).} 
\centering
\begin{tabular}{lcrrcrrrr@{\vrule height 9pt depth 5pt width 0pt}}
\hline\hline
  Species      &        Transition                      &  Frequency  & $E_{\rm u}/k$ & $A_{\rm ul}$ & $n_{\rm cr}^{a}$ & Telescope & HPBW & \textit{Herschel} \\
        &                    &    (GHz)    &   (K)         & (s$^{-1}$)   & (cm$^{-3}$)                &  /Instrument         & ($''$) & ObsID \\\hline
CH$^+$       & $J$\,$=$\,5--4                  &  4155.8719  &   599.6       &    1.05                 & 8$\cdot$10$^{10}$  &   \textit{Herschel}/PACS    & $\sim$12 & 342218571/2  \\
CH$^+$       & $J$\,$=$\,4--3                  &  3330.6297  &   400.1       &    0.53	               & 2$\cdot$10$^{10}$  &   /PACS         & $\sim$12 & 342218571/2\\
CH$^+$       & $J$\,$=$\,3--2                  &  2501.4404 &    240.3       &    0.22                 & 5$\cdot$10$^{9}$  &   /PACS         & $\sim$12 & 342218571/2 \\\hline
C$^+$ ($\dagger$)    & $^{2}P_{3/2}$-$^{2}P_{1/2}$ & 1900.5369  &   91.3 &  2.3$\cdot$10$^{-6}$        & 4$\cdot$10$^{3}$   &   \textit{Herschel}/HIFI         &   12.0 & 1342250412/4/5 \\
$o$-H$_2$O   & 3$_{12}$-2$_{21}$               & 1153.1268   &   215.2       &  2.7$\cdot$10$^{-3}$    & $\sim$6$\cdot$10$^{7}$ &  /HIFI         & 19.8 & 1342250465 \\
$^{12}$CO& $J$\,$=$\,10--9                     & 1151.9854   &   304.2       &  1.0$\cdot$10$^{-4}$    & 1$\cdot$10$^{6}$     &  /HIFI         & 19.8 & 1342250465 \\
HCN          & $J$\,$=$\,13--12                & 1151.4491   &   387.0       &  7.6$\cdot$10$^{-2}$    & 6$\cdot$10$^{9}$     & /HIFI         & 19.8 & 1342250465 \\
CH$^+$       & $J$\,$=$\,1--0                  &  835.1375   &    40.1       &  6.2$\cdot$10$^{-3}$    & 1$\cdot$10$^{8}$     &  /HIFI         & 27.3 & 1342250217 \\
HCO$^+$      & $J$\,$=$\,6--5               &  535.0616   &    89.9       &  1.2$\cdot$10$^{-2}$       & 3$\cdot$10$^{7}$     &  /HIFI         & 42.6 & 1342244307  \\
CH           & $N,J$=1,3/2--1,1/2             &  532.7933   &    25.7       &  4.1$\cdot$10$^{-4}$     & $\sim$3$\cdot$10$^{8}$             &  /HIFI         & 42.8 & 1342244307  \\
HCN          & $J$\,$=$\,6--5                  &  531.7164   &    89.3       &  7.2$\cdot$10$^{-3}$    & 7$\cdot$10$^{8}$     &  /HIFI         & 42.8 & 1342244307  \\
$A$-CH$_3$OH &  11$_{0,11}$--10$_{0,10}$ &  531.3193   &   153.1       &  6.8$\cdot$10$^{-4}$          & $\sim$5$\cdot$10$^{6}$ &  /HIFI         & 42.9 & 1342244307  \\\hline
HCO$^+$      & $J$=3-2                  &  267.558    &    25.7       &  1.4$\cdot$10$^{-3}$           & 4$\cdot$10$^{6}$ 	 &  IRAM30m/HERA         & 9.2 & \\
$^{12}$CO ($\ddagger$)   & $J$\,$=$\,2--1     &  230.538    &    16.6       &  6.9$\cdot$10$^{-7}$     & 1$\cdot$10$^{4}$ 	 &  /HERA         & 10.7 & \\
$^{13}$CO ($\ddagger$)   & $J$\,$=$\,2--1      & 220.399     &    15.9       &  6.1$\cdot$10$^{-7}$    & 1$\cdot$10$^{4}$  	 &  /HERA         & 11.2 & \\
C$^{18}$O    & $J$\,$=$\,2--1                  & 219.560     &    15.8       &  6.0$\cdot$10$^{-7}$    & 1$\cdot$10$^{4}$  	 &  /HERA         & 11.2 & \\
\hline
\end{tabular}
\tablefoot{$\dagger$Published in \citet{Goi15}. 
$\ddagger$Published in \citet{Berne14}.
$^a$Critical density, $A_{\rm ul}/\gamma_{\rm ul}(T_{\rm k})$, in collisions with H$_2$ at 100\,K.}
%\textbf{References.}...}
\end{table*}

 In summary, CH$^+$ has been historically considered  a unique, sometimes \mbox{exotic}, tracer of  harsh conditions \citep[][]{Black98}. \mbox{Despite} many theoretical efforts, there have not been any observational study on how the extended CH$^+$ \textit{emission} relates with the gas physical conditions and with the flux of FUV photons.
 In this paper we present the detection of \mbox{large-scale} 
 \mbox{CH$^+$~($J$\,$=$\,1--0}) emission toward Orion molecular cloud~1 \mbox{(OMC-1)}, complemented with velocity-resolved maps of other key diagnostics of the warm molecular gas that help to understand the role of stellar radiation at large spatial scales.
The paper is organized as follows. In Sect.~\ref{Sect:Observations} we 
introduce the region and describe the observations. In Sect.~\ref{Sect:Results} we present the main observational results. %(integrated intensity and velocity channel maps). 
In Sect.~\ref{Sect:Analysis} we analize the maps, focusing  on the CH$^+$ emission, and in Sec.~\ref{Sect:Discussion} we model and discuss the role of FUV radiation.

\section{Observations and Data Reduction}\label{Sect:Observations}

\subsection{\mbox{OMC-1} in Orion~A}

OMC-1, in the Orion~A  complex, lies behind the iconic Orion Nebula (M42) and the 
massive stars $\theta^1$\,Ori  of the Trapezium cluster 
\citep[with spectral types from O7V to B3V; e.g.,][]{Simon06} roughly at the center of the Orion nebula cluster (ONC) \mbox{\citep[e.g.,][]{Zuckerman73,Genzel89,Odell01}}.
At a distance\footnote{Recent estimations, from \textit{Gaia} DR2, suggest
that the distance to ONC is lower,
386$\pm$3\,pc \citep[e.g.,][and references therein]{Grossschedl18}. Here we still adopt
$\sim$414\,pc to be consistent with the  luminosities computed
in our previous works \citep[e.g.,][]{Goipacs15,Goi15}.} of $\sim$414\,pc \citep{Menten07}, \mbox{OMC-1}  is the closest region of on-going intermediate- and high-mass star formation. The two major star-forming sites
are the Becklin-Neugebauer/Kleinmann-Low region \citep[BN/KL;][]{Becklin67} and
 \mbox{Orion~South} (Orion~S; Fig.~\ref{fig:rgb}). 
BN/KL hosts a high-velocity wide-angle  outflow that contains the two brightest IR H$_2$ emission peaks in the sky: Peak~1~(P1) and Peak~2~(P2), produced by hot shocked gas \citep[marked with crosses in Fig.~\ref{fig:rgb}; see][]{Kwan76,Beckwith78,Gonzalez02,Goipacs15}. 
Both BN/KL and Orion~S show chemically rich spectra 
\citep[e.g.,][]{Blake87,Tercero10,Tahani16}.
\mbox{OMC-1} is directly exposed to the strong FUV radiation emitted by young hot stars in the Trapezium cluster, and in particular dominated by star \mbox{$\theta^1$ Ori C1}  \citep[\mbox{$M_{\star}$$\simeq$33.5\,$M_{\odot}$}; e.g.,][]{Karl18} located
at $\sim$0.3\,pc in front of the molecular cloud.
The FUV-irradiated surface of \mbox{OMC-1},  to a first approximation a large-scale face-on PDR,   copiously emits in FIR [\CII]\,158\,$\upmu$m and [\OI]\,63\,$\upmu$m fine-structure gas-cooling lines \citep{Stacey93,Herrmann97,Goi15}.

Velocity-resolved maps  over  tens of square arcmin are needed to constrain the kinematics and physical  properties of the extended, non star-forming gas in GMCs. A number of studies have provided such molecular line maps toward \mbox{OMC-1}. However, most of them focus on the lowest-energy rotational lines  that typically trace cold molecular gas \citep[e.g.,][]{Bally87,Goldsmith97,Hans97,Berne14,Kauffmann17}.
Indeed, the warm molecular gas, $T_k$\,$\approx$100\,K, predominantly emits in higher frequency lines.
\citet{Schmid-Burgk89} presented a pioneering velocity-resolved \mbox{CO~$J$\,$=$\,7--6} line map (\mbox{$\sim$48\,arcmin$^2$}) of \mbox{OMC-1} taken with \textit{KAO} at $\sim$100$''$ angular resolution. Improved angular resolution maps were later taken from ground-based telescopes in  \mbox{CO~$J$\,$=$\,7--6} \citep[\mbox{$\sim$8\,arcmin$^2$};][]{Wilson01} and up to \mbox{CO~$J$\,$=$\,8--7} \citep[\mbox{$\sim$30\,arcmin$^2$;}][]{Peng12}. The narrow CO line profiles ($\Delta$v$\lesssim$4~km\,s$^{-1}$) observed toward many positions   suggested
the widespread nature of warm (but not shocked) molecular gas. In addition, the spatial distribution
of the extended CN emission implied  an active photochemistry at the interface between the \HII~regions M42 and M43 and the molecular cloud \citep{Rodriguez98,Rodriguez01}.

\begin{figure*}[h]
\centering
\includegraphics[scale=0.125, angle=0]{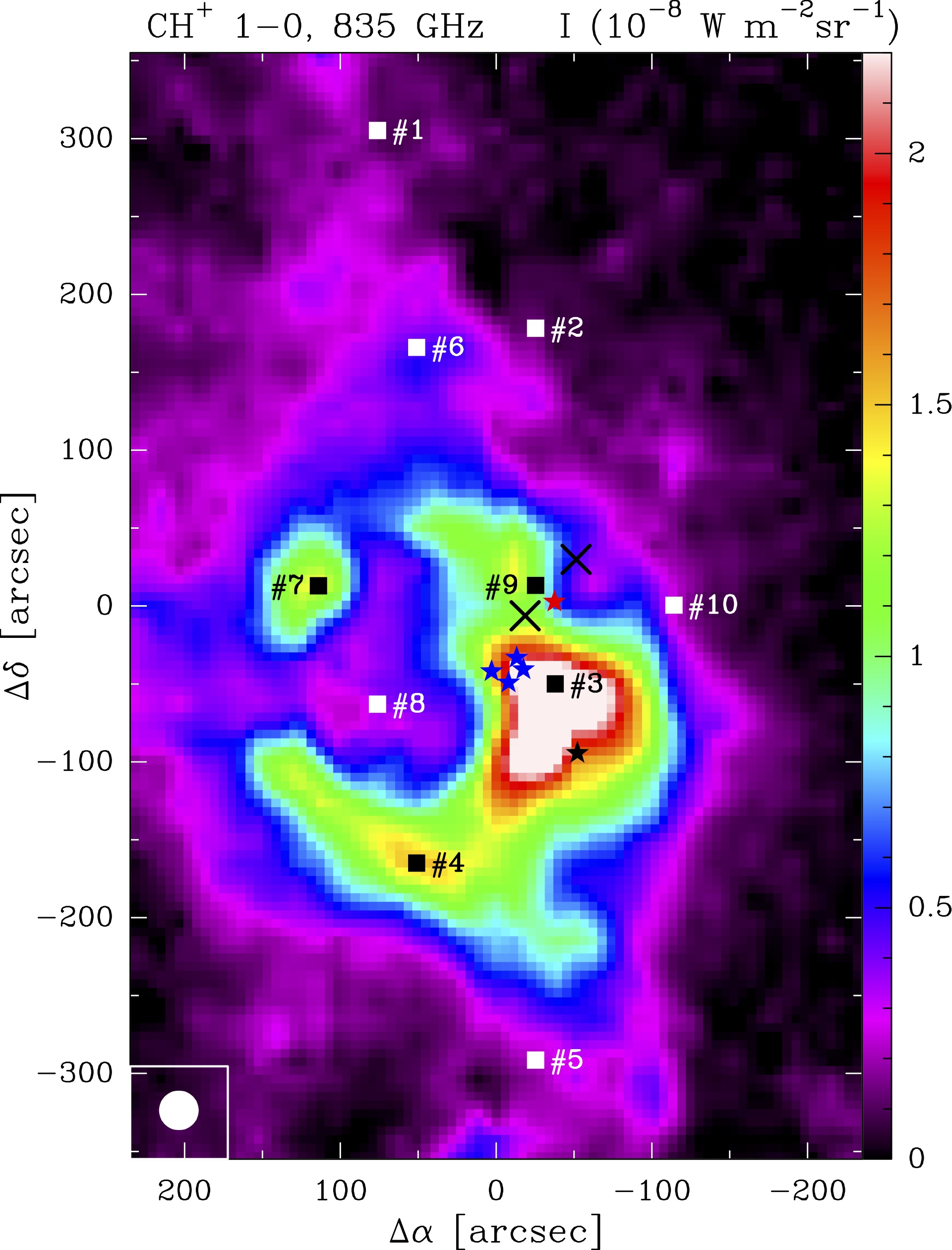}\hspace{0.5cm}
\includegraphics[scale=0.125, angle=0]{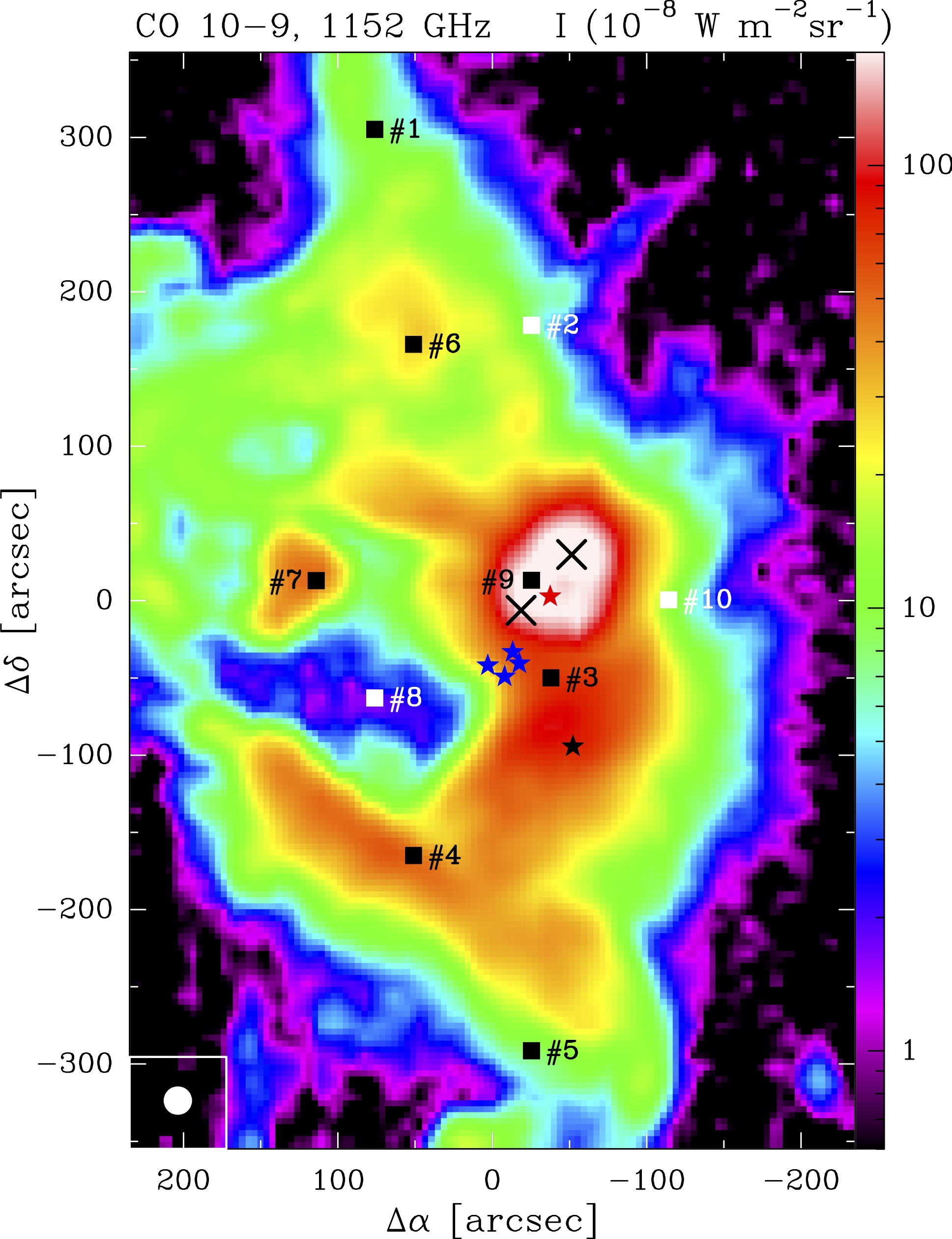}\vspace{0.5cm}
\includegraphics[scale=0.125, angle=0]{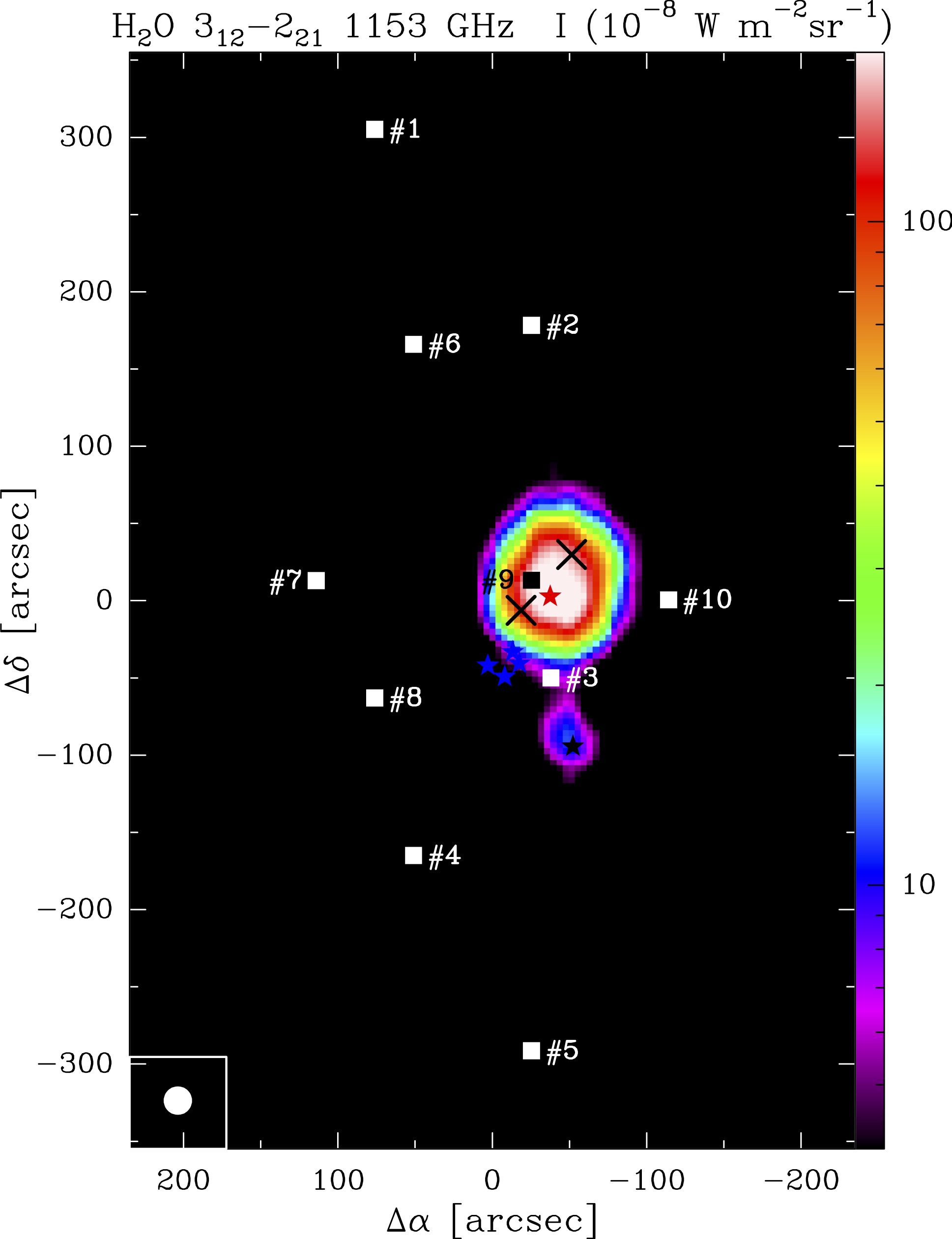}\hspace{0.3cm}
\includegraphics[scale=0.125, angle=0]{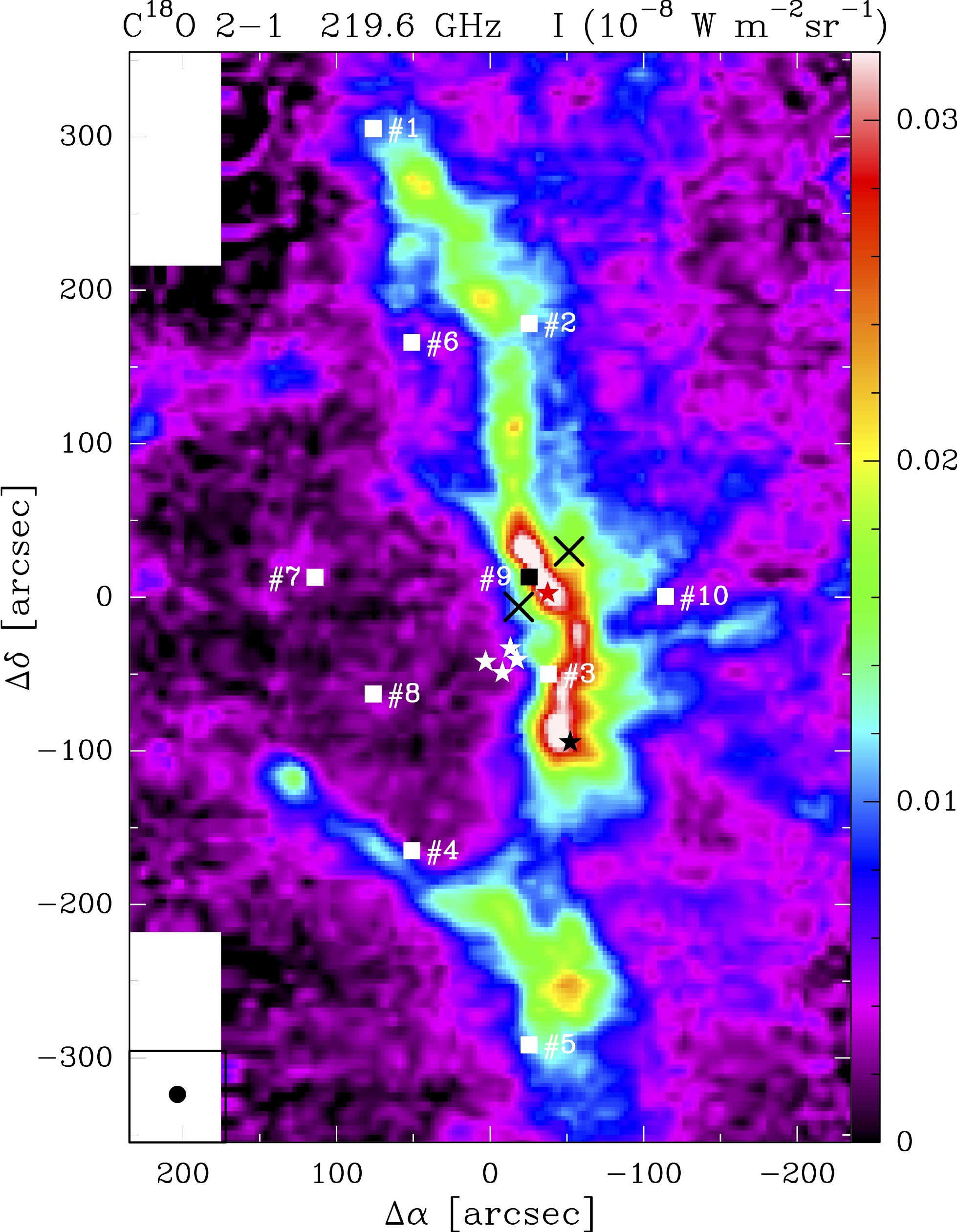}\\
\caption{\textit{Herschel}/HIFI and IRAM\,30m maps of different molecular emission lines toward \mbox{OMC-1}, all showing very different spatial distributions: 
\mbox{CH$^+$ $J$\,$=$\,1--0} (strongly FUV-irradiated gas), CO $J$\,$=$\,10--9 (extended warm gas), H$_2$O 3$_{12}$-2$_{21}$ (hot shocked gas), and C$^{18}$O $J$\,$=$\,2--1 (colder and more
\mbox{FUV-shielded} molecular gas). 
The color scale shows the  integrated line intensity in units of \mbox{W\,m\,$^{-2}$\,sr$^{-1}$}.
The native angular-resolution of each observation, the HPBW, is plotted in the bottom-left corner. 
Representative positions discussed in the text
are indicated with numbers.}\label{fig:Original_maps1}
\end{figure*}\clearpage

\subsection{Herschel/HIFI and PACS maps}

In this work we present new submm line maps of \mbox{OMC-1}
taken with the heterodyne instrument HIFI \citep{deGra10} on board \textit{Herschel}  space telescope \citep{Pilbratt10}. The maps have a size of 
\mbox{$\sim$85~arcmin$^2$} ($\sim$0.9\,pc\,$\times$1.4\,pc).
They belong to program \textit{OT1$\_$jgoicoec$\_$4}.
A [\CII]\,158\,$\upmu$m map from this program was first presented  in \citet{Goi15}.
\citet{Morris16} previosulsy presented smaller 
\mbox{($\sim$12~arcmin$^2$)} maps around Orion~BN/KL based on observations performed during the regular instrument calibration activities.
At the observed frequencies, HIFI employed  Superconductor-Insulator-Superconductor (SIS) mixers in two orthogonal polarizations. We used the Wide Band Acousto-Optical Spectrometer providing a spectral resolution of 1.1\,MHz and a bandwidth of 4~GHz.
We obtained on-the-fly (OTF) maps with half-beam sampling, and angular resolutions 
(half-power beam-widths, HPBW) ranging from 43$''$ in Band~1, to 20$''$ in Band 5. A reference-OFF position at 9$'$ was observed.
Total map integration times were $\sim$1.4\,hr at $\sim$535\,GHz, $\sim$2.2\,hr at $\sim$835\,GHz, and $\sim$4.6\,hr at $\sim$1152\,GHz. 
 We used the main beam temperature scale ($T_{\rm mb}$ in K) as opposed to $T_{\rm A}^{*}$.
 For semi-extended  emission sources, not infinite but larger than HIFI's HPBW, of uniform brightness temperature ($T_{\rm b}$), $T_{\rm mb}$
is the most appropriate intensity scale ($T_{\rm mb}$$\simeq$$T_{\rm b}$).
The achieved rms noise was typically 50\,mK at $\sim$531\,GHz,
$\sim$100\,mK at $\sim$835\,GHz and $\sim$1152\,GHz, 
and $\sim$800\,mK per  channel.

As part of a related \textit{Herschel}  program (\textit{KPGT$\_$ebergin$\_$1}), 
we used the PACS spectrometer \citep{Poglitsch10} to map a  $\sim$2$'$$\times$2$'$ region around Orion~BN/KL 
outflows and the Trapezium cluster area.  
These  $\sim$12$''$ angular resolution observations cover the shorter \mbox{$\lambda$=70--94~$\upmu$m} and \mbox{108--190~$\upmu$m} wavelength ranges where CH$^+$ rotationally excited lines appear. PACS map data calibration and line flux extraction are described in \citet{Goipacs15}.
Here we present the detection of  bright CH$^+$ \mbox{$J$\,$=$\,3--2}, \mbox{4--3},
and \mbox{5--4} lines toward a position  close to the Trapezium cluster: \mbox{position $\#$3}
(see Figs.~\ref{fig:rgb} and \ref{fig:Original_maps1}).

\subsection{IRAM\,30m maps}

We complemented our study with the analysis of unpublished  \mbox{C$^{18}$O ($J$\,$=$\,2--1)} and \mbox{HCO$^+$ ($J$\,$=$\,3--2)}  line maps, obtained by us with the multi-beam receiver HERA at the IRAM-30m  
telescope (Pico Veleta, Spain).
The HPBW ranges from $\sim$9$''$ to 11$''$.
We used the versatile spectrometer VESPA as a backend providing 320\,kHz of the spectral resolution (0.4\,km\,s$^{-1}$). These maps belong to a project to map a much larger area
of Orion~A in CO isotopologues. The data reduction is described
in \citet{Berne14}. The achieved rms noise is \mbox{$\simeq$0.2\,K (C$^{18}$O $J$\,$=$\,2--1)}
and $\simeq$1\,K (HCO$^+$ $J$\,$=$\,6--5) per  0.4\,km\,s$^{-1}$ channel.

\textit{Herschel} and \mbox{IRAM-30m} data cubes were processed with GILDAS. 
\mbox{Figure~\ref{fig:Original_maps1}} and also \mbox{Figs.~\ref{fig:Original_maps2}} and \ref{fig:Original_maps3} in the \mbox{Appendix} show the resulting emission maps
 at their native angular resolutions.  Offsets in arcsec  are given with respect
to the  maps center at \mbox{$\alpha_{2000}$:~5$^{\rm h}$35$^{\rm m}$17.0$^{\rm s}$}, \mbox{$\delta_{2000}$:~$-5^o$22$'$33.7$''$}.
 Table~\ref{table:freqs} summarizes the main spectroscopic parameters of the
observed lines.
To determine the total luminosity emitted by each line, we converted the integrated line intensity
 maps from \mbox{K\,km\,s$^{-1}$} units  to line surface brightness in \mbox{W\,m$^{-2}$\,sr$^{-1}$}. The conversion$^3$ scales with the cube of the line frequency. Hence, it greatly affects the comparison of FIR, submm and mm lines. Finally, to match the \mbox{CH$^+$ $J$\,$=$\,1--0} (\mbox{HCO$^+$ $J$\,$=$\,6--5}) angular resolution and carry out a combined analysis (e.g., line ratio maps), we also convolved the maps  to an uniform resolution of 27$''$ (43$''$).

\section{Results}\label{Sect:Results}
\subsection{Global spatial distribution of different  line tracers}

Figure~\ref{fig:rgb} shows a composite image of the integrated line emission in:
[\CII]\,158\,$\upmu$m (red) from the \mbox{FUV-irradiated} surfaces of \mbox{OMC-1};
HCO$^+$\,$J$\,$=$\,3--2 (green) from warm and dense molecular gas; and C$^{18}$O\,$J$\,$=$\,2--1 (blue) from colder and more \mbox{FUV-shielded} gas, mostly tracing the cloud interior \mbox{\citep[e.g.,][]{Bally87,Hacar17,Hacar18}}.
The strong flux of FUV photons from the Trapezium  illuminates  the surface of the molecular cloud and creates bright and dense PDRs. The most noticeable  \mbox{edge-on} \mbox{\HII/H/H$_2$}  interfaces are the \mbox{Orion Bar} and the \mbox{East} PDRs, but the entire illuminated skin of the cloud, bright in [\CII]\,158\,$\upmu$m emission (Fig.~\ref{fig:maps_Cp_CHp_CO_compa}), can be seen as a face-on  PDR \mbox{\citep[e.g.,][]{Stacey93,Goi15}}. 

\begin{figure}[t]
\centering
\includegraphics[scale=0.097, angle=0]{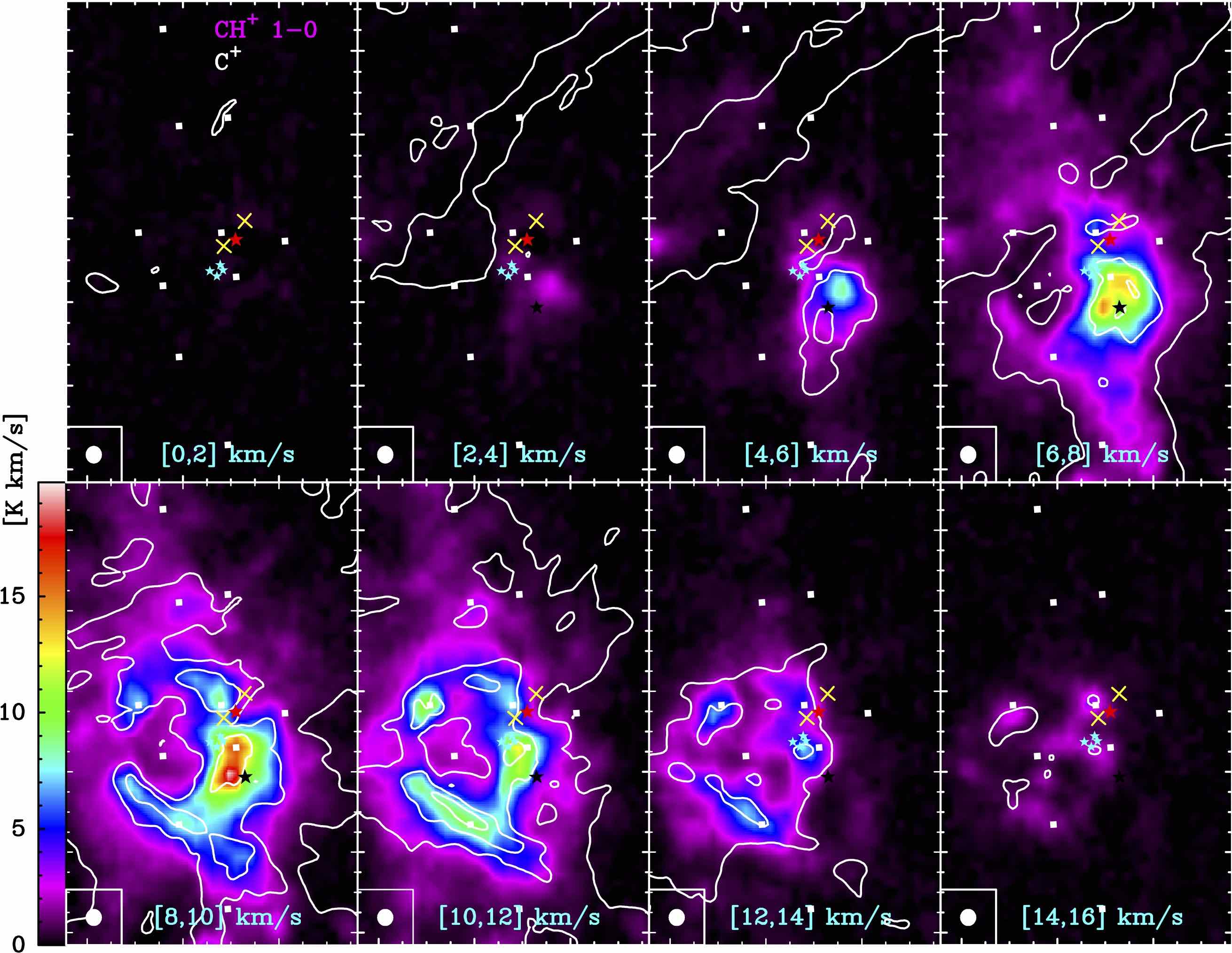}\vspace{0.35cm}
\includegraphics[scale=0.097, angle=0]{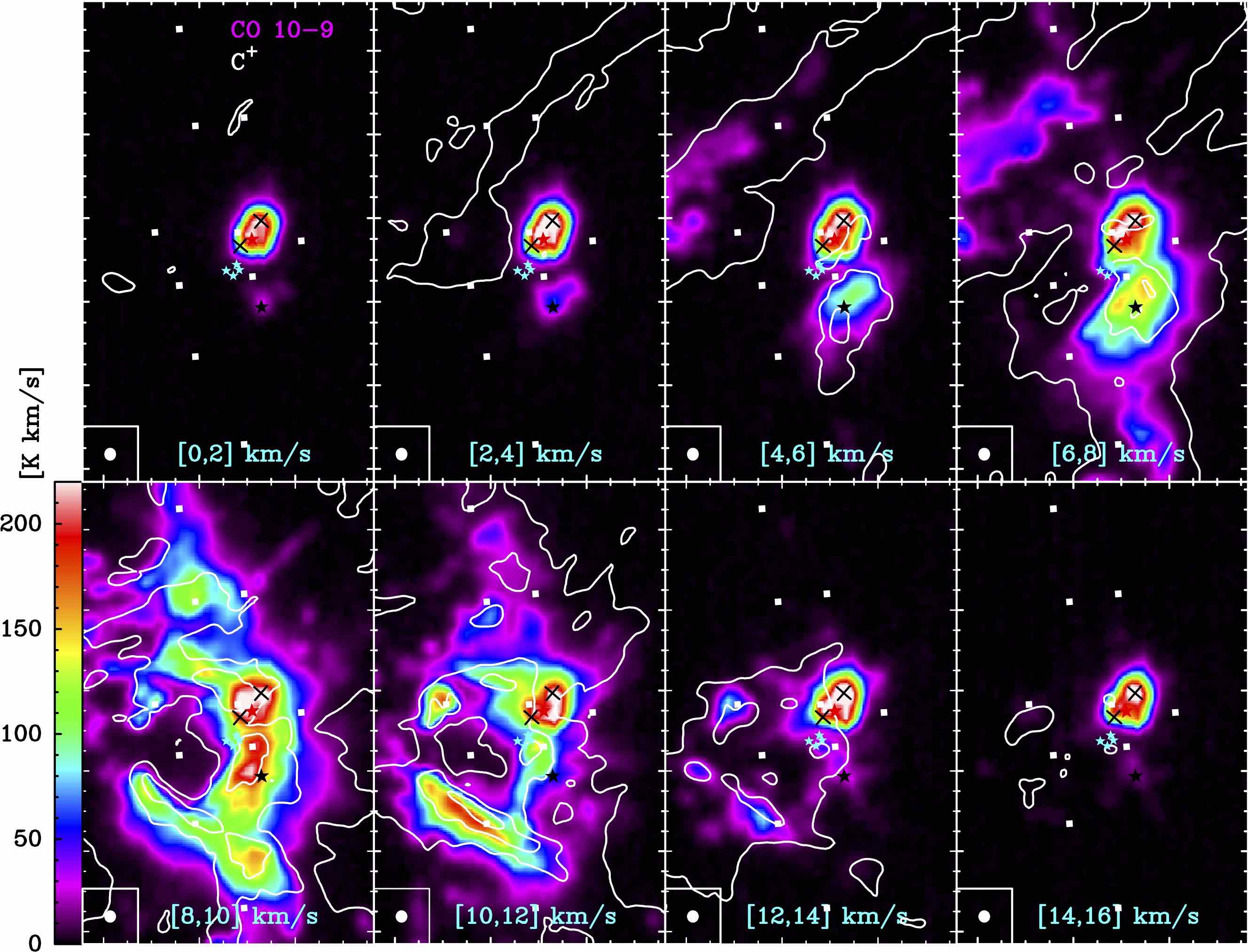}
\caption{CH$^+$ $J$\,$=$\,1--0 (\textit{top}) and CO $J$\,$=$\,10--9 (\textit{bottom}) velocity channel maps
from v$_{\rm LSR}$=0 to 16~km\,s$^{-1}$ in steps of 2~km\,s$^{-1}$. White contours show the
[\CII]\,158\,$\upmu$m line emission 
from 50 to 500~K\,km\,s$^{-1}$ in steps of 150~K\,km\,s$^{-1}$ \citep[][]{Goi15}.}\label{fig:vel_channel-chp}
\end{figure}

Figure~\ref{fig:Original_maps1} shows maps of the CH$^+$~$J$\,$=$\,1--0, CO~$J$\,$=$\,10--9, H$_2$O~3$_{12}$--2$_{21}$, and C$^{18}$O~$J$\,$=$\,2--1 lines. The four emission lines show remarkably different spatial distributions, emphasising the distinctive diagnostic power of different molecular species and lines.  Although much less abundant than C$^+$, the spatial distribution of  CH$^+$~($J$\,$=$\,1--0)  is similar to that of [\CII]\,158\,$\upmu$m (Figs.~\ref{fig:vel_channel-chp} and \ref{fig:maps_Cp_CHp_CO_compa}). 
In particular, the CH$^+$~$J$\,$=$\,1--0  intensity peaks appear very close to those of [\CII]\,158\,$\upmu$m. Hence, CH$^+$ also traces FUV-irradiated 
gas at the surface of the  molecular cloud.  However, while the brightest regions of [\CII]\,158\,$\upmu$m emission show uniform intensities, the \mbox{CH$^+$~$J$\,$=$\,1--0} line is brighter closer to the Trapezium, where the stellar FUV  flux is stronger. Of all the observed species, only the CH emission shows a similar spatial distribution to that of CH$^+$ (Fig.~\ref{fig:Original_maps2}). This is a consequence of their very related chemistry in PDR gas \citep[see][]{Morris16}.

Despite the high excitation requirements of the CO  rotational level $J$=10 ($E_{\rm u}$/$k$=304\,K), thus only populated at warm gas temperatures, the  \mbox{CO\,$J$\,$=$\,10--9} line emission is very extended. 
 Indeed, it is more spatially extended than that of \mbox{C$^{18}$O\,($J$\,$=$\,2--1)} 
 (Fig.~\ref{fig:Original_maps1}). Unlike CH$^+$ and CH,  the  CO\,$J$\,$=$\,10--9  emission peaks toward BN/KL. Even if it locally peaks  toward shocked gas 
in outflows, most of the line luminosity  arises from the 
extended cloud component. In addition, outside the BK/KL region, the spatial distribution of the  \mbox{CO\,$J$\,$=$\,10--9}   emission approximately follows that of CH$^+$ and [\CII]\,158\,$\upmu$m (see \mbox{Fig.~\ref{fig:vel_channel-chp}}).
 Together with the narrow CO  line-widths, $\Delta$v\,$\simeq$\,3\,km\,s$^{-1}$, this suggests that most of the \mbox{CO\,$J$\,$=$\,10--9} emission arises also from the warm \mbox{FUV-irradiated} surface of \mbox{OMC-1}.
 
Among the species studied in this work, 
the detection of broad (\mbox{$\Delta$v\,$>$\,30\,km\,s$^{-1}$}) line-wing \mbox{H$_2$O (3$_{12}$-2$_{21}$)} emission implies the presence of shocked gas activity \citep[e.g.,][]{vanDishoeck11,vD13}. Indeed, we only detect H$_2$O and CH$_3$OH rotationally excited lines toward Orion~S and BN/KL  (Figs.~\ref{fig:Original_maps1} and \ref{fig:Original_maps2}). Both species are abundant in the ice mantles that coat grains in cold dark clouds 
\citep[e.g.,][]{Gibb04}. After the onset of protostellar outflows, high-velocity shocks sputter these grain mantles and heat the gas to high temperatures. Both effects enhance the abundance of gas-phase H$_2$O and CH$_3$OH  \mbox{\citep[e.g.,][]{Draine95,Izaskun08}}. In \mbox{OMC-1}, 
the low- and high-velocity outflows from BN/KL  plunge into the ambient molecular cloud
\citep{Genzel89} producing  hot (from $T_{\rm k}$$\simeq$200 to 2000\,K) and dense 
(up to $n$(H$_2$)$\simeq$\mbox{10$^6$-10$^7$~cm$^{-3}$}) post-shocked gas \citep[e.g.,][]{Gonzalez02,Goipacs15}. In our HIFI maps, these extreme conditions can also be inferred from the moderately extended emission of the  \mbox{HCN~$J$\,$=$\,13--12} line (Fig.~\ref{fig:Original_maps3}), a rotational transition with a critical density 
close to 10$^{10}$\,cm$^{-3}$ (\mbox{$n_{\rm cr}$\,$=\,$$A_{\rm ul}/\gamma_{\rm ul}(T_{\rm k})$}, where
$\gamma_{\rm ul}$ is the collisional de-excitation rate coefficient in cm$^3$\,s$^{-1}$),  around
 BN/KL outflows. \mbox{Interestingly}, the observed \mbox{HCN to HCO$^+$~$J$\,$=$\,6--5} line intensity ratio is $\geq$2 toward BN/KL,  and $<$\,1 almost \mbox{elsewhere} \mbox{(see Fig.~\ref{fig:peaks}, right)}. This may
reflect a change in the chemistry between the extended PDR cloud component and
the shocked gas in BN/KL outflows. It may also reflect the much stronger mid-IR (MIR) radiation from
the BN/KL region that favors the radiative pumping of HCN through its vibrational levels
and enhances the high-$J$ rotational emission \citep[e.g,][]{Carroll81,Ziurys86}.
Finally, the maps show that both HCO$^+$ and HCN~$J$\,$=$\,6--5 lines display widespread emission outside the main star-forming sites (see Fig.~\ref{fig:Original_maps2}). This suggests that the gas density  of the extended cloud layers traced by HCO$^+$ and HCN~$J$\,$=$\,6--5 lines is 
moderately high.

\subsection{C$^+$, CO and CH$^+$ line luminosities: warm gas cooling}

For collisionally excited transitions and subsequent  optically thin emission, the
emitted line  \mbox{luminosity}\footnote{Line-luminosity ($L$)
computed in power units (W or $L_{\odot}$). The conversion from velocity-integrated   line intensities  $W$=$\int \Delta T_{\rm b}\, d$v\, (K\,km\,s$^{-1}$)  to line surface brightness $I$ (\mbox{in W\,m$^{-2}$\,sr$^{-1}$}) is $I$=2$k$\,$W$\,$\nu^3/c^3$, with $\nu$ the line frequency, and $\Delta T_{\rm b}$ the   
continuum-subtracted brightness temperature.} provides a measure of its gas cooling power. The [\CII]\,158\,$\upmu$m line dominates the cooling of FUV-irradiated gas. Indeed, it is the brightest of the observed lines toward \mbox{OMC-1} (\mbox{see Table~\ref{table:luminosity}}), with
a surface luminosity of \mbox{$L_{\rm CII}$/$A$\,$\simeq$\,140\,$L_{\odot}$\,pc$^{-2}$}  \citep{Goi15}. This is more than 3 orders of magnitude more luminous than the widely observed \mbox{low-$J$ CO} lines \citep[][]{Bally87,Berne14}. 
 Despite being a trace species, the luminosity emitted by the \mbox{CH$^+$ $J$\,$=$\,1--0} line in \mbox{OMC-1} is similar to that of \mbox{CO $J$\,$=$\,2--1}. Remarkably,  
the \mbox{CO\,$J$=10-9} line, with \mbox{$L_{\rm CO\,10-9}$/$A$\,$\simeq$\,4\,$L_{\odot}$\,pc$^{-2}$}, 
is 35 times brighter than the \mbox{CO\,$J$=2-1} line along the mapped region, and
it is more than a hundred times brighter toward specific dense PDRs like the Orion Bar.
This means that the
CO line cooling of  warm molecular gas is dominated by the \mbox{mid-$J$} lines.
In addition, the average $^{12}$CO/$^{13}$CO~$J$\,$=$\,2--1 line luminosity ratio in the map is $\sim$7; and it is lower than $\sim$15 in most of the region. This is about an order of magnitude lower than the $^{12}$C/$^{13}$C isotopic ratio in Orion \citep[$\sim$67;][]{Langer90} and implies that the $^{12}$CO~$J$\,$=$\,2--1 line is optically thick at $\sim$1\,pc$^2$ scales. On the other hand, the high $^{13}$CO/C$^{18}$O~$J$\,$=$\,2--1 luminosity ratio, $\simeq$9, indicates that both lines are optically thin at large scales. 

Only in shocked and post-shocked gas, such as H$_2$ Peak~1 and Peak~2 regions in BN/KL outflows, the hot molecular gas cooling is dominated by emission from FIR  high-$J$ CO ($J$\,$>$\,15) and water vapor lines, and from MIR H$_2$ lines \citep[e.g.,][]{Gonzalez02,Goipacs15}.

\begin{table}[t]
\caption{Total$^a$ line integrated luminosities in \mbox{OMC-1}. \label{table:luminosity}} 
\centering
\begin{tabular}{lcc@{\vrule height 10pt depth 5pt width 0pt}}
\hline\hline
Line & $L$ ($L_{\odot}$) & Surface $L$ ($L_{\odot}$\,pc$^{-2}$)  \\ \hline
[\CII] 158\,$\upmu$m    & 1.7$\cdot 10^{2}$     & 1.4$\cdot 10^{2}$ \\ 
CO $J$\,$=$\,10--9             & 5.1$\cdot 10^{0}$     & 4.0$\cdot 10^{0}$ \\ 
CO $J$\,$=$\,2--1              & 1.5$\cdot 10^{-1}$    & 1.2$\cdot 10^{-1}$ \\ 
CH$^+$ $J$\,$=$\,1--0         & 1.4$\cdot 10^{-1}$    & 1.1$\cdot 10^{-1}$ \\ 
$^{13}$CO $J$\,$=$\,2--1       & 2.1$\cdot 10^{-2}$    & 1.7$\cdot 10^{-2}$ \\ 
C$^{18}$O $J$\,$=$\,2--1       & 2.2$\cdot 10^{-3}$    & 1.7$\cdot 10^{-3}$ \\ 
\hline
\end{tabular}
\tablefoot{$^a$Within the mapped area of $\sim$7.5$'$$\times$11.5$'$ ($\sim$0.9\,pc\,$\times$1.4\,pc).}
\end{table}

\subsection{Kinematics of the FUV-irradiated Gas: velocity maps}

Figure~\ref{fig:vel_channel-chp}
shows CH$^+$~($J$\,$=$\,1--0) and CO~($J$\,$=$\,10--9) velocity channel emission maps, both in color scale, compared to those of [\CII]\,158\,$\upmu$m  in white contours. Both panels display the emission from v$_{\rm LSR}$=0 to 16~km\,s$^{-1}$ in steps of 2~km\,s$^{-1}$. The line centroid of 
the molecular cloud emission lies between v$_{\rm LSR}$$\simeq$8.5 and 10.5~km\,s$^{-1}$ \citep[e.g.,][]{Bally87}. At the angular resolution of our observations, the	\mbox{CH$^+$~$J$\,$=$\,1--0} and
\mbox{CO~$J$\,$=$\,10--9} emission appears at these velocities too. In most channels, there is a very good agreement between the \mbox{CH$^+$\,($J$\,$=$\,1--0)}  and [\CII]\,158\,$\upmu$m emission structures
\citep[C$^+$ main spectral component; see][]{Goi15}. 
The brightest CH$^+$ emission peaks arise from 
the interface between the \HII~region around the Trapezium  and \mbox{OMC-1}. 
The spatial distribution of these bright peaks suggest a spherical shell structure, at least in projection, where the \HII~region is confined by the background dense cloud.
The red-shifted CH$^+$ and C$^+$ structures (v$_{\rm LSR}$=12 to 16~km\,s$^{-1}$;
Fig.~\ref{fig:vel_channel-chp}) are likely compressed structures that are being pushed away from the source of FUV radiation, the Trapezium stars,  and toward the molecular cloud.

In most areas, the gas kinematics revealed by the  \mbox{CO~$J$\,$=$\,10--9}  and \mbox{CH$^+$~$J$\,$=$\,1--0} lines is similar.
Toward the more \mbox{edge-on} \HII/\mbox{OMC-1} interfaces, the bright Orion Bar and East PDRs for example, the maps convolved to the same angular resolution show that the CO~$J$\,$=$\,10--9 line peaks slightly deeper inside the molecular cloud \mbox{(Figs.~\ref{fig:vel_channel-chp} 
and \ref{fig:maps_Cp_CHp_CO_compa})}.  Below v$_{\rm LSR}$$\simeq$4\,km\,s$^{-1}$ and above $\simeq$16\,km\,s$^{-1}$,  the velocity channel maps show \mbox{CO~$J$\,$=$\,10--9} emission only  from Orion~S and BN/KL outflows (Fig.~\ref{fig:vel_channel-chp}). 
In agreement with previous  observations \mbox{\citep[e.g.,][and references therein]{Tahani16}} there is  \mbox{CH$^+$~$J$\,$=$\,1--0} and  \mbox{CO~$J$\,$=$\,10--9} emission component between v$_{\rm LSR}$$\simeq$6 and 8\,km\,s$^{-1}$ that traces the strongly  irradiated surface of the cloud that envelops Orion~S. In addition to lie closer to the gravitational center of the ONC \citep{Hacar17},  Orion~S seems to be a condensation detached from \mbox{OMC-1} and embedded in the \HII~region itself \citep{Odell09}. 
\mbox{CH$^+$~($J$\,$=$\,1--0)} is brighter toward Orion~S than toward BN/KL.

\begin{figure}[t]
\centering
\includegraphics[scale=0.46, angle=0]{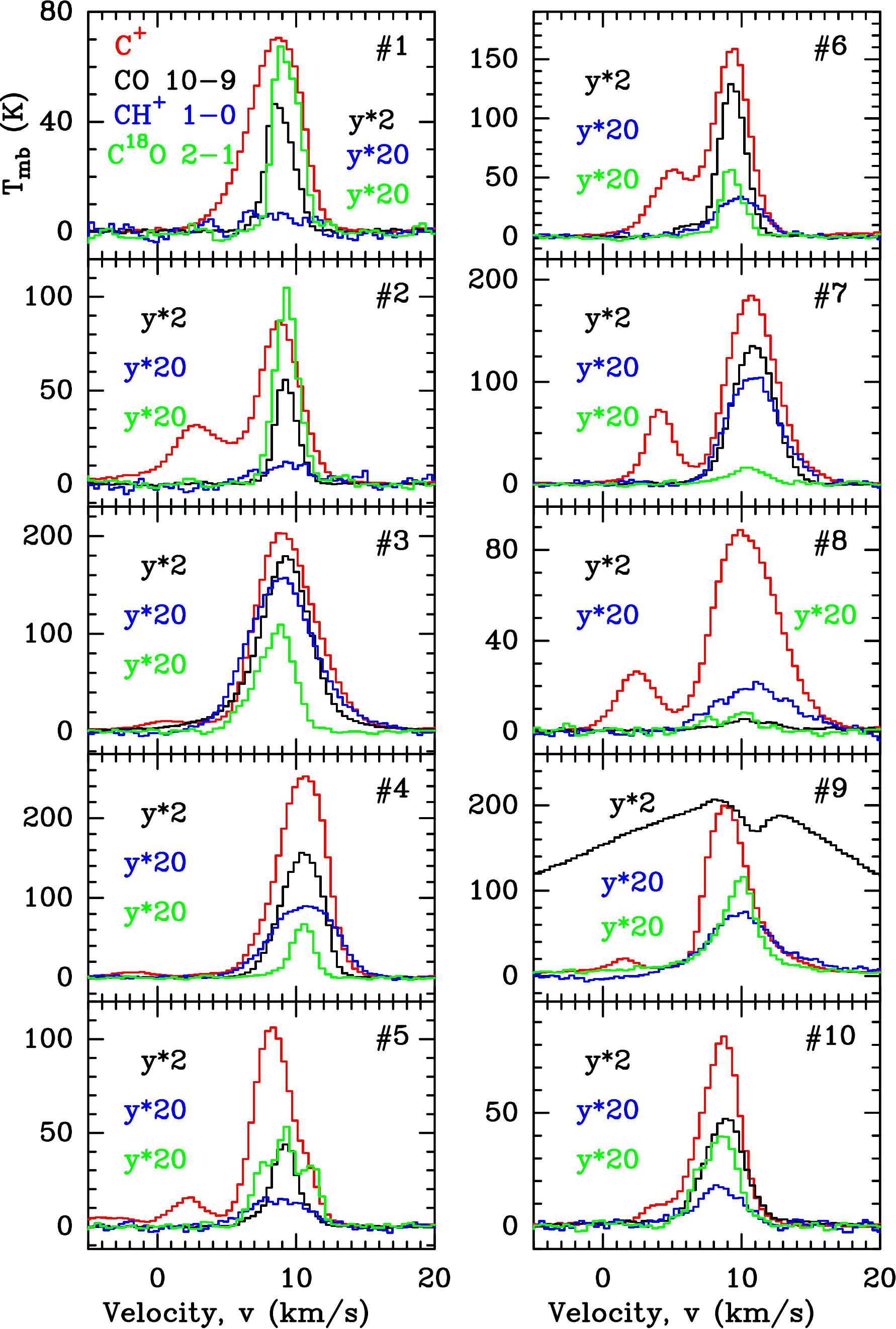}\\
\caption{Spectra at representative positions (see Table \ref{table:positions}). Lines were extracted from maps convolved to the same angular resolution of 27$''$.}\label{fig:spectra_27}
\end{figure}

Outside these bright emission areas, there is still widespread but fainter CH$^+$~($J$\,$=$\,1--0) emission from the extended cloud. We note that some of the spatial distribution differences  in the \mbox{CH$^+$~$J$\,$=$\,1--0} and \mbox{CO~$J$\,$=$\,10--9} velocity channel maps are due to the systematically broad \mbox{CH$^+$~$J$\,$=$\,1--0}  line-profiles (see next section). Only the emission from hydrogen and helium recombination lines,
 arising from ionized gas in the \HII~region, appears at negative LSR velocities, blue-shifted by $\sim$10-20~km\,s$^{-1}$ with respect to the molecular gas emission. This is the signature of flows of ionized gas  that photoevaporate from  \mbox{OMC-1} and toward the observer \citep[e.g.,][]{Genzel89,Goi15}.

\section{Analysis}\label{Sect:Analysis}

\begin{figure*}[t]
\centering
\includegraphics[scale=0.85, angle=0]{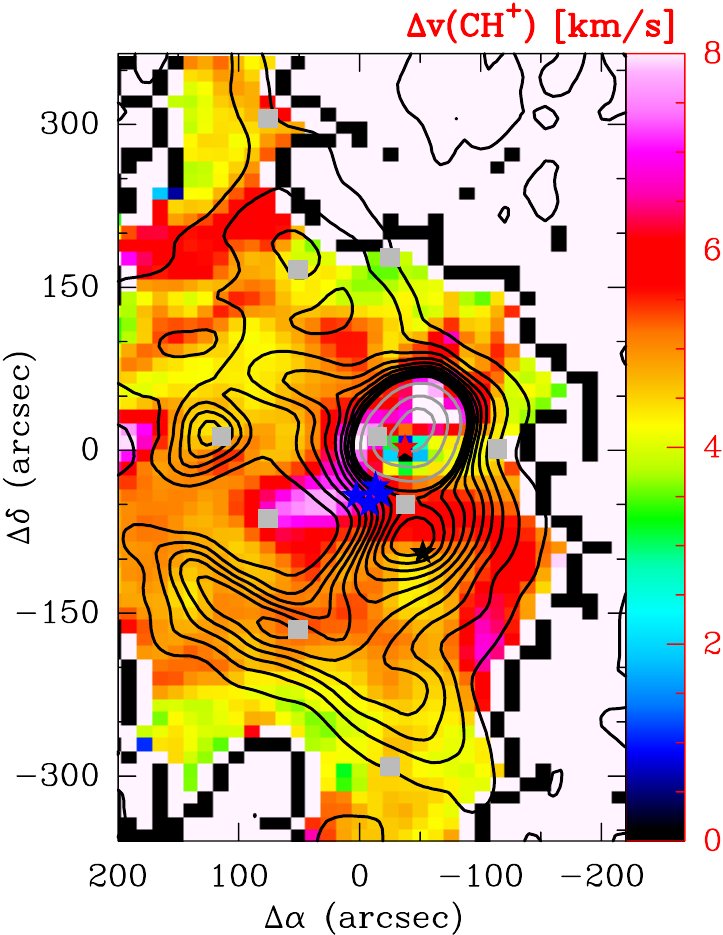}\hspace{3cm}
\includegraphics[scale=0.85, angle=0]{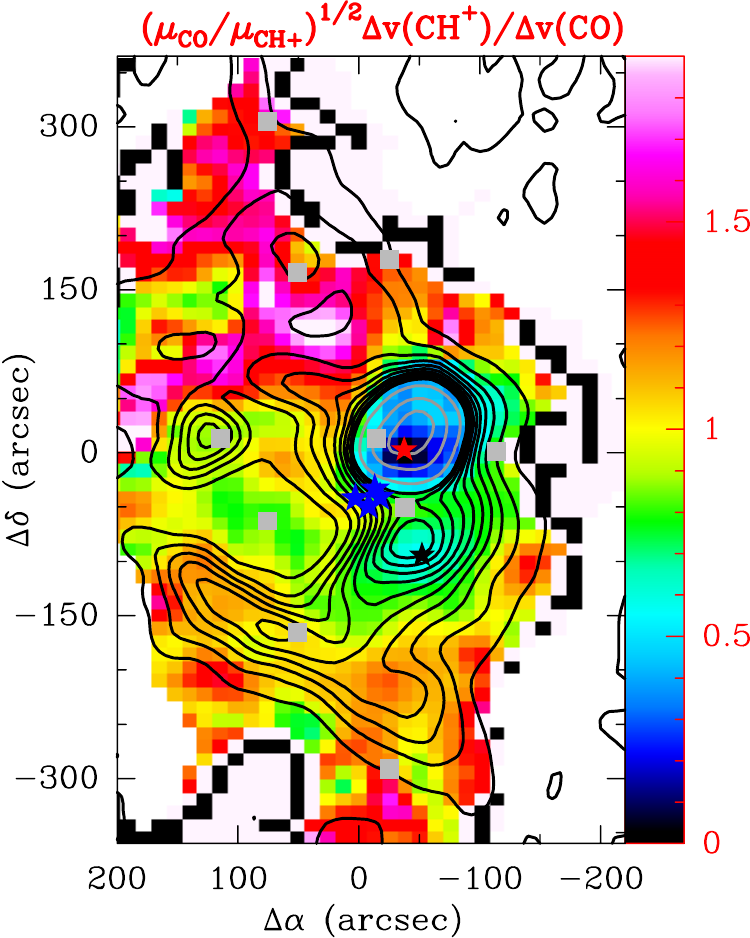}\\
\caption{\textit{Left}: Map of the CH$^+$ $J$\,$=$\,1--0 line-widths in km\,s$^{-1}$
shown in color scale.
\textit{Right}: Map of the \mbox{CH$^+$ ($J$\,$=$\,1--0) to CO ($J$\,$=$\,10--9)} \mbox{line-width} 
ratio, in color scale, corrected by the different molecular masses. 
In both panels, black and grey contours show the CO $J$\,$=$\,10--9 line integrated intensity. All maps were convolved to the same 27$''$ angular resolution.
}\label{fig:widths}
\end{figure*}

\subsection{Line Profiles: broad CH$^+$ profiles at large-scales}\label{sec-profiles}

Based on our knowledge of \mbox{OMC-1} \citep{Genzel89,Stacey93,Bally2008} and  to ease the interpretation of our maps, we select ten representative positions that we analyze in more detail \mbox{(Table~\ref{table:positions})}. These positions
are  marked in the figures with a \mbox{symbol $\#$}.
\mbox{Figure~\ref{fig:spectra_27}} shows [\CII]\,158\,$\upmu$m (red), \mbox{CO~$J$\,$=$\,10--9} (black), \mbox{CH$^+$~$J$\,$=$\,1--0} (blue), and \mbox{C$^{18}$O~$J$\,$=$\,2--1} (green) line spectra toward these positions, extracted from maps convolved to a uniform angular resolution of 27$''$
($\sim$0.05\,pc). In  \mbox{Appendix~\ref{App-more-figs}} we show \mbox{HCO$^+$~$J$\,$=$\,6--5}, \mbox{HCN~$J$\,$=$\,6--5}, and \mbox{C$^{18}$O~$J$\,$=$\,2--1} spectra from maps convolved to 43$''$  (Fig.~\ref{fig:spectra_43}). 

Toward most positions, the observed emission is characterized by a single, roughly Gaussian line profile centered at the LSR velocities of \mbox{OMC-1}. The only exception is the [\CII]\,158\,$\upmu$m line, that shows a second emission component between v$_{\rm LSR}$=$-$2 and $+$5~km\,s$^{-1}$ \citep{Goi15}.  This emission  arises from 
the near side of a 2\,pc sized expanding shell of gas driven by the strong winds from
star \mbox{$\theta^1$ Ori C1} (Pabst et al. 2019, \mbox{submitted}). 
This foreground component, also known as Orion's Veil, has no, or little, molecular line emission.
 \mbox{Finally}, all species that are abundant in shocked gas (CO, H$_2$O, HCO$^+$, HCN, etc.) 
 display high-velocity wings toward 
Orion~S and BN/KL (see spectra of position $\#$9 in \mbox{Figs.~\ref{fig:spectra_27}} and \ref{fig:spectra_43}).

\begin{table}[t]	
\caption{Representative positions discussed in the text.} 
\centering
\begin{tabular}{lcl@{\vrule height 9pt depth 5pt width 0pt}}
\hline\hline
Position & Offset & Comments\\ \hline
$\#1$        &  (${76}",{305}"$)   &  Northern ridge\\ 
$\#2$        &  (${-25}",{178}"$)  &  North-west ridge edge \\ 
$\#3$        &  (${-38}",{-50}"$)  &  CH$^+$ peak near Trapezium\\ 
$\#4$        &  (${51}",{-165}"$)  &  Orion Bar PDR\\ 
$\#5$        &  (${-25}",{-291}"$) &  Southern edge\\ 
$\#6$        &  (${51}",{166}"$)   &  Northern lane\\ 
$\#7$        &  (${114}",{13}"$)   &  East PDR \\ 
$\#8$        &  (${76}",{-63}"$)   &  Toward dense \HII\,region\\ 
$\#9$        &  (${-25}",{13}"$)   &  BN-KL outflows\\ 
$\#10$       &  (${-114}",{1}"$)   &  Western edge\\ 
\hline \label{table:positions}
\end{tabular}
%\tablefoot{}
\end{table}

C$^+$ and CH$^+$ line profiles  show a slightly different behaviour among them. While the  [\CII]\,158\,$\upmu$m line is narrow toward BN/KL outflows, the \mbox{CH$^+$~$J$\,$=$\,1--0} profile displays a weak red-wing emission. This demonstrates that the outflows around BN/KL  are illuminated by a moderate FUV radiation field
\citep[][]{Chen14,Goipacs15,Melnick15}. It has been suggested that the available C$^+$ in irradiated dense shocked gas must be quickly converted into CH$^+$ \citep{Morris16}. Nevertheless, the \mbox{CH$^+$~$J$\,$=$\,1--0} wing is  faint compared to  CO or H$_2$O lines, and is only restricted to low velocities.

Table~\ref{table:vlsr} tabulates the velocity centroid, obtained from Gaussian fits, toward the ten representative positions. 
 All lines show, within fit errors and  spectral resolution, similar velocity centroids. 
Table~\ref{table:widths} tabulates the line-width for the same set of lines and positions;
there is more scatter in this case.
In principle, the gas velocity dispersion is determined by the gas temperature, through the thermal broadening caused by \mbox{elastic} collisions with other species, and by gas macroscopic motions: turbulence, outflows, etc. 
The narrowest measured widths (\mbox{$\Delta$v\,$<$\,3~km\,s$^{-1}$}) are those of C$^{18}$O lines, predominantly arising from \mbox{FUV-shielded} and colder gas.  At large spatial scales, the \mbox{CO~$J$\,$=$\,10--9} line also displays narrow line-widths (\mbox{$\Delta$v\,$\simeq$\,3~km\,s$^{-1}$}) except toward BN/KL and \mbox{Orion\,S} outflows. This confirms that the bulk of  the \mbox{mid-$J$ CO emission} arises from relatively \mbox{quiescent}  gas and not from fast shocks \citep[see also][]{Wilson01,Peng12}.    

Assuming optically thin emission and negligible non-thermal line broadening, the observed  \mbox{CO~$J$\,$=$\,10--9} line-width sets an upper limit value to the 
emitting-gas temperature of $T_{\rm k}$$\lesssim$450\,K.
Taking a more realistic non-thermal velocity dispersion of $\sigma_{\rm nth}$=1~km\,s$^{-1}$ ($\Delta$v$_{\rm nth, FHWM}$=2.355$\cdot$$\sigma_{\rm nth}$, where FHWM refers to the full width at half maximum) for the \mbox{mid-$J$ CO-emitting} PDR gas (see next Section), one obtains $T_{\rm k}$$\lesssim$150\,K. These values are consistent with the lower limit to the gas temperature
provided by the \mbox{CO~$J$\,$=$\,2--1}  line intensity peak  ($T_{\rm peak}$ in K; see map in Fig.~\ref{fig:peak_co}). Because the  \mbox{CO~$J$\,$=$\,2--1} emission is optically thick in most of the field, one obtains $T_{\rm peak}$\,$\simeq$\,$T_{\rm rot}$\,(CO 2-1)\,$\leq$\,$T_{\rm k}$.

The observed [\CII]\,158\,$\upmu$m line-peak temperatures, however, suggest that the \mbox{C$^+$-emitting} layers, the atomic PDR, are \mbox{hotter} \mbox{\citep[$T_{\rm k}$$\ge$300\,K;][]{Goi15}} than the 
 \mbox{CO-emitting} layers located slightly deeper into the PDR
(see  model predictions in Fig.~\ref{fig:PDR_mod}). This difference is the signature of sharp temperature  gradients: from the \HII~region, the atomic PDR and molecular PDR,  to the shielded cloud interior.

\begin{figure*}[h]
\centering
\includegraphics[scale=0.097, angle=0]{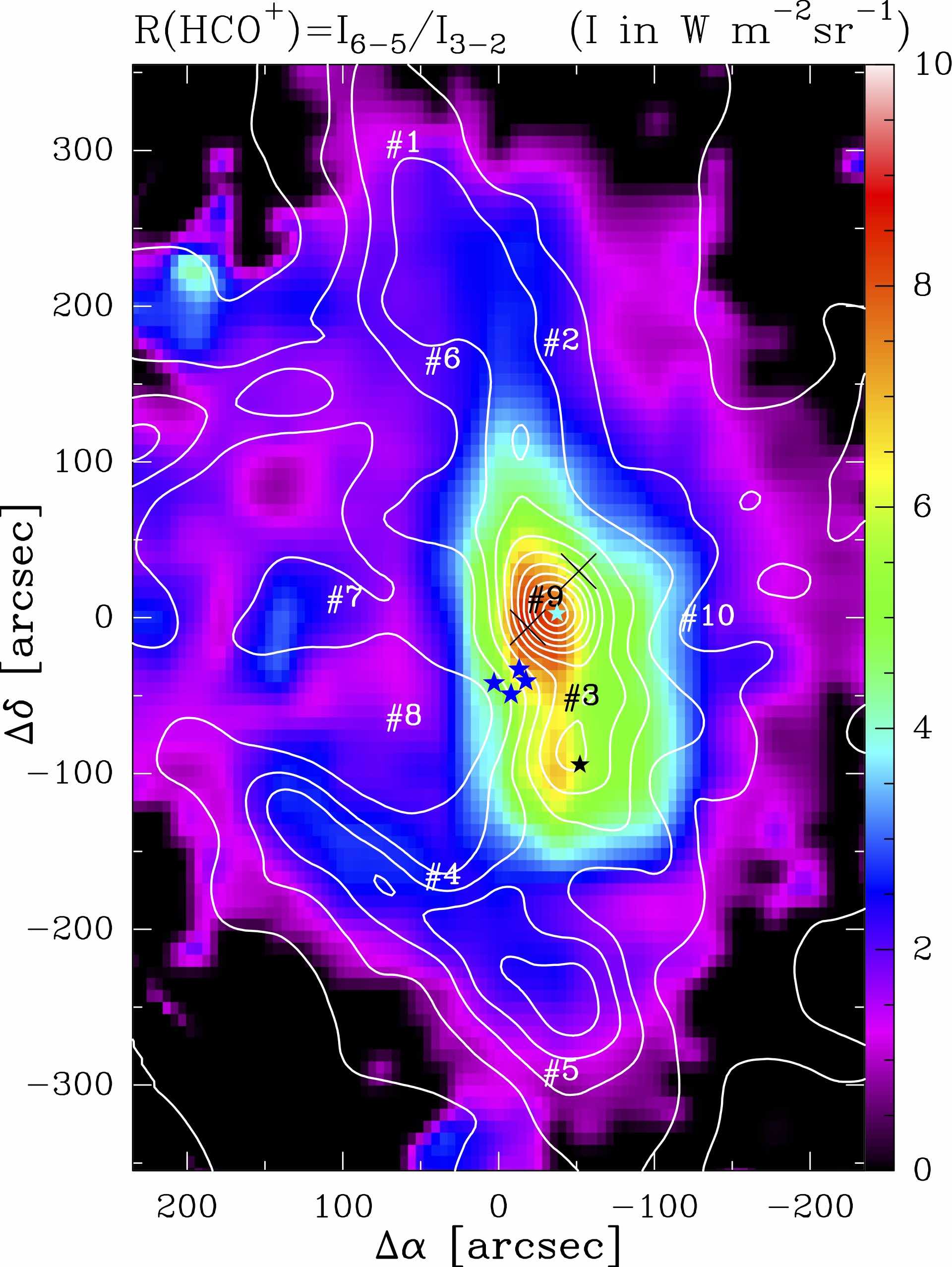}\hspace{2cm}
\includegraphics[scale=0.097, angle=0]{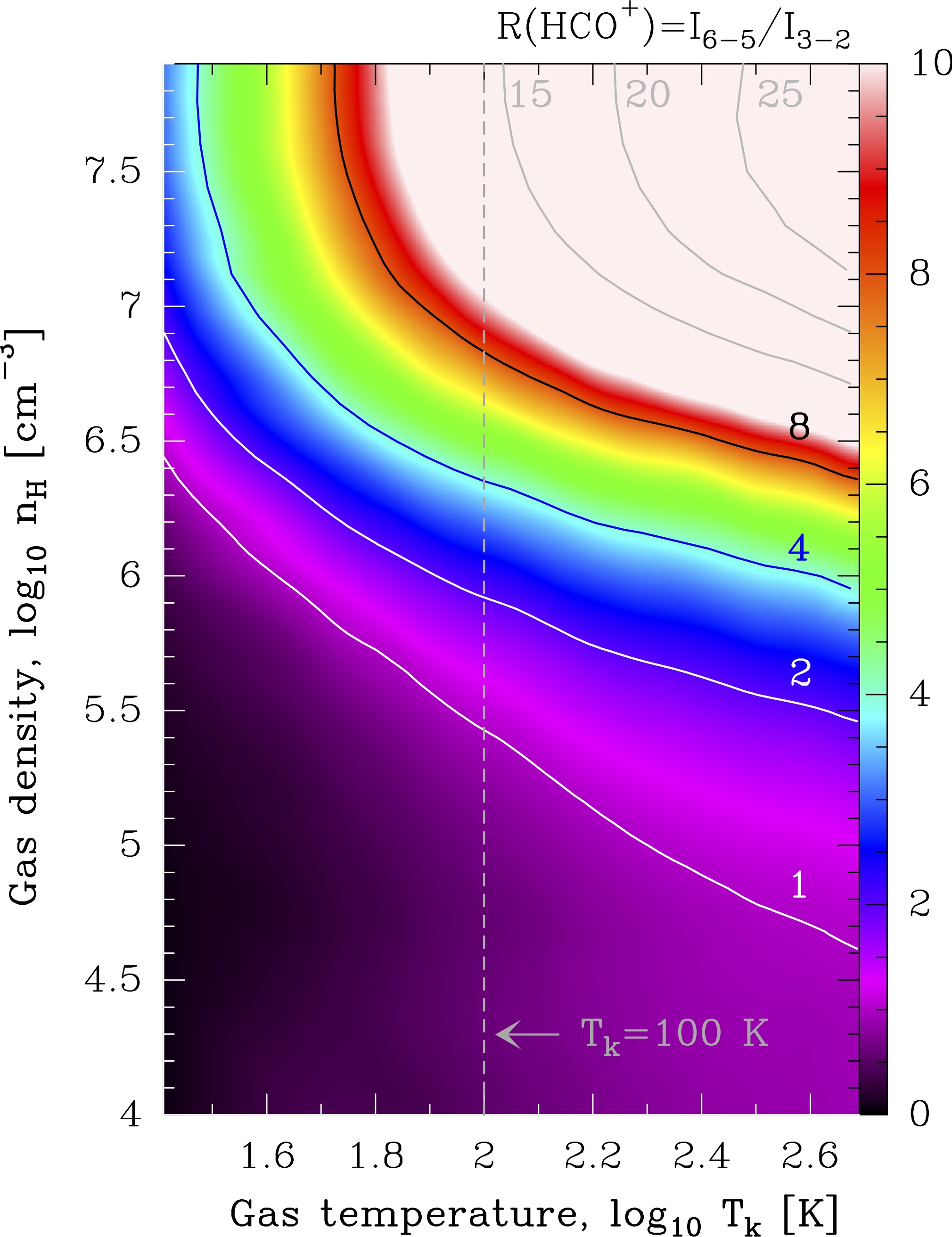}\vspace{0.5cm}
\includegraphics[scale=0.097, angle=0]{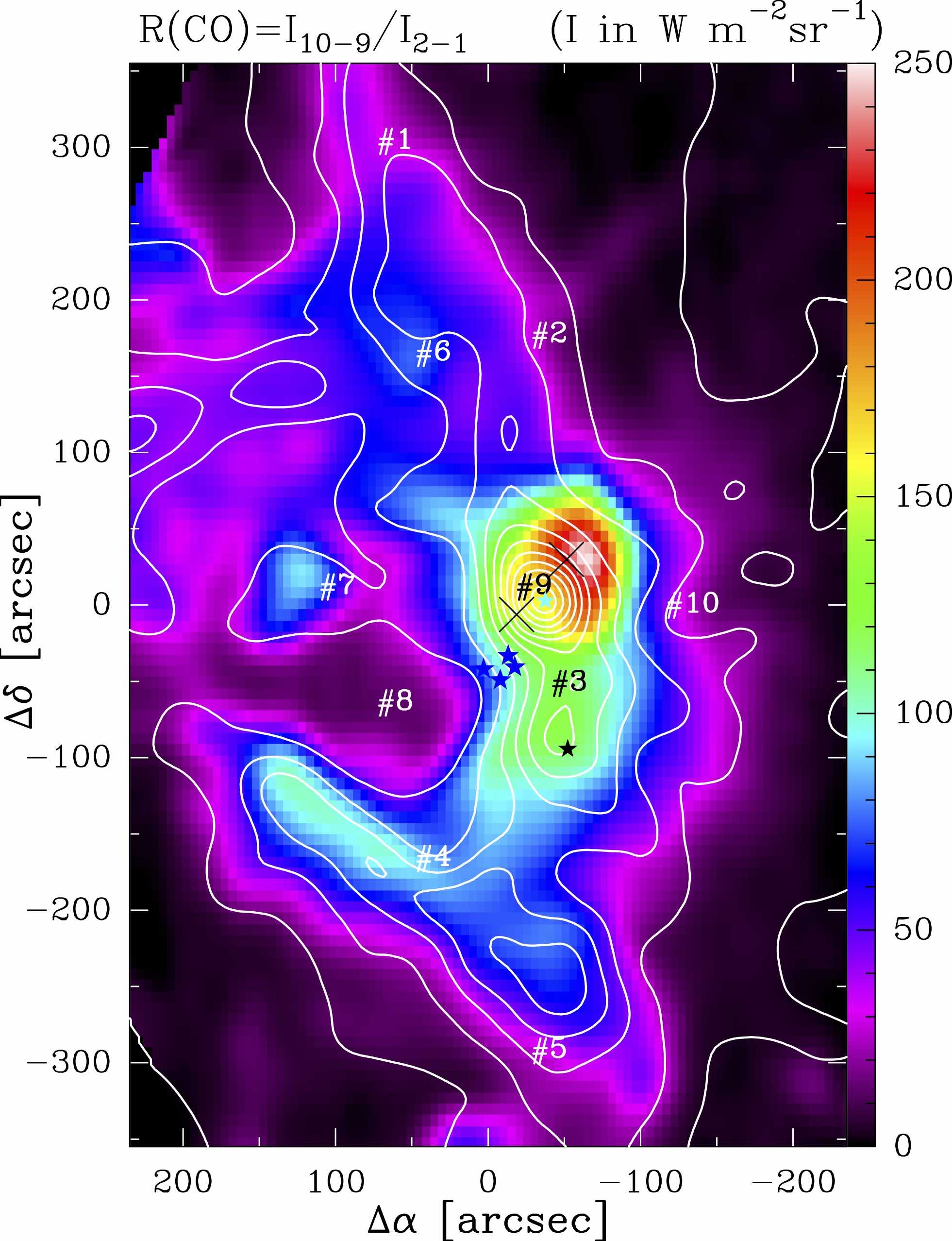}\hspace{2cm}
\includegraphics[scale=0.097, angle=0]{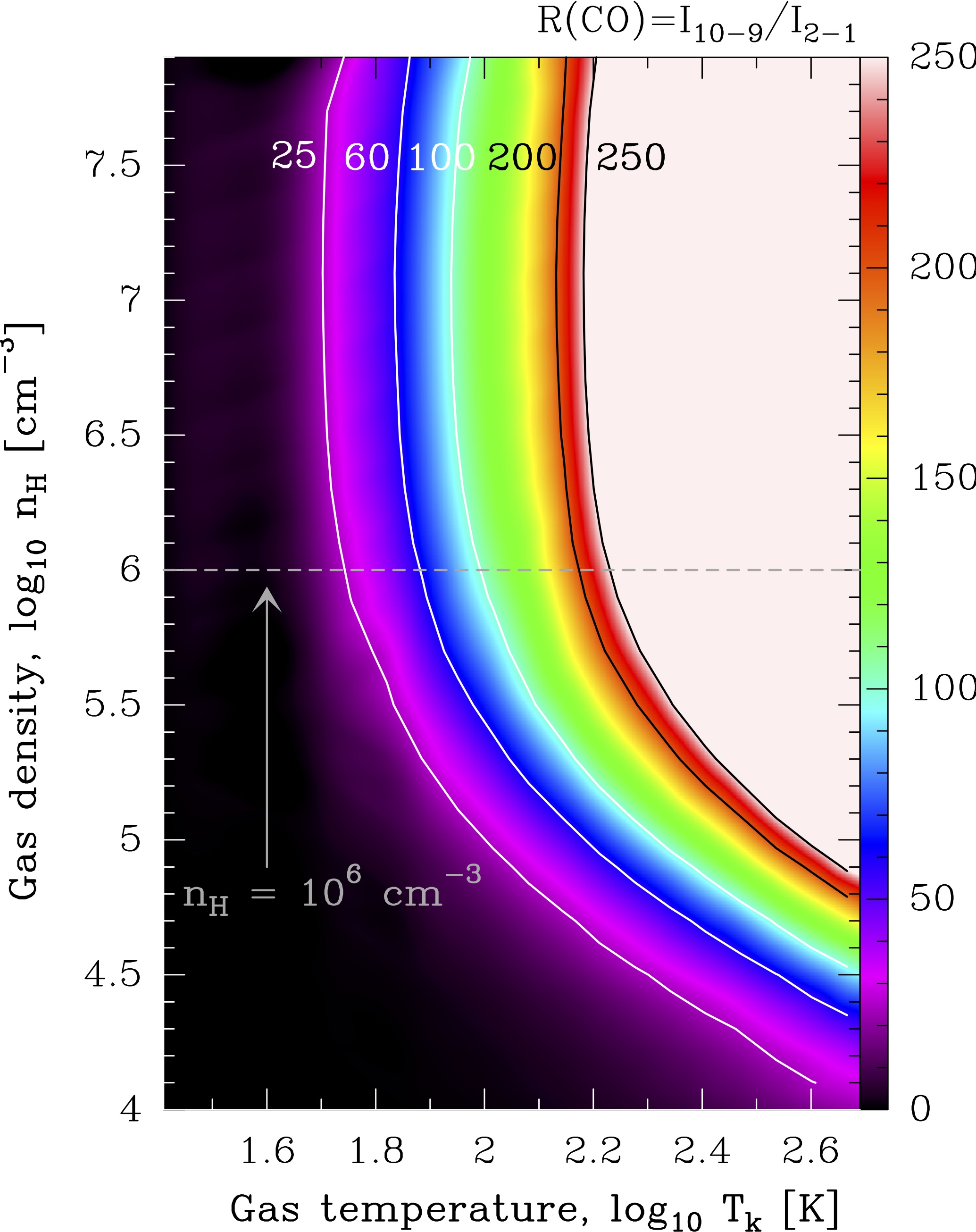}
\caption{\textit{Top left}: Map of the HCO$^+$~6--5/3--2 intensity ratio, at 43$''$ resolution, sensitive to gas density variations.
\textit{Bottom left}: Map of the \mbox{CO~10--9/2--1} intensity ratio, at 27$''$ resolution, sensitive to gas temperature variations.
In both maps, the white contours represent the $^{13}$CO $J$\,$=$\,2--1 optically thin emission.
\textit{Right}: Grid of nonlocal and non-LTE excitation models and predicted line intensity ratios.}\label{fig:ratios-mtc}
\end{figure*}

CH$^+$ ($J$\,$=$\,1--0) shows the broadest line-widths  over the observed field, broader than [\CII]\,158\,$\upmu$m and \mbox{CO~$J$\,$=$\,10--9} lines.  Figure~\ref{fig:widths} (left) shows a map of the \mbox{CH$^+$~$J$\,$=$\,1--0} line-widths in the region \mbox{($\Delta$v$\simeq$5-6~km\,s$^{-1}$)}. 
Unlike other species, the broad CH$^+$~$J$\,$=$\,1--0 line-profiles are quite uniform along the mapped area. \mbox{Figure~\ref{fig:widths}} (right) shows a map of the  \mbox{CH$^+$~($J$\,$=$\,1--0) to CO~($J$\,$=$\,10--9)} line-width ratio corrected by their different molecular masses, \mbox{$(\mu_{\rm CO}/\mu_{\rm CH^+})^{1/2}$}=1.47. Only toward BN/KL and Orion~S,  the ratio is much lower than one. In the extended cloud component, the line-width ratio is always greater than one.
These maps thus confirm the broader CH$^+$ line profiles, not only toward local bright and dense
\HII/PDR interfaces \citep{Nagy13,Morris16,Parikka_2017} but at large-spatial scales, where the \mbox{CH$^+$ ($J$\,$=$\,1--0)} emission
is certainly optically thin and not affected by opacity broadening.
 Taking $\sigma_{\rm nth}$=1~km\,s$^{-1}$, the observed broad \mbox{CH$^+$~$J$\,$=$\,1--0} line-widths would imply unrealistically high gas temperatures, from $T_{\rm k}$\,$\simeq$\,5500 to 8600~K. As previously discussed in the literature  \citep{Black98,Nagy13,Godard13,Goico17}, this broadening is likely related to the high reactivity of the ion, and to the  exothermic route (\mbox{$\Delta E/k$\,$\geq$\,5360~K})  that forms CH$^+$ from reaction C$^+$+H$_2$~($v$\,$\geq$\,1). The CH$^+$  lifetime
in dense PDRs  is so short   that the newly formed molecular ion does not have time to thermalize its motions through elastic collisions to a Maxwellian velocity distribution at $T_{\rm k}$. 
In this interpretation, part of the formation exothermicity  goes into excitation and translational motion. Hence,  the broad CH$^+$ line-widths would be related to the excess of energy upon formation \mbox{\citep[][]{Nagy13}}. We see that in high-mass star forming regions like \mbox{OMC-1}, 
sources of strong FUV radiation,
 this mechanism operates at large-spatial scales.

\subsection{Physical conditions of the extended warm gas}\label{sec-props} 

In this section we estimate the average physical conditions of the extended warm molecular gas traced by 
the observed submm lines.
For  high-dipole molecules such as HCO$^+$, having high critical density transitions, the intensity ratio of two rotational lines is a  good tracer of gas density variations. On the other hand,  for low-dipole moment molecules 
such as CO, and for  gas densities comparable or higher than the critical density, line intensity ratios trace gas temperature  variations.

Figure~\ref{fig:ratios-mtc} shows maps of the \mbox{HCO$^+$ 6-5/3-2} and 
\mbox{CO 10-9/2-1} line intensity ratios, obtained from integrated intensities in units
of W\,m$^{-2}$\,sr$^{-1}$. To invert the observed ratios into a range of beam-averaged gas temperatures, $T_{\rm k}$, and densities,
\mbox{$n_{\rm H}$=$n$(H)+2$n$(H$_2$)}, we compare the observed line intensities with predictions of a grid of nonlocal and non-LTE excitation  models \citep[Monte Carlo radiative transfer code of][]{Goicoechea06}.
The model accounts for line trapping, collisional excitation, and radiative excitation by absorption  of the 2.7~K cosmic background and by the dust continuum emission.
In FUV-illuminated gas, the high electron abundance, 
up to \mbox{$x_e=n_e/n_{\rm H}\lesssim  x$(C$^+$)$\simeq$10$^{-4}$} for standard cosmic ray ionization rates, can play an important role in the collisional excitation of high-dipole moment molecules as the collisional rates compete with those of H$_2$ and H  \citep[e.g.,][]{Tak12,Goldsmith17}.
Here we assume an ionization fraction of $x_e$\,$=$\,10$^{-4}$ (roughly the C$^+$ abundance
in PDR gas), which means that our derived $n_{\rm H}$ values are lower limits if $x_e$ is not that high. We used \mbox{HCO$^+$--H$_2$} and \mbox{HCO$^+$--$e^-$} inelastic collisional rates from \citet{Flower1999}, \citet{Faure2001}, and \citet{Fuente2008}.
For CO, we just used the  rates for \mbox{CO--H$_2$} collisions from \citet[][]{Yang2010}.

Regarding radiative excitation, we modelled the frequency-dependent dust continuum  emission as a modified  black body with an effective dust grain temperature of \mbox{$T_{\rm d}$\,$=$\,55 K}, spectral emissivity index \mbox{of 2}, and a dust opacity $\tau_{\rm d}$\,$=$\,0.05 at a reference wavelength of 160\,$\upmu$m. These FIR and submm illumination conditions are typical of the Orion Bar PDR \citep{Arab_2012} but we note that  submm pumping plays a negligible role in the excitation of the extended  CO and HCO$^+$ emission. This is, however, not the case of H$_2$O or OH excitation toward BN/KL, where the dust thermal continuum is very strong \citep[e.g.,][]{Gonzalez02,Melnick10,Goipacs15}. Our model includes thermal, turbulent, and line opacity broadening. The non-thermal velocity dispersion that reproduces the observed 
line-widths is typically \mbox{$\sigma_{\rm nth}$$\simeq$1~km\,s$^{-1}$}. We adopted  column densities of
\mbox{$N$(CO)=5$\cdot$10$^{17}$\,cm$^{-2}$} and 
\mbox{$N$(HCO$^+$)=5$\cdot$10$^{13}$\,cm$^{-2}$}  extracted from our line survey toward the edge of the Orion Bar 
\mbox{\citep[e.g.,][]{Cuadrado15,Goico17}}.

Figure~\ref{fig:ratios-mtc} (right panels) shows synthetic line ratios
in the plane \mbox{log\,($T_{\rm k}$)--log\,($n_{\rm H}$)}, with same color scale as in the  maps, for 
a grid of models with \mbox{$n_{\rm H}$ ranging from 10$^4$ to 10$^8$\,cm$^{-3}$}, and
\mbox{$T_{\rm k}$ from 25 to 500\,K}. The lowest measured \mbox{HCO$^+$ 6--5/3--2} intensity
ratio in the map is $\sim$1 (magenta regions, e.g., positions~$\#$1 and $\#$5). This ratio imposes a minimum gas density of a few 10$^5$\,cm$^{-3}$ for $T_{\rm k}$$\lesssim$150\,K. Bright \mbox{edge-on} 
\mbox{\HII/OMC-1} interfaces such as the Orion Bar \mbox{(position~$\#$4)} or
the East PDR (position~$\#$7) require higher gas temperatures and densities
(\mbox{$T_{\rm k}$$\simeq$100\,K} and \mbox{$n_{\rm H}$$\simeq$10$^6$\,cm$^{-3}$}).
Closer to the Trapezium cluster and around \mbox{Orion~S} \mbox{(position~$\#$3)},  densities increase (reaching $\sim$10$^7$\,cm$^{-3}$). 
The shocked gas associated with Orion~BN/KL  (position~$\#$9) shows the highest excitation conditions in both $T_{\rm k}$ and $n_{\rm H}$,  but we stress again that most of the line integrated emission in the map arises from the irradiated cloud surface.
This is an extended but thin gas layer (see PDR models in  Sect.~\ref{sub-sec:PDR-mods})  characterized by high thermal pressures, in the range \mbox{$P_{\rm th}=T_k \cdot n_{\rm H} \approx 10^7-10^9$~cm$^{-3}$\,K}
according to our models.
The presence of a layer of high-density gas close to the interface between the
\HII~region and the molecular cloud was anticipated by 
\citet{Rodriguez98,Rodriguez01} from observations of the CN radical, which predominantly arises,
but not only \citep[e.g.,][]{Pety17}, from PDR gas. They
also estimated  gas densities ranging from $\sim$10$^5$\,cm$^{-3}$ toward the extended cloud, to  several $\sim$10$^6$\,cm$^{-3}$ toward the Trapezium region. They  could not, however, constrain the gas temperatures.

Using our CO maps, we estimate the mass contained in the high pressure molecular PDR layer, 
$M_{\rm mPDR}$(H$_2$), along \mbox{OMC-1}.
Assuming \mbox{$T_{\rm rot}$(10--9)\,$\simeq$\,$T_{\rm rot}$(2--1)} (see map in Fig.~\ref{fig:peak_co}) and that the \mbox{CO\,$J$\,=\,10--9} emission is optically thin and only arises from the cloud surface, we convert the CO\,$J$\,=\,10--9 integrated intensity map into a CO column density map, $N$(CO), and then into a total column density map as $N_{\rm H}$\,=\,$x$(CO)$\cdot$$N$(CO)\,$\simeq$10$^{-4}$$\cdot$$N$(CO) 
(see \mbox{Appendix~\ref{App-masses}} for details). With these assumptions,
we obtain $M_{\rm mPDR}$(H$_2$)\,$\simeq$\,150\,$M_{\odot}$. Allowing a opacity correction factor toward lines of sight with $\tau_{\rm 10-9}$\,$\simeq$\,1,  we then derive $M_{\rm mPDR}$(H$_2$)\,$\simeq$\,300\,$M_{\odot}$. This warm molecular gas mass is similar to the
$\sim$200\,$M_{\odot}$ mass that arises from the [\CII]\,158\,$\upmu$m emitting gas 
\citep[the atomic PDR;][]{Goi15}. It represents between $\sim$5\,$\%$ and $\sim$10\,$\%$ of the
total gas mass in \mbox{OMC-1} \citep{Goi15}.
Finally, using \mbox{$A_{\rm V}$/$N_{H}$\,$=$\,3.5$\cdot$10$^{-22}$\,mag\,cm$^2$}, appropriate to Orion, we derive that the average extinction thickness traced by the \mbox{CO\,$J$\,=\,10--9} line (roughly the high pressure gas layer) is  $<A_{\rm V}>$\,$\simeq$\,3--6\,mag.

\begin{figure*}[t]
\centering
\includegraphics[scale=0.11, angle=0]{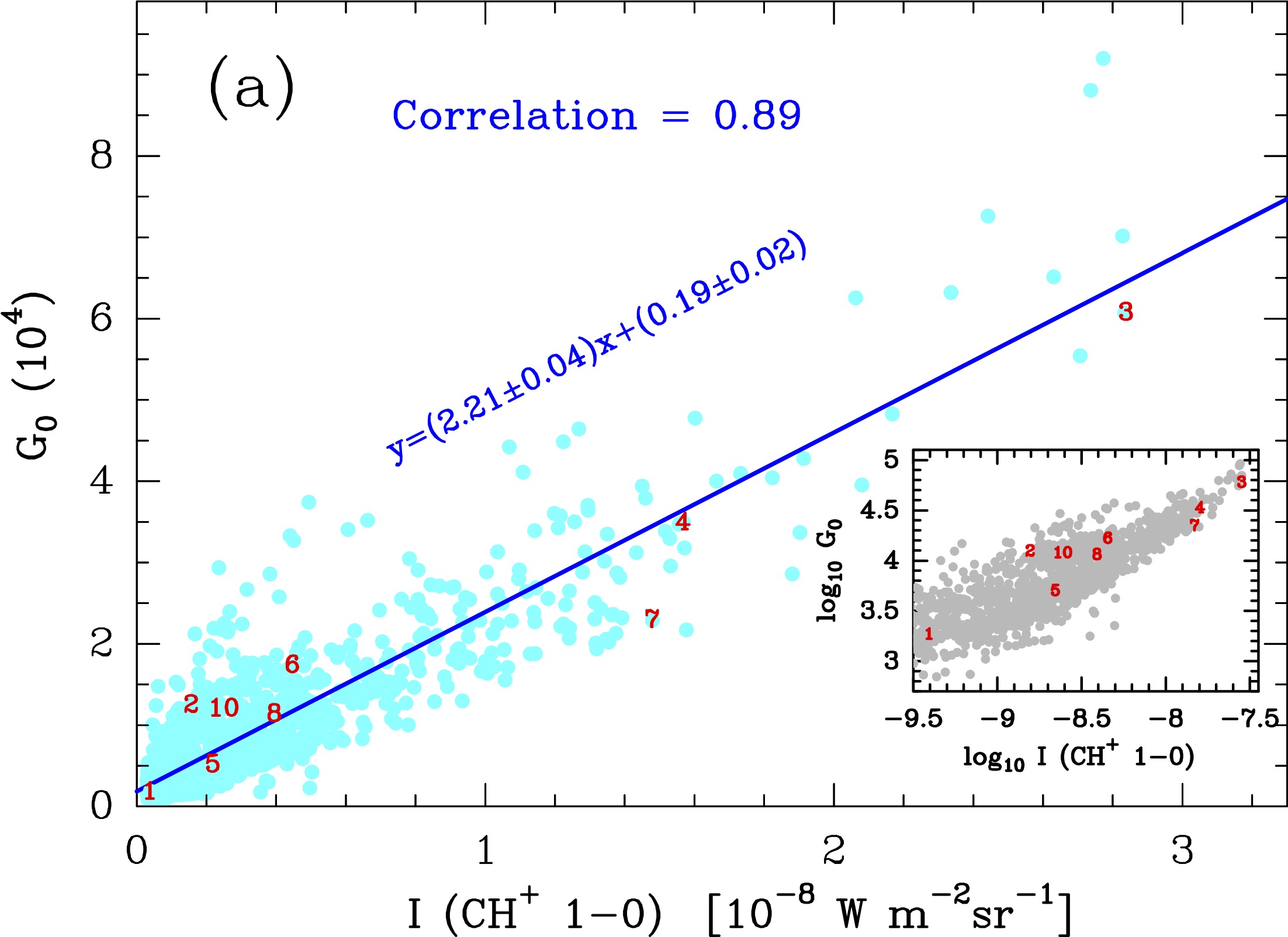}\hspace{1cm}
\includegraphics[scale=0.11, angle=0]{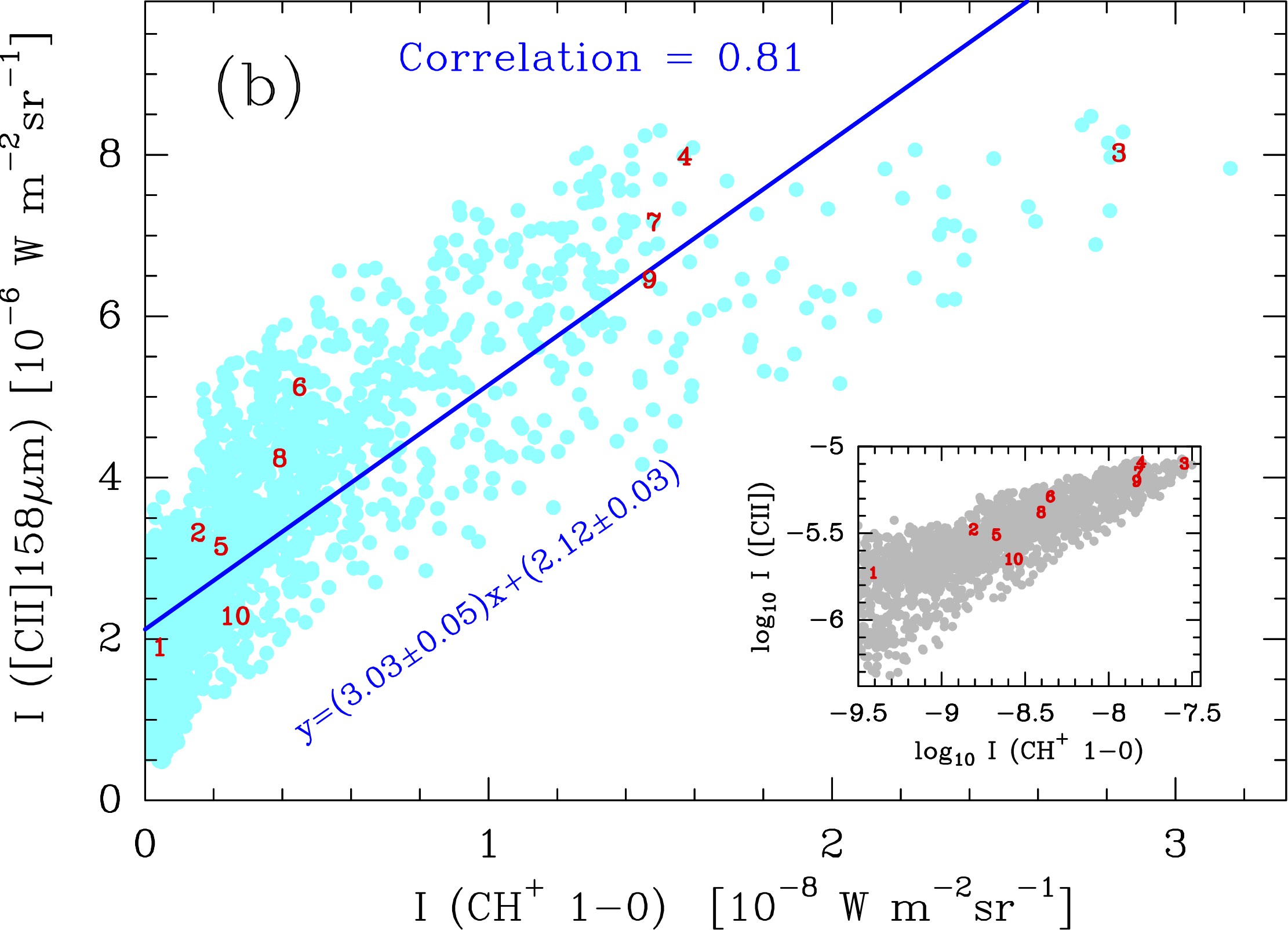}\\\vspace{0.4cm}
\includegraphics[scale=0.11, angle=0]{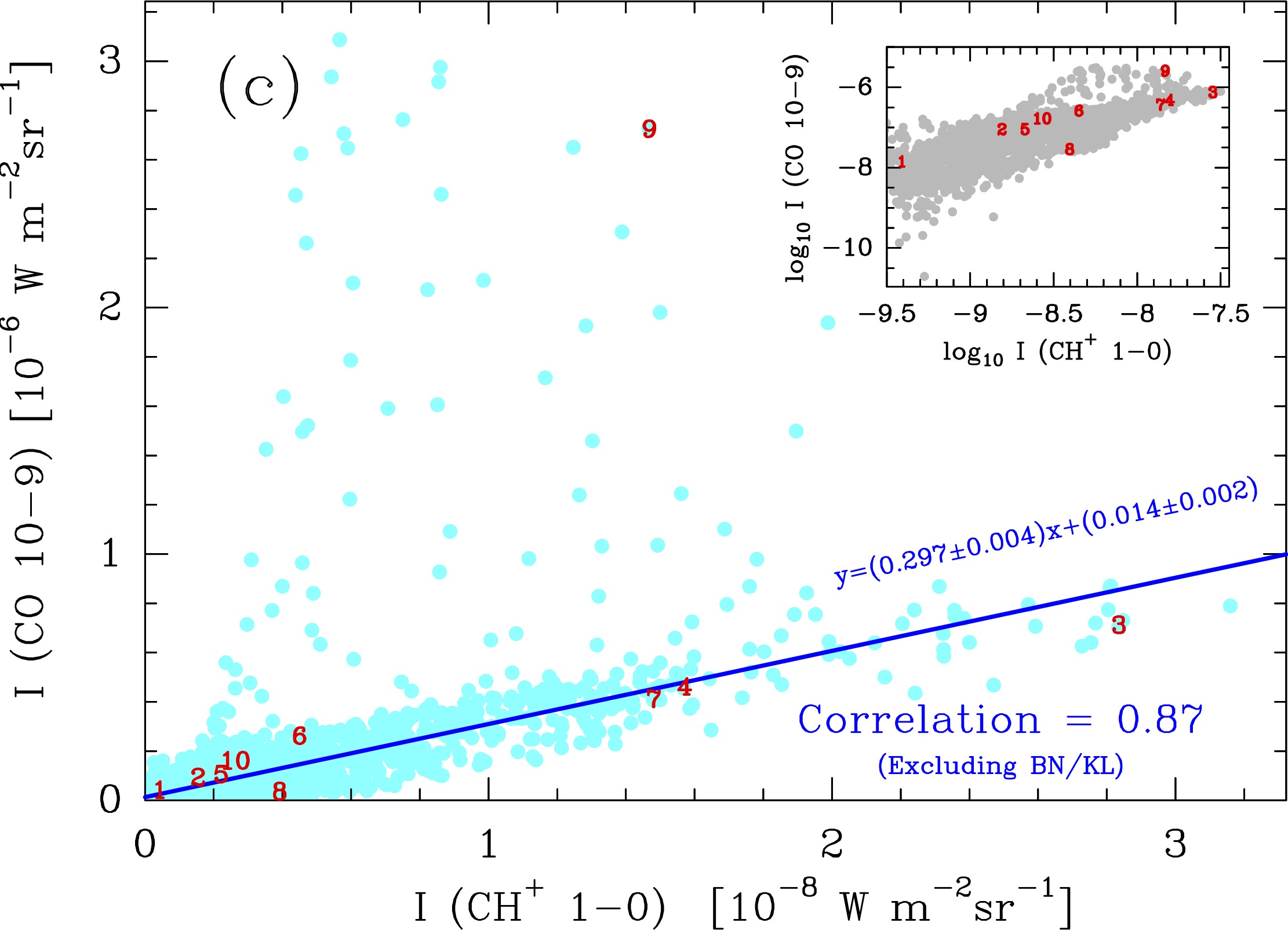}\hspace{1cm}
\includegraphics[scale=0.11, angle=0]{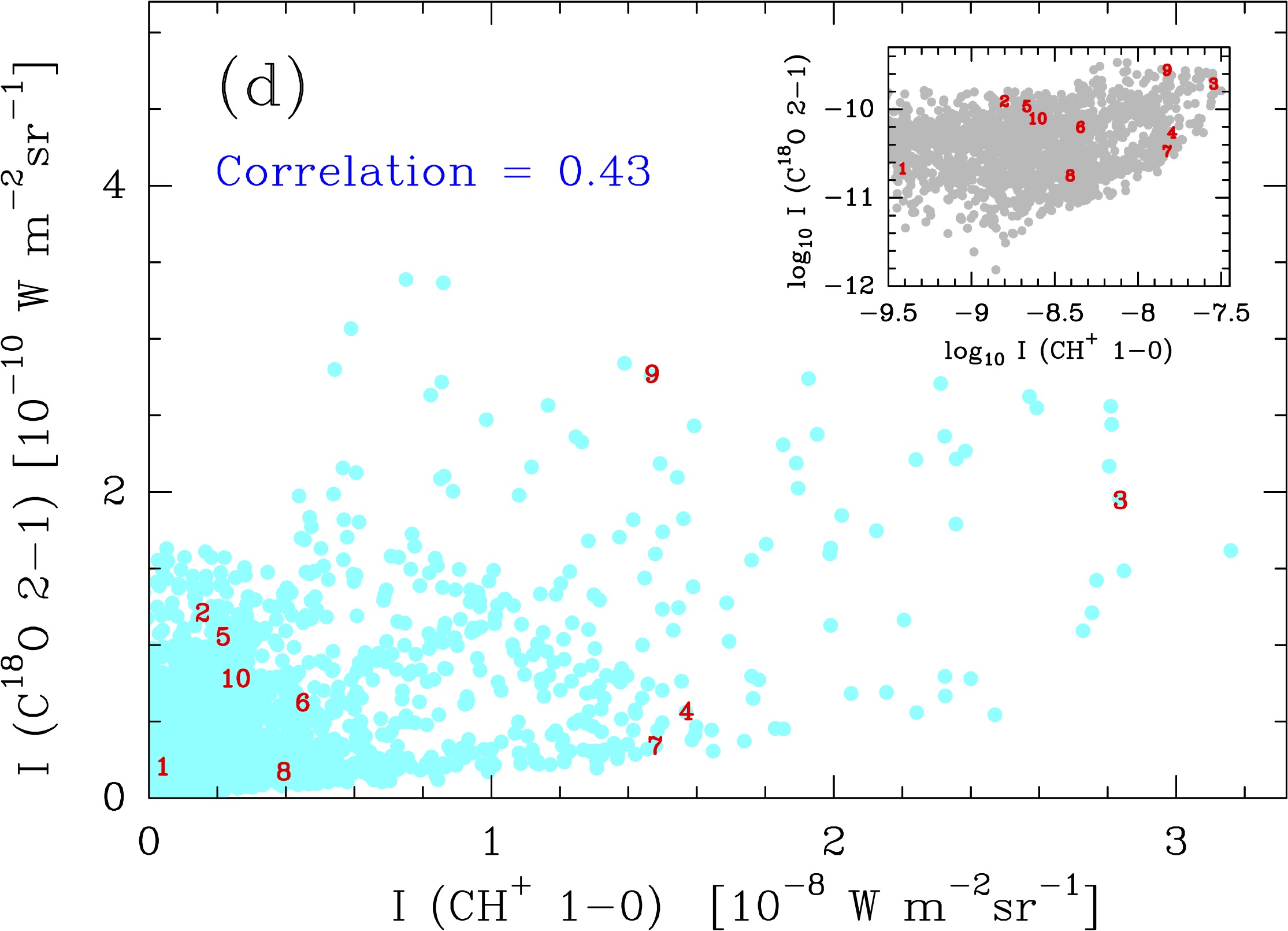}\\
\caption{CH$^+$~$J$\,$=$\,1--0 line intensity correlation plots extracted from maps convolved to a uniform  resolution of 27$''$. Data points are shown in linear scale (cyan points) and in logarithmic scale (gray points). Results and dispersions from a linear fit, $y=(m\pm\Delta m)\,x+(b\pm\Delta b)$, are shown in blue. The location of selected positions \# in the map (Table~\ref{table:positions})
are indicated with red numbers.
 (a)~\mbox{FUV-radiation} field flux (G$_0$) estimated from FIR luminosities \citep[][]{Goi15}, (b)~\mbox{[\CII]\,158\,$\upmu$m}, (c)~\mbox{CO~$J$\,$=$\,1--0}, and 
(d)~\mbox{C$^{18}$O~$J$\,$=$\,1--0} line intensities.}\label{fig:CHp_correlation}
\end{figure*}

\begin{figure*}[h]
\centering  
\includegraphics[scale=1.06, angle=0]{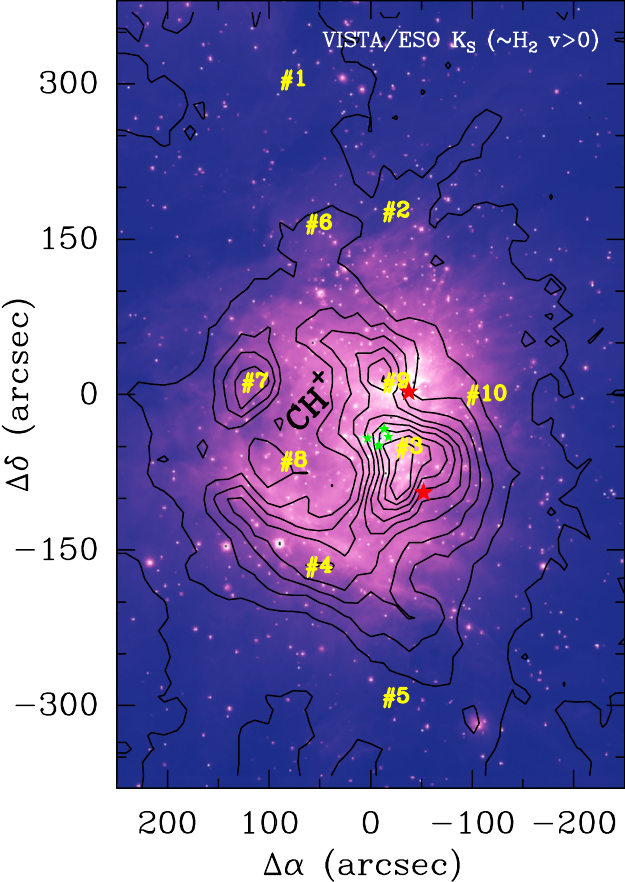}\hspace{2.5cm}
\includegraphics[scale=1.06, angle=0]{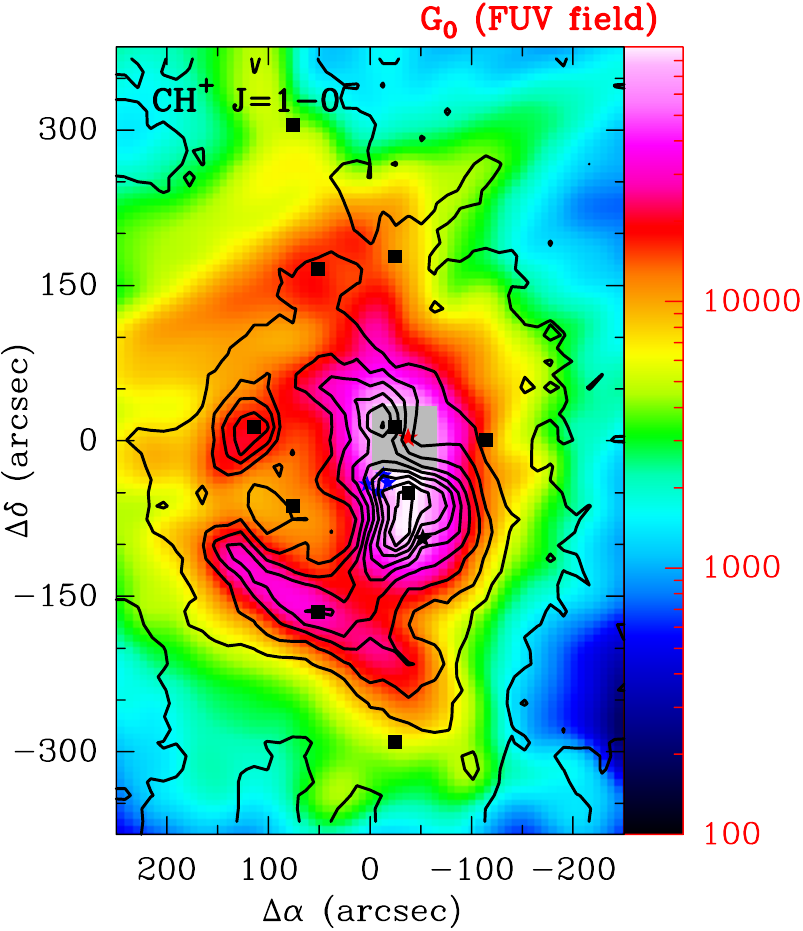}\\
\caption{\textit{Left}: NIR  VISTA image of M42 in the $K_S$ band dominated
by interstellar emission from vibrationally excited H$_2$.
The intensity scale is chosen to enhance the very  bright NIR emission from
gas illuminated by the strong FUV field from the Trapezium stars; and locally by thermally-excited shocked gas in outflows  (Fig.~\ref{fig:Vista_large} shows a larger image that
enhances the much more extended NIR emission). Black contours show the CH$^+$~$J$\,$=$\,1--0 intensities from 1 to 46\,K\,km\,s$^{-1}$ in steps of 5\,K\,km\,s$^{-1}$. \textit{Right}: FUV radiation field, $G_0$, estimated from FIR luminosities \citep{Goi15}. The gray area around BN/KL indicates positions where the \textit{Herschel} photometric data is saturated.} \label{fig:G0}
\end{figure*}

\subsection{CH$^+$ J=1-0 intensity correlation plots}\label{sec-correlations}

Figure~\ref{fig:CHp_correlation} shows correlation plots between the CH$^+$~$J$\,$=$\,1--0
line intensity  and other lines or quantities mapped along \mbox{OMC-1}: 
(a)~the strength  of the FUV radiation flux, $G_0$,  estimated by Goicoechea et al. (2015b) from FIR  luminosities in the region, (b)~[\CII]\,158\,$\upmu$m, (c)~\mbox{CO~$J$\,$=$\,10--9}, 
and (d)~\mbox{C$^{18}$O~$J$\,$=$\,2--1}  line intensities (all in  W\,m$^{-2}$\,sr$^{-1}$). The correlation plots are shown in a linear scale (cyan points). The results from a linear fit are shown in blue. The small inset in each panel shows the same data in log--log scale (gray points). These plots use all the line intensity measurements  contained in the maps  and thus provide robust clues  to the origin of the CH$^+$ emission at large scales. 
 
 Owing to the proximity of the Trapezium stars, $G_0$ is very high along \mbox{OMC-1} ($G_0$$\simeq$10$^3$-10$^5$). Under these irradiation conditions, the \mbox{CH$^+$~$J$\,$=$\,1--0} line intensity appears tightly correlated with $G_0$ (Fig.~\ref{fig:CHp_correlation}a). 
 The \mbox{CH$^+$ ($J$\,$=$\,1--0)} emission also scales with [\CII]\,158\,$\upmu$m line intensity (Fig.~\ref{fig:CHp_correlation}b). This shows again  that CH$^+$ arises very close to the edge of the molecular cloud. Indeed, the CH$^+$ emission is much less correlated with
 the C$^{18}$O~$J$\,$=$\,2--1 emission that, at first order, traces the column density in
 the colder cloud interior  (Fig.~\ref{fig:CHp_correlation}d). 
Toward positions such as the Orion Bar PDR, where the [\CII]\,158\,$\upmu$m line is very bright:  \mbox{$I$([\CII])$\gtrsim$8$\cdot$10$^{-6}$~W\,m$^{-2}$\,sr$^{-1}$}, the
\mbox{C$^+$ versus CH$^+$} correlation starts to decline. Observations of the $^{13}$C$^+$ fine-structure lines show that the [\CII]\,158\,$\upmu$m line actually becomes optically thick toward 
very bright PDRs
\citep[][]{Ossenkopf13,Goi15}. This may partially explain that the \mbox{C$^+$ versus CH$^+$} relation becomes less linear when the [\CII]\,158\,$\upmu$m emission is very bright and opaque.
In addition, for moderate densities \mbox{($n_{\rm H}\gtrsim 10^4$~cm$^{-3}$)} and FUV radiation fields \mbox{($G_0 \lesssim 10^5$)}, the intensity of the  [\CII]\,158\,$\upmu$m line is governed by the
$G_0/n_{\rm H}$ ratio \citep[e.g.,][]{Kaufman99}. This implies that, in this range of paramenters, the
 \mbox{CH$^+$\,($J$\,$=$\,1--0)/[\CII]\,158\,$\upmu$m} intensity ratio would scale with
 $n_{\rm H}$. Figure~\ref{fig:peaks} (\textit{left}) shows
 a map of this intensity ratio. Indeed, the brightest regions in the plot are associated with the highest
 density PDR layers in the region, $n_{\rm H}\gtrsim 10^6$~cm$^{-3}$ (also revealed by the HCO$^+$ 6--5/3--2 map in Fig.~\ref{fig:ratios-mtc}). Hence, the decline of the \mbox{[\CII]\,158\,$\upmu$m}  versus \mbox{CH$^+$\,($J$\,$=$\,1--0)} emission trend is likely driven by the increase of gas density toward these cloud edges.

Finally, if one excludes the bright CO~$J$\,$=$\,10--9 line emission from shocked gas in outflows, that is,
positions with \mbox{$I_{\rm CO\,10-9}$\,$\gtrsim$\,10$^{-6}$~W\,m$^{-2}$\,sr$^{-1}$} in our maps, there is also a good correlation between \mbox{CH$^+$~$J$\,$=$\,1--0} and \mbox{CO~$J$\,$=$\,10--9} line intensities (Fig.~\ref{fig:CHp_correlation}c).
This is another proof of the \mbox{FUV-irradiated} cloud edge origin of the large scale \mbox{CO~$J$\,$=$\,10--9} emission.

\subsection{CH$^+$ and the extended FUV-pumped H$_2$ emission}

Right panel in Fig.~\ref{fig:G0} shows the tight spatial correlation between the distribution of the \mbox{CH$^+$\,($J$\,$=$\,1--0)} emission (black contours) and that of $G_0$ (in color scale). The strongest FUV fluxes ($G_0$$\simeq$10$^5$) appear near Orion~S, close to the Trapezium cluster. The Orion Bar and East PDRs are also strongly illuminated ($G_0$ of a few  10$^4$). At the edges of the mapped area, the FUV radiation field is still high ($G_0$$\approx$10$^3$).
 Therefore, an intense stellar FUV  flux reaches parsec scales in \mbox{OMC-1} \mbox{\citep{Stacey93,Goi15}}.  
These FUV photons radiatively pump H$_2$ molecules to vibrationally excited states  \mbox{\citep[e.g.,][and references therein]{Hollenbach97}} 
 at the edge of the irradiated cloud, leading to bright near-IR (NIR) H$_2$\,($v$\,$\geq$\,1) emission that is detected  at large-spatial scales \citep[][]{Luhman94}. In particular,  H$_2$ lines from  vibrational levels up to $v$=10 (or \mbox{$E_{\rm u}/k\approx$50,000 K}) have been  detected toward the Orion Bar
\citep[e.g.,][]{Kaplan17}. 

Among the brightest NIR H$_2$\,($v$\,$\geq$\,1) lines is
the H$_2$ \mbox{$v$=1-0 $S$(1)} ro-vibrational line. Figure~\ref{fig:G0} (left) shows a photometric image  of \mbox{OMC-1} taken with  the ESO’s Visible and Infrared Survey for Astronomy (VISTA) in the $K_S$ band \mbox{\citep[][]{Meingast16}}. The  $K_S$  filter is centered at $\lambda$=2.15~$\upmu$m and has a width of $\Delta\lambda$=0.3~$\upmu$m; and thus covers the H$_2$ \mbox{$v$=1-0} $S(1)$ (2.12~$\upmu$m) and \mbox{$v$=2-1} $S(1)$ (2.24~$\upmu$m) lines.  Hence, in addition to the hundred of NIR  (proto)stellar point-sources in the field, this image is sensitive to extended emission of interstellar H$_2$\,($v$\,$\geq$\,1). Hot shocked gas from protostellar outflows also produce bright, collisionally excited, and nearly thermal  H$_2$\,($v$\,$\geq$\,1) emission locally
\citep[e.g.,][]{Rosenthal00}.
The most obvious example in Fig.~\ref{fig:G0} (left) are \mbox{H$_2$ Peaks 1 and 2}
regions in BN/KL, around position \#9 in the maps  
\citep[P1 and P2 in Fig.~\ref{fig:rgb}; see also
][and references therein]{Bally11}.
At large scales, however, the $K_S$ band image is dominated by extended emission from
FUV-pumped H$_2$\,($v$\,$\geq$\,1) \citep[e.g.,][]{Luhman94}.

PDR models predict that at moderate densities, \mbox{$n_{\rm H}$$>$10$^5$~cm$^{-3}$}, the 
\mbox{FUV-pumping} contribution to the \mbox{H$_2$ $v$=1-0 $S(1)$}   line intensity 
scales with $G_0$ \citep[e.g.,][]{Burton1990}. Only if the gas temperature is
$T_{\rm k}$$>$1000~K, collisions will also contribute to populate the vibrational
\mbox{level $v$=1}, so we expect that thermal excitation does not dominate the large-scale IR H$_2$\,($v$\,$\geq$\,1)  emission. Hence, in a first approximation, the extended interstellar emission shown in the $K_S$ image (Fig.~\ref{fig:G0}, left panel) 
should also reflect variations of $G_0$ along the \mbox{FUV-irradiated} surface of \mbox{OMC-1}.
The two images in Fig.~\ref{fig:G0} display the CH$^+$ $J$\,$=$\,1--0 integrated intensity map in black contours.
Although the NIR and submm observations have very different angular resolutions, one clearly infers the presence of  CH$^+$ everywhere 
H$_2$\,($v$\,$\geq$\,1) emits. Indeed, the brightest \mbox{CH$^+$ ($J$\,$=$\,1--0)} emission peaks,  associated to regions of high $G_0$ values, are also bright in NIR H$_2$\,($v$\,$\geq$\,1) emission (Fig.~\ref{fig:G0}).
The observed spatial correlations reflect the tight connection between  both 
the CH$^+$ rotational and  H$_2$\,($v$\,$\geq$\,1) ro-vibrational emission with
the flux of stellar FUV photons.

\subsection{Rotationally warm CH$^+$ emission toward the Trapezium}
\begin{figure}[t]
\centering
\includegraphics[scale=0.36, angle=0]{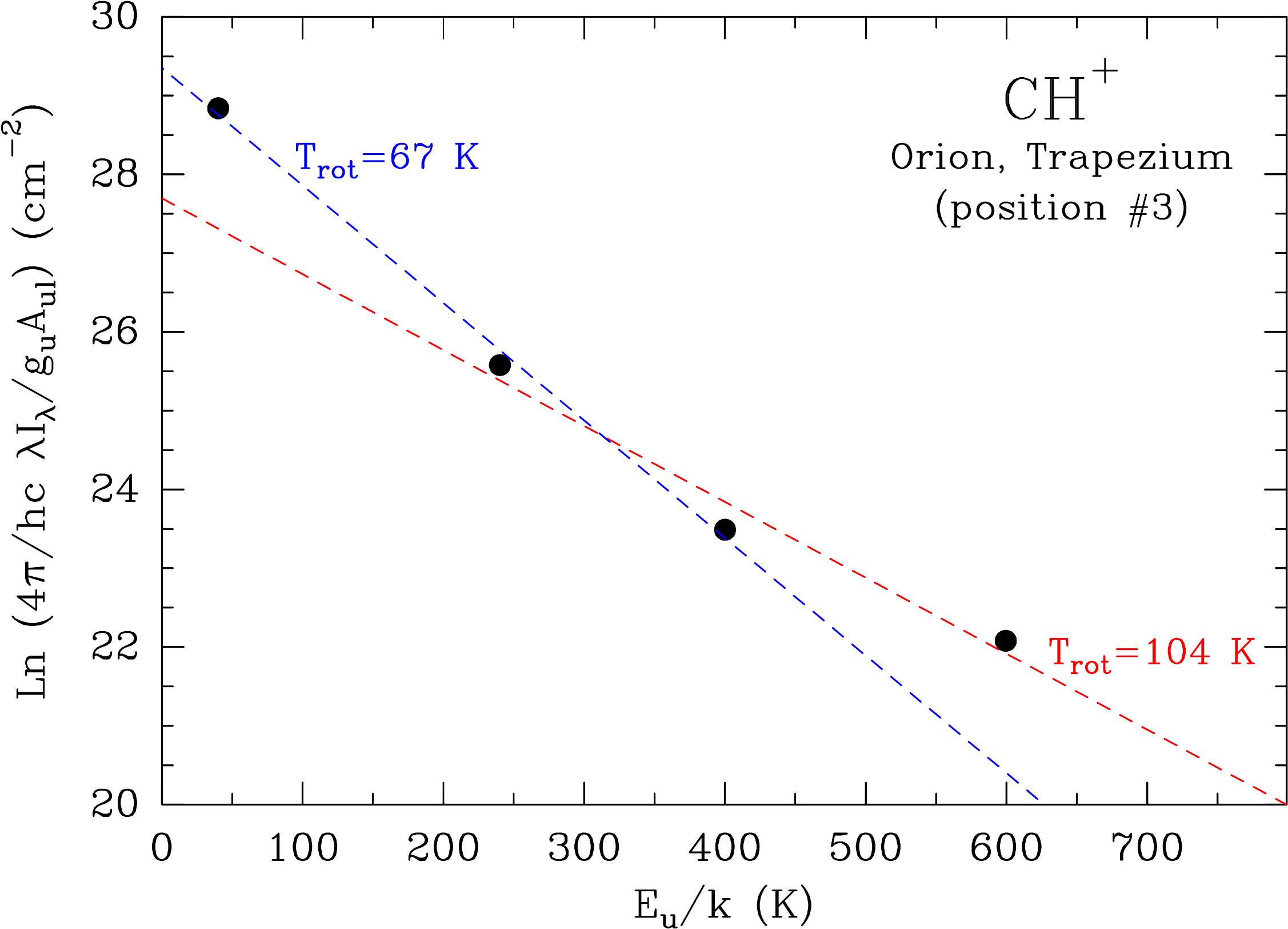}\\
\caption{CH$^+$ rotational population diagram ($J_{\rm u}$=1 to 5) obtained from \textit{Herschel}/PACS and HIFI observations toward \mbox{position\,\#3} near the Trapezium region. The straight lines and associated rotational temperatures are fits to the emission lines in different $J$ ranges. }\label{fig:chp_DR}
\end{figure}

In terms of FUV irradiation, the most extreme conditions are those in the regions between the Trapezium and Orion~S, with $G_0$$\simeq$10$^5$ (Fig.~\ref{fig:G0}, right). These are
even  harsher conditions than in the prototypical PDR the Orion Bar (position $\#$4). Indeed, with 
\mbox{\textit{Herschel}/PACS} we detect  CH$^+$ rotational lines up to 	\mbox{$J$\,$=$\,5--4} 
\mbox{(or $E_{\rm u}/k$\,$\simeq$\,600~K)} toward position $\#$3. The intensities of these rotationally excited CH$^+$ lines (a few \mbox{10$^{-7}$~W\,m$^{-2}$\,sr$^{-1}$}) are a factor of \mbox{$\sim$4 to 8} brighter than toward the Orion Bar  \citep{Nagy13,Joblin18}. 
CH$^+$ rotational lines
have very high critical densities, several \mbox{10$^9$-10$^{10}$~cm$^{-3}$}. Hence, at the gas densities we infer for the edge of \mbox{OMC-1}
\mbox{(Section~\ref{sec-props})}, one would expect very subthermal excitation;
$n_{\rm H}\ll n_{\rm cr}$ implies weak collisional excitation, in other words,
 very low rotational temperatures, $T_{\rm rot}\ll T_{\rm k}$, similar to those inferred for reactive ions such as CO$^+$ or HOC$^+$ \citep[\mbox{$T_{\rm rot}$$\simeq$10-30~K} in the Orion~Bar; see][]{Fuente03,Nagy13,Goico17}.
Figure~\ref{fig:chp_DR} shows the observed CH$^+$ line intensities toward
position \#3 in the form of a rotational population diagram. Due to the observed curvature of the diagram, the measured  line intensities are incompatible with a single rotational temperature.
Just for reference, we note that a simple two-temperature fit provides 
\mbox{$T_{\rm rot}{\rm(CH^+)}$$\simeq$67 and 104~K}, significantly higher than  $T_{\rm rot}$ of other reactive ions. Hence, CH$^+$ is  \mbox{\textit{rotationally warm}}. 
This is consistent with the very short-life of  CH$^+$ in dense gas  and with the exothermic formation pathway from the reaction of C$^+$  with H$_2$\,($v$\,$\geq$\,1). In particular, CH$^+$ 
can be excited by radiation many times during its short lifetime, and during its mean-free-time for inelastic collisions, so that it remains rotationally warm while it emits \citep[][]{Black98,Goico17}.
This chemical formation pumping enhances the populations of CH$^+$ excited levels 
 \citep[especially for \mbox{$J$\,$\geq$\,3}; see][]{Godard13}.

\section{Discussion}\label{Sect:Discussion}

ALMA observations of the Orion Bar have spatially resolved the compressed PDR layers
at the edge of the  molecular cloud, very close to the H to H$_2$ transition zone. 
This layer is characterized by high thermal pressures
 \citep[\mbox{$P_{\rm th}$=$n_{\rm H}$$\cdot$$T_{\rm k}$\,$\approx$\,10$^8$~cm$^{-3}$\,K},][]{Goico16,Joblin18}, 
it is the source of H$_2$\,($v$\,$\geq$\,1) emission, and the only layers where reactive ions such as CH$^+$ and SH$^+$ can \mbox{efficiently} form, and indeed are observed \citep{Nagy13,Goico17,Parikka_2017,Joblin18}. 
Observations of different PDRs in the Milky Way further suggest a  correlation between $G_0$ and the thermal pressure in these compressed layers. In particular,
many PDRs seem to lie in the range  
\mbox{$P_{\rm th}/G_0\approx[5\cdot 10^3-8\cdot10^4]$~cm$^{-3}$\,K}
\citep[][]{Joblin18,Wu18}. Interestingly, this
correlation is predicted, almost independently of the initial gas density, by non-stationary hydrodynamical models of photoevaporative PDRs  \citep[][]{Bron18}.
The high pressures inferred at the PDR surface, often unbalanced by those
of the surrounding environment, would then have a \mbox{dynamical} origin:
cloud edge heating, compression, and photoevaporation. These proceses greatly depend on
the strength and shape of the stellar UV radiation field  \citep[][]{Hill78,Bertoldi89,Bertoldi96,Storzer98,Hosokawa06,Pellegrini_2009,Bron18}.
ALMA observations of the Orion Bar PDR do suggest the presence of photoevaporative flows
of neutral gas \citep{Goico16}.

\subsection{PDR modeling}\label{sub-sec:PDR-mods}

We have used the \textit{Meudon} PDR  code \citep[e.g.,][]{LePetit_2006,Goicoechea07,Bron14} 
 to model the cloud edge layers where the observed submm line emission arises.
 The model simulates a \mbox{stationary} PDR; the penetration of FUV photons, the thermal balance, and the chemistry are computed self-consistently. 
We adopted an extinction to color-index ratio, $R_{\rm V}=A_{\rm V}/E_{\rm B-V}$, of 5.5 consistent with the flatter extinction curve toward Orion \citep{Cardelli89}. Regarding CH$^+$ formation, these models include an H$_2$ state-dependent  treatment of reaction~(1)  \citep[see][]{Agundez10,Zanchet13,Herraez14,Faure17}. 
In particular, the CH$^+$ formation rate is computed by summing over all formation rates for each specific state of H$_2$. 

\begin{figure}[t]
\centering
\includegraphics[scale=0.47, angle=0]{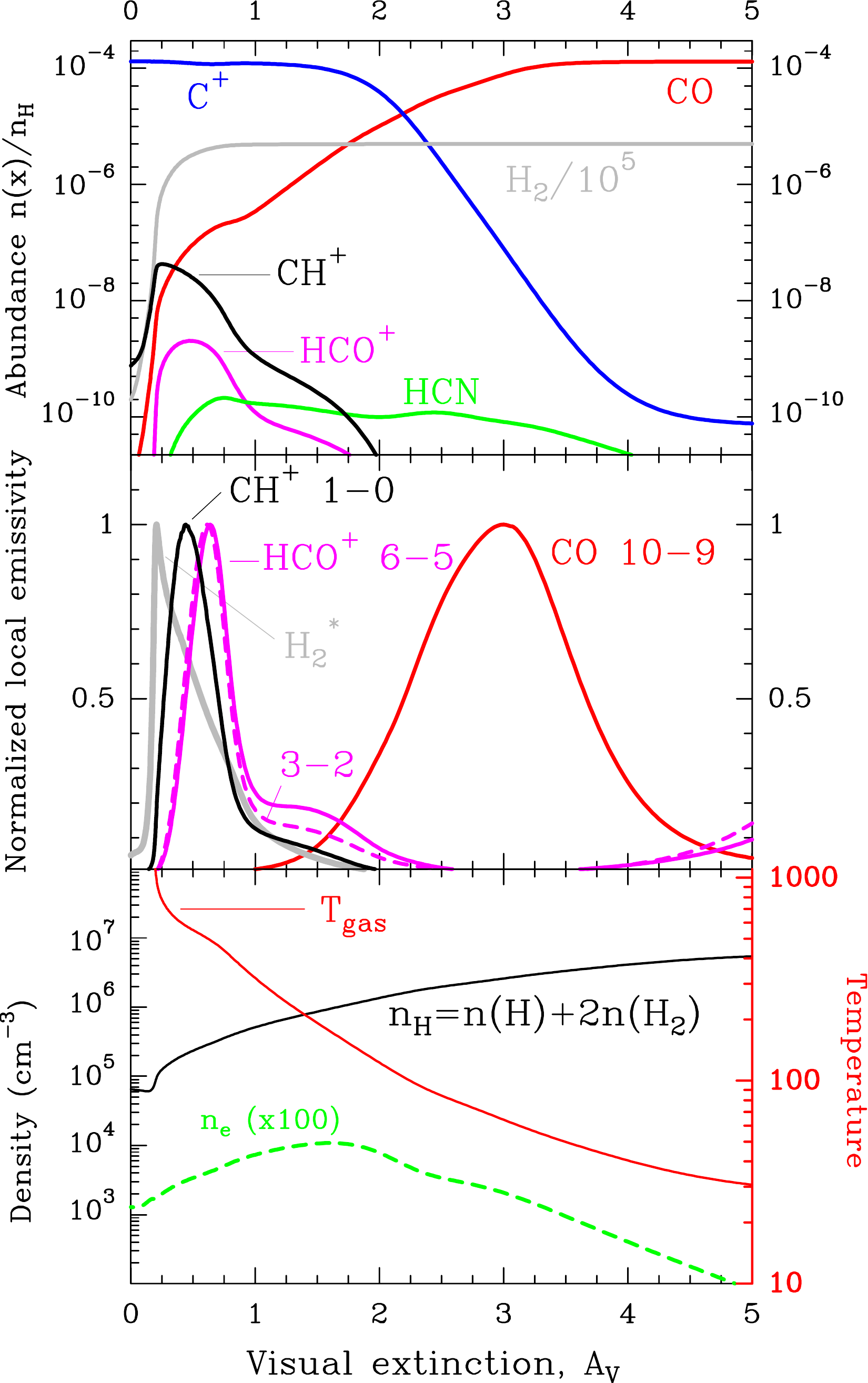}\\
\caption{Isobaric PDR model with \mbox{$P_{\rm th}$\,=\,10$^8$~cm$^{-3}$\,K}
and \mbox{$G_0$\,=\,7.5$\cdot$10$^3$}, the mean $G_0$  in \mbox{OMC-1}.
\textit{Top panel}: Fractional abundances as a function of visual extinction 
($A_{\rm V}$) into the molecular cloud. \textit{Middle}: Normalized local line emissivities.
H$_2^{*}$ stands for the H$_2$ $v$=1-0 $S$(1) line at 2.12~$\upmu$m.
\textit{Bottom}: Predicted gas temperature ($T_{\rm gas}$), hydrogen nuclei gas density ($n_{\rm H}$) and electron density ($n_{\rm e}$) profiles.
}\label{fig:PDR_mod}
\end{figure}

 \citet{Bron18} concluded that, in the frame of \mbox{stationary} PDR models, a better description of these quasi-isobaric layers, typically between \mbox{$A_{\rm V}$\,=\,0} and $\sim$5~mag  into the cloud,  is obtained by constant pressure  models instead of the constant density models often used in the literature.  With this in mind, in \mbox{Fig.~\ref{fig:PDR_mod}} we show a representative stationary PDR model adapted to the  irradiated edge of \mbox{OMC-1}.
This is a high-pressure, \mbox{$P_{\rm th}$=10$^8$~cm$^{-3}$\,K}, \mbox{isobaric} PDR illuminated by the mean FUV radiation flux in the region mapped by HIFI ($<G_0>\,\simeq7.5\cdot$10$^3$).  
The bottom panel in \mbox{Fig.~\ref{fig:PDR_mod}} shows the predicted physical structure, from 
\mbox{$A_{\rm V}$=0 to 5~mag} of visual extinction into the molecular cloud.
 The upper panel shows fractional abundances of H$_2$, C$^+$, CH$^+$, CO, HCO$^+$  and HCN
 with respect to H nuclei. We note that CH$^+$ and HCO$^+$ abundances peak at the edge of the PDR, before the C$^+$ to CO transition layers,
 that is, closer to the \HII~region. Therefore, both CH$^+$ and HCO$^+$ are 
 abundant\footnote{HCO$^+$ is also abundant deep inside the shielded molecular cloud, mainly formed by the reaction CO + H$_{3}^{+}$ $\rightarrow$ HCO$^+$ + H$_2$.} in layers where the ionization fraction, driven by the photoionization of carbon atoms, is high, $x$($e^-$)$\approx$10$^{-4}$. The high electron density in these layers, \mbox{$n_e$$\simeq$30} to 100~cm$^{-3}$, contributes to the collisional excitation of their rotational levels.  The \mbox{middle} panel in \mbox{Fig.~\ref{fig:PDR_mod}} shows the predicted local emissivities
 of the observed molecular lines, normalized by their emission peak. 
In the optically thin limit, the actual line intensities will follow the same spatial distribution. 

Because of the efficient chemical formation of HCO$^+$ from reactive ions CH$^+$ and CO$^+$   
at the PDR surface, before the C$^+$ to CO transition \citep[e.g.,][]{Sternberg_1995,Goico16}, the 
\mbox{CH$^+$ $J$\,$=$\,1--0} and  \mbox{HCO$^+$ $J$\,$=$\,6--5} lines are predicted to trace hotter gas (\mbox{$T_{\rm k}$$\simeq$500\,K} in this model), than the \mbox{CO\,$J$\,$=$\,10--9} emission. 
The \mbox{CO\,$J$\,$=$\,10--9} line, traces cooler and an order of magnitude denser gas, \mbox{$n_{\rm H}$$\simeq$10$^6$\,cm$^{-3}$}, that arises from slightly deeper inside the PDR. 
Given the relatively high gas densities in these warm PDR layers, the observed submm molecular lines arise from a very thin layer: 
\mbox{$\sim$1.6$\cdot$10$^{16}$\,cm$^{-2}$}\mbox{$\simeq$5$\cdot$10$^{-3}$\,pc}\mbox{$\simeq$1000\,AU} in this particular model. This is not far from previous claims,
based on \mbox{low-angular} resolution observations, 
for the dense molecular gas confining the \HII~region M42 \citep[e.g.,][]{Rodriguez01}. At the distance to Orion, these spatial scales imply angular thicknesses of only several arcseconds. 
In other words, they cannot be spatially resolved by single-dish telescopes.
This explains the similar  morphology  of the \mbox{CH$^+$\,$J$\,$=$\,1--0} and 
\mbox{CO\,$J$\,$=$\,10--9} extended emission observed by \textit{Herschel}.
ALMA images of SH$^+$ and \mbox{HCO$^+$\,($J$\,=\,3--2)} emission 
do resolve the compressed, warm layers of the Orion Bar PDR
\mbox{\citep{Goico16,Goico17}}. 

\begin{figure}[t]
\centering
\includegraphics[scale=0.33, angle=0]{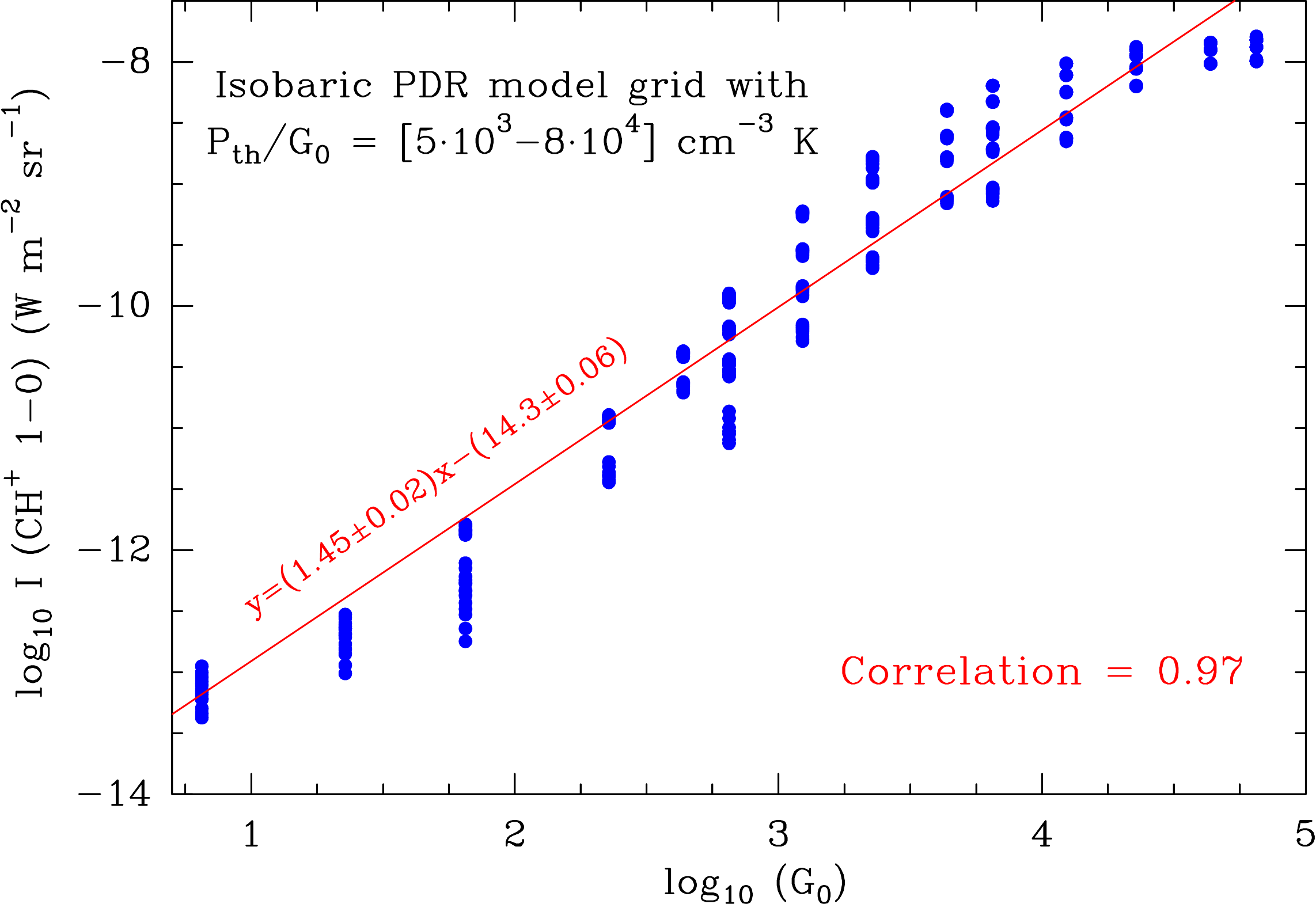}\\
\caption{Results from a grid of isobaric PDR models with the 
$P_{\rm th}$/$G_0$ ratio ranging from 5$\cdot$10$^3$ to 8$\cdot$10$^4$~cm$^{-3}$\,K (see text). 
The plot shows a correlation between incident FUV flux, $G_0$, and  CH$^+$\,($J$\,$=$\,1--0)  intensity. }\label{fig:PDR_grid}
\end{figure}

We also analyzed the results of a grid of $\sim$1300 stationary, isobaric PDR models,  also run with the \textit{Meudon} code, and searched for a theoretical validation of the observed \mbox{$I$\,(CH$^+$\,1-0)}
versus \mbox{$G_0$} correlation (Section~\ref{sec-correlations}).
\mbox{Figure~\ref{fig:PDR_grid}} shows results from models that cover a wide parameter 
space in \mbox{$G_0$ and $P_{\rm th}$}, and satisfy the condition 
$P_{\rm th}$/$G_0$ from 5$\cdot$10$^3$ to \mbox{8$\cdot$10$^4$~cm$^{-3}$\,K},
the same range obtained from photoevaporative PDR models \citep{Bron18}
and also suggested by observations \citep[e.g.,][]{Joblin18,Wu18}.
In this broad parameter space of illumination conditions, models do predict a  correlation between incident FUV flux and  \mbox{CH$^+$ $J$\,$=$\,1--0} line intensity. The slope of the predicted correlation, in logarithmic scale, is $m_{\rm mod}$=1.45$\pm$0.02
(Fig.~\ref{fig:PDR_grid}). This value is only slightly above the one from \mbox{OMC-1} observations, $m_{\rm obs}$=1.0$\pm$0.1, but covering regions of higher FUV fluxes, $G_0$$\gtrsim$10$^3$.
This satisfactory agreement suggests that, for a wide range of \mbox{$G_0$} values, the stationary  isobaric PDR model does  capture the global properties of the compressed
PDR layers at the irradiated edges of \mbox{OMC-1}. It also supports the observational result that the \mbox{CH$^+$ $J$\,$=$\,1--0} emission  traces the strength of the incident FUV flux.

\subsection{CH$^+$, an unambiguous tracer of FUV-irradiated gas}

The [\CII]\,158\,$\upmu$m line emission is the most luminous tracer of the extended \mbox{FUV-irradiated} neutral gas in the Milky Way \mbox{\citep[e.g.,][]{Bennett94}}. In some circumstances, however,  the interpretation of [\CII]\,158\,$\upmu$m observations is not trivial: uncertain contribution from \HII~ionized gas,
multiple velocity components along a given line-of-sight, optical depth effects toward bright and dense PDRs, etc. \citep[see e.g.,][]{Ossenkopf13,Pabst17}. 
The detection of rotational emission from reactive molecular ions 
such as CH$^+$ or SH$^+$ is an alternative and  unambiguous signature of  FUV-irradiated 
dense\footnote{We refer to \textit{dense} molecular gas, $n_{\rm H} \gtrsim 10^5$\,cm$^{-3}$, as opposed to the detection of CH$^+$\,($J$\,$=$\,1--0) absorption from \mbox{low-density}, 
\mbox{$n_{\rm H} \lesssim 100$\,cm$^{-3}$}, diffuse interstellar clouds 
\citep[e.g.,][]{Godard12}} molecular gas
 \mbox{\citep[][]{Black98,Nagy13,Goico17}}. 
 We have shown that the intensity of the \mbox{CH$^+$~$J$\,$=$\,1--0} line scales with $G_0$. Despite the elevated endothermicity of reaction \mbox{C$^+$ + H$_2$\,($v$=0) $\rightarrow$ CH$^+$ + H}, a high \mbox{FUV photon flux} enhances the gas temperature, the column density of FUV-pumped H$_2$\,($v$\,$\geq$\,1), and that of CH$^+$ \citep{Sternberg_1995,Agundez10,Faure17}.
 We conclude that reaction~(1) is not only efficient  locally, in bright and dense PDR/\HII~interfaces such as the Bar. It is also efficient at large spatial scales 
where \mbox{CH$^+$~($J$\,$=$\,1--0)} traces an extended but thin layer of 
\mbox{FUV-irradiated} molecular gas.

\subsection{Observational tracers of the radiative feedback from massive stars
and extragalactic link}

Compared to the lowest-energy \mbox{``$J$\,$=$\,1--0''} rotational lines of  CO, HCN, HCO$^+$, or N$_2$H$^+$  typically mapped from  \mbox{ground-based} telescopes
\citep[e.g.,][]{Pety17,Hacar17,Kauffmann17}, 
the more excited submm  \mbox{``mid--$J$''} lines are
better diagnostics of the high-pressure gas in star-forming regions (SFRs).
Because \mbox{FUV radiation} and \mbox{shocks} dominate the heating of the 
 warm gas in molecular clouds, \mbox{mid-$J$} lines are more sensitive probes of the radiative and mechanical feedback from stars and protostars. 
This warm gas component can dominate the
 line luminosity emitted by SFRs in the Milky Way, and from distant \mbox{star-forming galaxies} as a whole.

Most  spectroscopic observations carried out by \textit{Herschel} toward nearby SFRs  \citep[e.g.,][]{vanDishoeck11} targeted  low-mass protostars and their outflows \citep[e.g., ][]{Nisini10,Kristensen12,Herczeg12,Goicoechea12,Karska13,Manoj13}, as well as 
more distant high-mass star-forming cores \citep[e.g.,][]{Bergin10,Etxaluze13,Tak13,Karska14,Goipacs15,Indriolo17}.
These are relatively compact sources, at \textit{Herschel}'s angular resolution, producing intense FIR \mbox{high-$J$ CO}
 and H$_2$O line emission that arises from  hot molecular  gas, \mbox{$T_{\rm rot}$(CO)\,$\gtrsim$\,300~K}, associated to  protostellar outflows.
The observed  CO rotational emission ladder toward these sources typically peaks at \mbox{$J_{\rm peak}$\,$>$\,15}  
and shows detectable CO emission at \mbox{$J$\,$\gtrsim$\,30}. Although at the spatial scales of an entire GMC, the filling factor of these protostellar outflows is small,
the shape of their CO spectral line energy distribution (SLED)
 is representative of hot, shock-heated molecular gas.
In a broader context, this shock-heated   component may dominate the observed  FIR \mbox{high-$J$ CO} emission from active galactic nuclei (AGN) galaxies in which these lines are readily detected \mbox{\citep[e.g.,][]{Sturm10,Hailey12}}.

The  CO line emission measured by \textit{Herschel} toward less extreme star-forming galaxies typically  peaks
at $J_{\rm peak}$\,$\lesssim$\,10 \citep[e.g.,][]{Kamenetzky12,Kamenetzky14,Indriolo17}. Hence, most of
the \mbox{mid-$J$ CO} luminosity emitted at kpc scales likely arises from non-shocked warm molecular gas. In normal (\mbox{Milky Way} type) and starburst galaxies, the observed mid-$J$ emission possibly arises from  widespread  PDR gas, similar to that in the extended irradiated surface of  \mbox{OMC-1}, that is, relatively dense molecular gas exposed to FUV radiation from \mbox{OB-type} stars.
As in any local PDR, this  warm gas component  copiously emits at FIR and submm wavelengths  \mbox{\citep[][]{Hollenbach97}}. This emission includes bright CO lines with a SLED \mbox{peaking} at \mbox{$J_{\rm peak}$\,$\simeq$\,10 to 15} for strongly irradiated Orion Bar-like PDRs with $G_0$\,$>$\,10$^3$ \citep{Joblin18}, at \mbox{$J_{\rm peak}$$\gtrsim$7--6} for lower \mbox{FUV-radiation} fields, $G_0$\,$<$\,10$^3$, such as the extended envelope around Sgr\,B2 cloud
\mbox{\citep[][]{Etxaluze13}}, or the Horsehead PDR, with $G_0$\,$\simeq$\,10$^2$.

Using \mbox{OMC-1} as template to quantify the radiative impact of \mbox{young massive} stars on their natal cloud, we see that the \mbox{FUV} flux is still strong enough to drive
the gas heating and chemistry of an extended gas component of warm,
 \mbox{$T_{\rm k}$$\approx$100-150\,K}, molecular gas
at pc distances from massive stars. This is seen from the  larger extent of the   \mbox{CO~$J$\,$=$\,10--9} emission compared to that of  \mbox{C$^{18}$O~$J$\,$=$\,2--1} \mbox{(Fig.~\ref{fig:Original_maps1})}. 
The HCO$^+$ and HCN~$J$\,$=$\,6--5 maps reveal that the gas 
density at the cloud edges are relatively high, 
\mbox{$n_{\rm H}$$\gtrsim$10$^5$-10$^6$\,cm$^{-3}$}, so the thermal pressures are high.  
Such densities may be related to a particularly dense condensation that gave rise to the \mbox{Trapezium stars}; they may be related to the cloud evolution itself, as a consequence of the global gravitational collapse of \mbox{OMC-1} \citep[e.g.,][]{Hartman07,Hacar17}; or  produced by the \mbox{FUV-induced} compression of the  cloud surfaces \citep[e.g.,][]{Goico16,Bron18}.

Square-degree maps of the GMCs  Orion~A and B in the HCN~$J$\,$=$\,1--0 line emission have revealed that, despite the high critical density of this transition, several 10$^6$\,cm$^{-3}$, most of the  emission arises from relatively low-density gas \citep[\mbox{$n_{\rm H}$$\approx$10$^3$\,cm$^{-3}$};][]{Pety17,Kauffmann17}. 
Therefore,  the mere observation of a low-energy  but high critical density  line does not immediately imply that the emission arises from cold and dense gas, the fuel that ultimately forms stars,
 as often assumed in the interpretation of
extragalactic observations \mbox{\citep[e.g.,][]{Gao04,Usero15}}.
The very extended \mbox{HCN~($J$\,$=$\,1--0)} emission detected along GMCs of the Milky Was  just reveals emission from  weakly collisionally-excited gas
\mbox{($T_{\rm ex}$(1--0)\,$\ll$\,$T_{\rm k}$)}, perhaps assisted by electron excitation if the ionization fraction is high enough due to FUV irradiation \mbox{\citep[e.g.,][]{Goldsmith17}}. This may be the case of many galaxies too. Our detection of extended \mbox{HCN $J$\,$=$\,6--5} emission and large \mbox{HCO$^+$ 6--5/3--2} intensity ratios in  \mbox{OMC-1}, however, does demonstrate the presence warm and dense$^5$ molecular gas
at 1~pc$^2$ scales. 

The \mbox{HCN to HCO$^+$ $J$\,$=$\,6--5} line ratio~($R$) map of \mbox{OMC-1} shown in 
\mbox{Fig.~\ref{fig:peaks}} further differentiates the extended
gas illuminated by FUV radiation ($R$$<$0.5), from the very high IR-luminosity regions,
such as BN/KL with  $R$$\simeq$2. These are regions hosting on-going massive star-formation and powerful protostellar outflows still
buried in large column densities of FIR-emitting dust. 
In the extragalactic context, ultraluminous infrared galaxies (ULIRGs), that is, merger galaxies
characterized by very high star-formation rates and massive nuclear outflows 
\citep[e.g.,][and references therein]{Gonzalez18} also display elevated 
$R$$\simeq$1-2 luminosity ratios  \citep[e.g.,][and references therein]{Krips08,Imanishi18}.
Pure starburst galaxies, however, show \mbox{$R$ ratios} smaller than one \citep[e.g.,][]{Salas14,Aladro15}, similar to our observed value
 for the extended gas in \mbox{OMC-1}.
 
The recent ALMA detection of \mbox{CH$^+$\,($J$\,$=$\,1--0)} emission and absorption toward  ULIRGs 
 at $z$$\sim$2 \citep{Falgarone17} demonstrates the surprisingly widespread nature of CH$^+$, also in distant and  extreme galaxies.
The absorbing CH$^+$ presumably arises from a turbulent diffuse gas reservoir.
The   emission component, emitted at \mbox{kiloparsec scales}, is interpreted as denser  gas affected by a collection of multiple \mbox{FUV-irradiated} shocks (needed to explain the broad line profiles) individually propagating at velocities of \mbox{$v_{\rm shock}\sim40$~km\,s$^{-1}$}, and driven by galactic winds \mbox{\citep{Falgarone17}}.
Our detection of parsec scale \mbox{CH$^+$\,($J$\,$=$\,1--0)} emission toward \mbox{OMC-1}, a star-forming cloud core irradiated by FUV photons from just a few young massive stars, provides an example of  extended CH$^+$ emission from \mbox{non-shocked gas} ($\Delta {\rm v} \sim 5$\,km\,s$^{-1}$). 
Because the lifetime of CH$^+$ in dense gas is so short, this CH$^+$ emission traces the instantaneous feedback from massive stars. This may be the case of many  starburst galaxies producing detectable CH$^+$
rotational emission too.

\section{Summary and conclusions}\label{Sect:Summary}

We have presented \mbox{$\sim$85~arcmin$^2$} \mbox{($\sim$0.9\,pc\,$\times$\,1.4\,pc)} velocity-resolved  maps of the  
\mbox{CH$^+$~($J$\,$=$\,1--0}), \mbox{CO~($J$\,$=$\,10--9)}, 
\mbox{H$_2$O (3$_{12}$-2$_{21}$)}, CH$_3$OH, \mbox{CH~($N$,\,$J$\,$=$1,\,3/2--1,\,1/2)}, HCN (\mbox{$J$\,$=$\,6--5} and \mbox{13--12}), and \mbox{HCO$^+$~($J$\,$=$\,6--5)} lines, obtained with the
\mbox{heterodyne} instrument \textit{Herschel}/HIFI toward the closest high-mass star-forming region \mbox{OMC-1}.
In combination with archival NIR photometric images tracing the
H$_2$\,($v$\,$\geq$\,1) emission, and new 
\mbox{HCO$^+$ $J$\,$=$\,3--2} and \mbox{C$^{18}$O $J$\,$=$\,2--1} line maps taken
with the IRAM\,30\,m telescope, we obtained the following results:
  
$\bullet$~The rotational emission from reactive ion CH$^+$ is very extended.
 The \mbox{CH$^+$~($J$\,$=$\,1--0})  emission  spatially correlates with that of [\CII]\,158\,$\upmu$m and of \mbox{FUV-pumped H$_{2}$}. The observed correlations  imply that the reaction of  C$^+$ ions with  H$_2$\,($v$\,$\geq$\,1) molecules  dominates the formation of CH$^+$, not only locally toward dense PDR/\HII~interfaces, but also at parsec  scales.
 
$\bullet$~The extended CO\,($J$\,$=$\,10--9) and HCO$^+$\,(mid-$J$)  narrow-line emission 
(\mbox{$\Delta$v\,$\simeq$\,3\,km\,s$^{-1}$}) 
 traces a thin layer  of warm  gas at the surface of the molecular cloud
 ($A_{\rm V}$\,$\simeq$\,3-6\,mag or $\sim$10$^{16}$\,cm in thickness)
  illuminated by \mbox{FUV radiation} from massive stars in the Trapezium cluster. 
 This layer, with a  mass density of \mbox{120-240\,$M_{\odot}$\,pc$^{-2}$}, accounts for $\sim$5\,\% to $\sim$10\%~of the total gas mass in \mbox{OMC-1}.
 Using \mbox{non-LTE} excitation  models we infer high thermal pressures, 
\mbox{$P_{\rm th}=T_k \cdot n_{\rm H} \gtrsim 10^7-10^9$~cm$^{-3}$\,K}, for this gas component. This is consistent with the expected dynamical response of molecular clouds to strong \mbox{FUV} radiation: the cloud edge is heated and compressed, and
photoevaporates if the high pressures in the PDR are not balanced by those of the  environment \citep[e.g.,][]{Bertoldi96,Bron18}.

$\bullet$~Photoevaporative PDR models predict that the quasi-constant thermal pressure in the compressed   PDR layers scales with the strength of the \mbox{FUV photon} flux  impinging the cloud
\citep{Bron18}. CH$^+$ turns as a unique tracer of these narrow  layers. In this work we have found a spatial correlation between the intensity of the \mbox{CH$^+$\,($J$\,$=$\,1--0)}  line and $G_0$, ranging from $G_0$$\sim$10$^3$ to $\sim$10$^5$  in \mbox{OMC-1}. The observed correlation is supported, and enlarged to lower $G_0$ values, by isobaric   models in the parameter space \mbox{$P_{\rm th}/G_0$\,$\approx$\,[5$\cdot$10$^3$--8$\cdot$10$^4$]~cm$^{-3}$\,K} where many  PDRs 
seem to lie \citep[][]{Joblin18,Wu18}. This correlation also implies that for  moderate gas densities, \mbox{$n_{\rm H}$$>$10$^5$ cm$^{-3}$},  the \mbox{CH$^+$\,($J$\,$=$\,1--0})/[\CII]\,158\,$\upmu$m intensity ratio 
traces gas density variations at the irradiated  cloud surfaces.

$\bullet$~We detect bright and  rotationally warm, $T_{\rm rot}$$\simeq$100\,K,  CH$^+$ emission toward the most irradiated regions facing the Trapezium cluster.
The detection of excited CH$^+$  lines, up to $J$\,$=$\,5--4, is consistent with its exothermic formation  from H$_2$\,($v$\,$\geq$\,1), and with the formation pumping
mechanism that enhances the population of CH$^+$ rotationally excited levels
\mbox{\citep{Godard13,Faure17}}. The broad $J=$\,1--0 line-widths are consistent with the high reactivity of the molecule: CH$^+$ can be excited by radiation many times during its short lifetime (a few hours), and during its mean-free-time for elastic and inelastic collisions, so that it remains kinetically hot (large velocity dispersion) and rotationally warm (high $T_{\rm rot}$) while it emits \citep[e.g.,][]{Black98,Goico17}.\\

The detection of extended, parsec-scale CH$^+$ ($J$\,$=$\,1--0) and narrow-line \mbox{mid-$J$ CO} emission in \mbox{OMC-1},
both lines arising from  PDR gas and not from fast shocks, probes the radiative interaction between young massive stars and their natal molecular cloud. This radiative feedback alters the \mbox{dynamics}, physical conditions, and chemistry of the most exposed neutral cloud layers. In turn, although not the most massive cloud core component, these PDR layers dominate the line luminosity emitted by GMCs.
Similar processes must take place in other clusters  hosting  more numerous
and more massive stars \mbox{\citep[e.g.,][]{Wareing18}}. They likely occur in starburst galaxies as well. There, the PDR line emission from C$^+$, \mbox{mid-$J$ CO}, and reactive ions such as CH$^+$ must reach  larger spatial scales, and can dominate the emitted
line luminosity too.

\begin{acknowledgements}

We thank the Spanish MICIU for funding support under grants AYA2016-75066-C2-1-P and AYA2017-85111-P, and the ERC for support under grant ERC-2013-Syg-610256-NANOCOSMOS.

\end{acknowledgements}

%-------------------------------------------------------------------

\bibliographystyle{aa}
\bibliography{references}

%\clearpage

\begin{appendix}\label{Sect:Appendix}

\section{Complementary Tables}\label{App-more-tables}

\begin{table*}	
\caption{Line centroid LSR velocities (in km\,s$^{-1}$) extracted from Gaussian line fits toward selected positions. \label{table:vlsr}} 
\centering
\begin{tabular}{lrrrrrrr@{\vrule height 8pt depth 5pt width 0pt}}
\hline\hline
v$_{LSR}$ (km\,s$^{-1}$)$^a$: &  Line:       &          &          &            &            &             &         \\ 
Position   &  CH$^+$ 1--0  &[\CII]\,158\,$\upmu$m & CO 10--9  & CO 2--1  & $^{13}$CO 2--1    &  C$^{18}$O 2--1 & HCO$^{+}$ 3--2 \\\hline
$\#1$      &  8.3(3)    & 8.5(1)   & 8.8(1)     & 9.2(1)     & 9.3(1)     & 9.3(1)     & 9.3(1)        \\ 
$\#2$      &  9.2(3)    & 8.7(1)   & 9.3(1)     & 9.2(1)     & 9.3(1)     & 9.3(1)     & 9.3(1)        \\ 
$\#3$      &  8.9(1)    & 9.2(1)   &  9.2(1)    & 9.0(1)     & 8.4(1)     & 8.5(1)     & 8.6(1)        \\ 
$\#4$      &  10.6(1)   & 10.4(1)  & 10.7(1)    & 10.1(1)    & 10.2(1)    & 10.3(1)    & 10.3(1)        \\ 
$\#5$      &  8.7(1)    &  8.4(1)  & 9.2(1)     & 9.3(1)     & 9.4(1)     & 9.2(1)     & 9.0(1)        \\ 
$\#6$      & 9.8(1)     & 9.3(1)   &  9.4(2)    & 9.4(1)  & 9.2(1)     & 9.3(1)     & 9.4(1)          \\ 
$\#7$      &  10.8(1)   & 10.7(1)  & 10.9(1)    & 10.4(1)    & 10.6(1)    & 10.4(1)    & 10.6(1)         \\ 
$\#8$      &  10.9(1)   & 10.3(1)  & 10.5(9)    &  9.9(1)    & 9.7(2)     & 9.4(2)     & 9.3(1)        \\ 
$\#9$      &  10.2(1)   &  9.0(1)  & 7.8(1)         &  9.0(1)    & 9.8(1)     & 9.8(1)     & 9.6(1)        \\ 
$\#10$     &  8.3(1)    & 8.5(1)    &  8.8(1)   & 8.4(1)     & 8.5(1)     & 8.4(1)     & 8.2(1)        \\ 
\hline
Average$^b$ & 10(3) & 9.3(9)  &  9.6(8)   & 9.4(6)     & 9.4(7)     & 9.3(6)     & 9.3(7)        \\
\hline
\end{tabular}
\tablefoot{$^a$From maps convolved to the angular resolution of the CH$^+$ 1-0 line (27$''$). 
$^b$We have not used position $\#9$ for the averages.
Parentheses indicate the uncertainty obtained by the Gaussian fitting programme. The fit uncertainty is in units of the last significant digit.}
\end{table*}

\begin{table*}
\caption{Line-widths (in km\,s$^{-1}$) extracted from Gaussian line fits toward selected positions. \label{table:widths}} 
\centering
\begin{tabular}{lrrrrrrr@{\vrule height 8pt depth 5pt width 0pt}}
\hline\hline
$\Delta$v (km\,s$^{-1}$)$^a$: &  Line:       &            &         &             &          &              &         \\ 
Position      &  CH$^+$ 1--0      &[\CII]\,158\,$\upmu$m &CO 10--9  & CO 2--1    & $^{13}$CO 2--1    &  C$^{18}$O 2--1 & HCO$^{+}$ 3--2 \\\hline
$\#1$                     &  4.9(7)    & 4.4(1)    & 2.3(1)    & 3.9(1)    & 2.6(1)  & 2.2(1)  &  2.7(1)        \\ 
$\#2$                     &  4.4(6)    & 4.1(2)    & 1.8(1)    & 4.0(1)    & 2.5(1)  & 2.0(1)  &  2.5(1)    \\ 
$\#3$                     &  5.7(1)    & 4.9(1)    & 4.4(1)    & 4.9(1)    & 3.8(1)  & 3.2(2)  &  3.8(1)    \\ 
$\#4$                     &  5.3(1)    & 4.0(1)    &  3.4(1)   & 4.1(1)    & 2.6(1)  & 2.1(1)  &  3.2(1)       \\ 
$\#5$                     &  4.8(2)    & 3.6(1)    & 2.3(1)    &  4.8(1)   & 3.8(1)  & 4.1(1)  &  3.4(1)       \\ 
$\#6$                     &  4.1(1)    & 2.9(1)    & 2.2(1) &  3.3(1)   & 2.1(1)  & 1.8(1)  &  2.3(1)      \\  
$\#7$                     &  4.4(1)    & 4.2(1)    & 3.5(1)     & 4.6(1)    & 3.5(1)  & 3.2(2)  &  3.0(1)       \\ 
$\#8$                     &  6.5(1)    & 5.5(1)    & 4.7(2)    &  4.6(1)   & 3.2(1)  & 4.4(3)  &  5.3(2)       \\
$\#9$                     &  6.8(2)    & 3.6(1)    & 38.6(1)      &  24.0(4)  & 6.3(1) & 3.5(1)  &  6.4(1)       \\
$\#10$                    &  4.4(2)    & 3.4(1)    &3.5(1)     &  5.0(1)    & 3.3(1) & 3.3(1)  &  3.5(1)       \\ 
\hline
Average$^b$               &  4.9(7)    & 4.1(7)     & 3.0(1)    & 4.4(5)    & 3.0(6)  & 2.9(9)  & 3.3(8)        \\
\hline
\end{tabular}
\tablefoot{$^a$From maps convolved to the angular resolution of the CH$^+$ 1-0 line (27$''$). 
$^b$We have not used position $\#9$ for the averages. Parentheses indicate the uncertainty obtained by the Gaussian fitting programme. The fit uncertainty is in units of the last significant digit.}
\end{table*}

\begin{table*}
\caption{Integrated-line intensities (in main brightness temperature, K\,km\,s$^{-1}$) extracted from Gaussian line fits toward selected positions. \label{table:areas}} 
\centering
\begin{tabular}{lrrrrrrr@{\vrule height 8pt depth 5pt width 0pt}}
\hline\hline
$W$(K\,km\,s$^{-1}$)$^a$: &  Line:          &            &              &            &           &               &         \\ 
Position                  &  CH$^+$ 1--0  & [\CII]\,158\,$\upmu$m  & CO 10--9	  & CO 2--1   & $^{13}$CO 2--1    &  C$^{18}$O 2--1 & HCO$^{+}$ 3--2 \\\hline
$\#1$    &  1.7(2)  	    & 332.1(4)  	& 57.0(2)     &  209.8(1)    &   64.7(1)   & 8.2(1)        & 11.3(3)  \\ 
$\#2$    &  2.2(3)   	    & 349(12)  	    & 53.5(2)  	  &  216.3(1)    &   67.1(2)   & 10.7(1)       & 22.5(3)   \\ 
$\#3$    &  47.2(1)  	    & 1048(3)    	& 408(1) 	  &  446(3)      &  111.0(1)  & 17.6(1)       & 75.5(2)   \\ 
$\#4$    &  26.5(1)  	    & 1081(2)  	    & 324.6(3) 	  &  384.0(7)    &   56.5(1)   & 5(1)          & 27.6(2)   \\ 
$\#5$    &  3.8(1)   	    & 392(4)   		& 53.5(2)  	  &  268.3(1)    &   72.6(1)   & 9.8(1)        & 10.0(2)    \\ 
$\#6$    & 7.1(1)   	    & 492(1)		& 153.0(2)  & 255.6(2)   &   57.6(5)   & 5.5(1)        &  18.8(3)   \\ 
$\#7$    &  24.8(1)  	    & 810(17)   	& 255.5(4) 	  &  363(2)      &   40.0(2)   & 2.6(1)        & 16.9(3)     \\
$\#8$    &  6.7(1)   	    & 508(8)   		& 10.5(4)  	  & 175.3(1)     &   17.6(1)   & 1.6(1)        & 7.9(3)     \\ 
$\#9$    &  25.2(4)  	    & 752(5)   		& 3626.3    	  & 1920(20)     &   181(3)    & 18.5(3)       & 102(1)     \\ 
$\#10$   &  4.0(1)   	    & 290(1)  	 	&  87.1(5)    & 265.8(2)     &   52.8(1)   & 6.8(1)        & 20.5(3)   \\ 
\hline
Average$^b$ &  14(15)       & 589(292)   	& 156(133)    & 287(85)     & 60(24)      & 8(5)           & 23(19)        \\
\hline
\end{tabular}
\tablefoot{$^a$From maps convolved to the angular resolution of the CH$^+$ 1-0 line (27$''$). 
$^b$We have not used position $\#9$ for the averages. Parentheses indicate the uncertainty obtained by the Gaussian fitting programme. The fit uncertainty is in units of the last significant digit.}
\end{table*}

\clearpage

\section{Complementary Figures}\label{App-more-figs}

\begin{figure*}
\centering
\includegraphics[scale=0.125, angle=0]{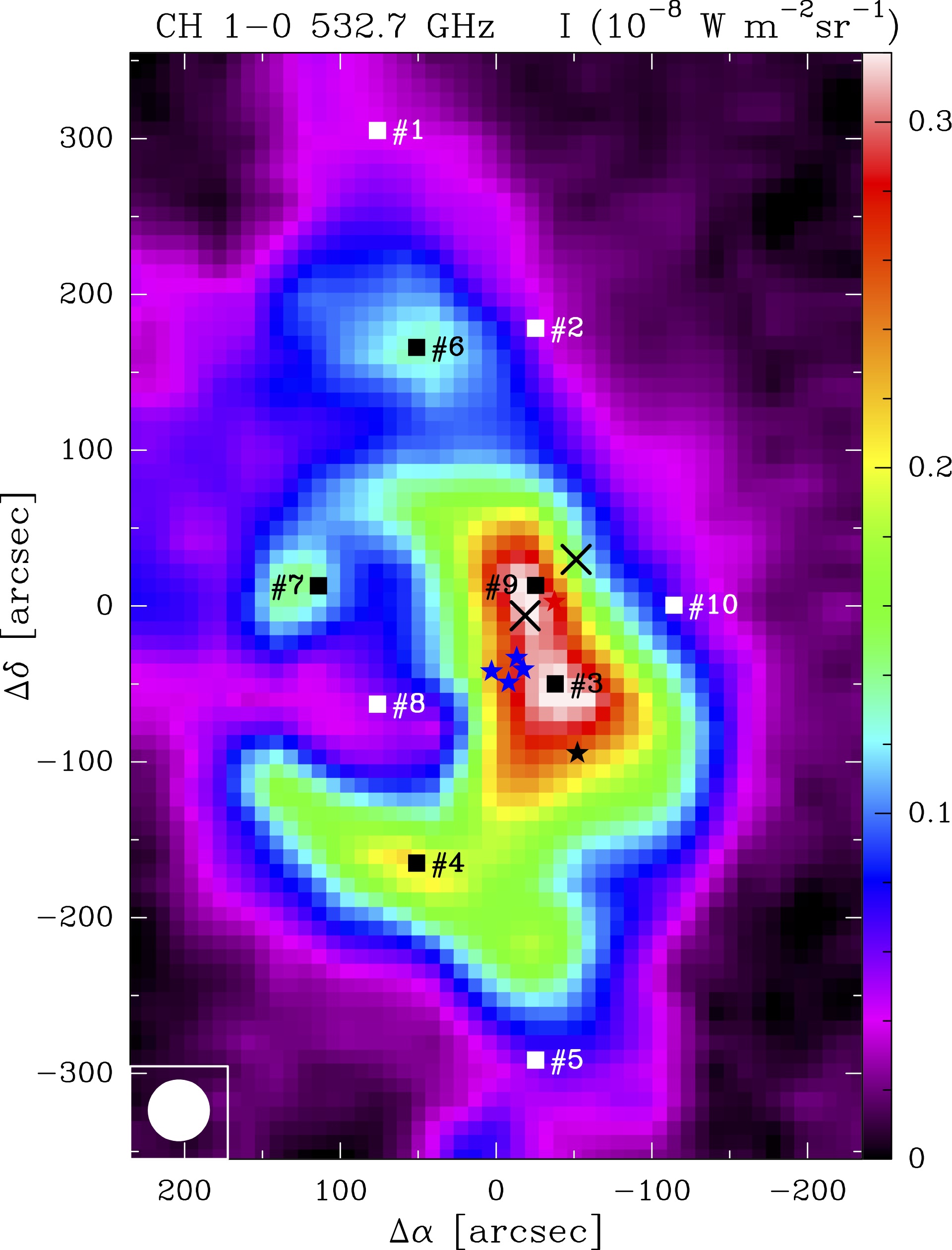}\hspace{0.5cm}
\includegraphics[scale=0.157, angle=0]{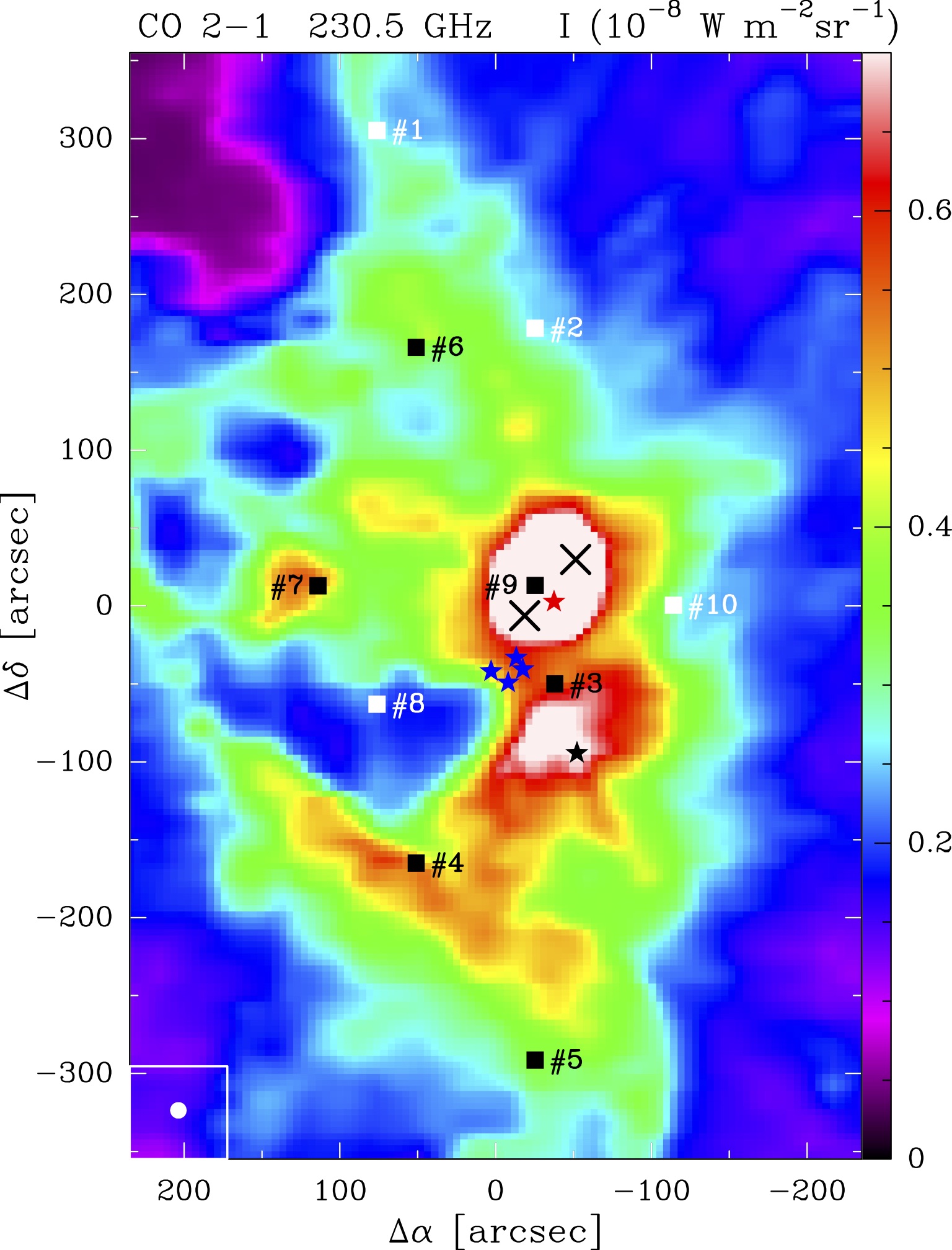}\vspace{0.5cm}
\includegraphics[scale=0.125, angle=0]{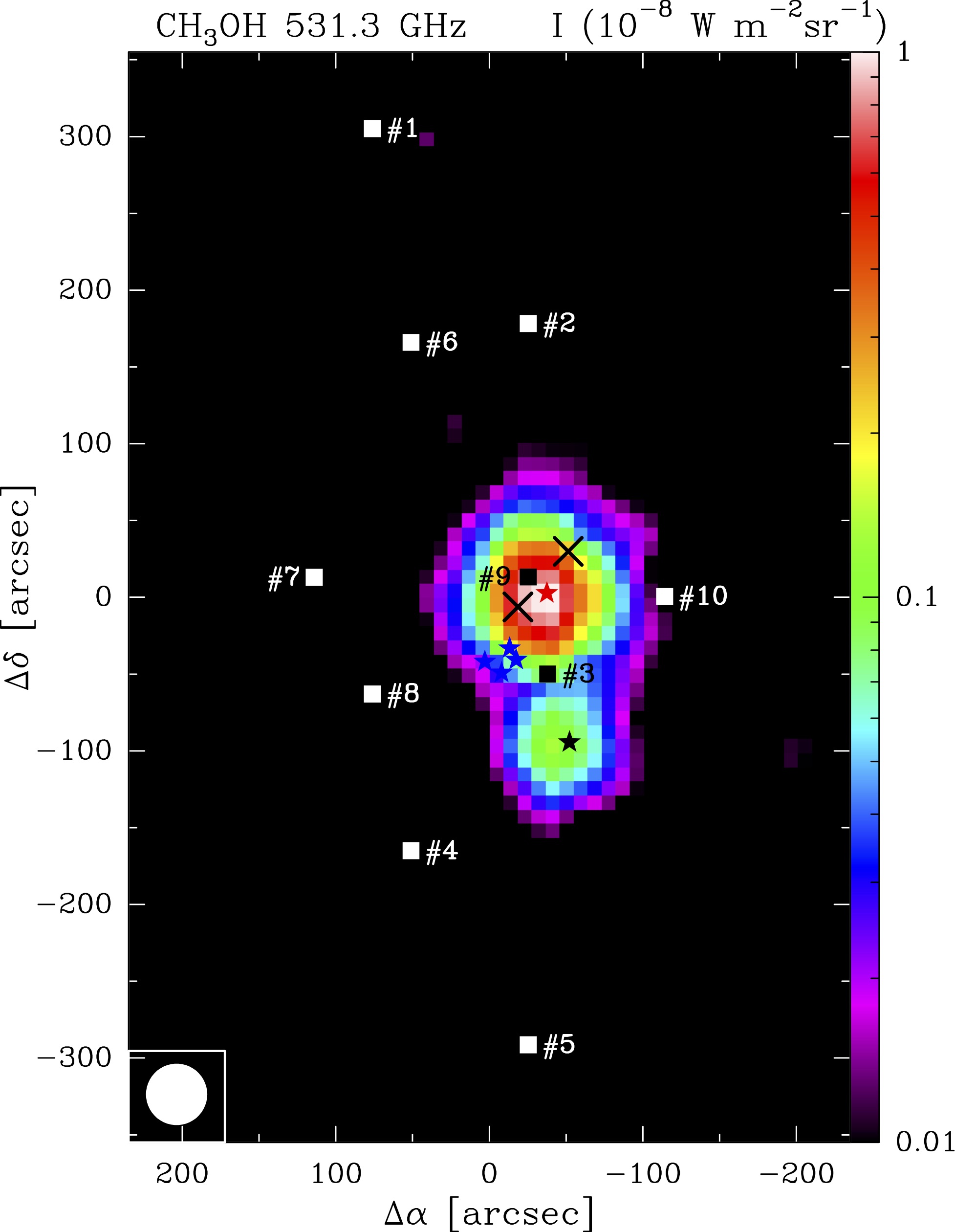}\hspace{0.5cm}
\includegraphics[scale=0.157, angle=0]{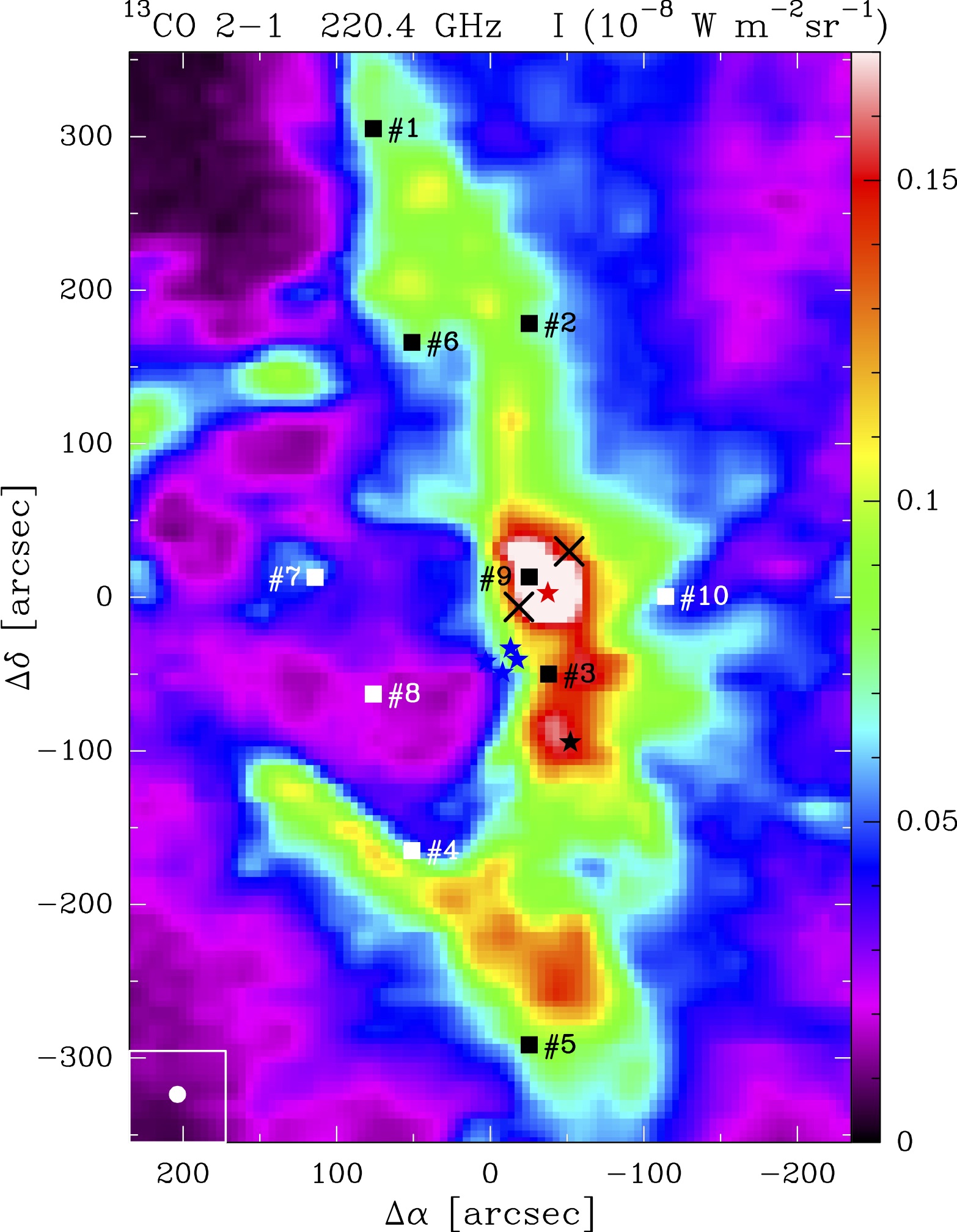}\\
\caption{\textit{Herschel}/HIFI and IRAM-30m maps of different molecular emission lines toward \mbox{OMC-1}. 
The color scale shows the  integrated line intensity in \mbox{W\,m\,$^{-2}$\,sr$^{-1}$}.
The native angular-resolution of each observation, the HPBW, is plotted in the bottom-left corner. 
Representative positions discussed in the text
are indicated with numbers (see Sect.~\ref{sec-profiles}).}\label{fig:Original_maps2}
\end{figure*}

\begin{figure*}
\centering
\includegraphics[scale=0.125, angle=0]{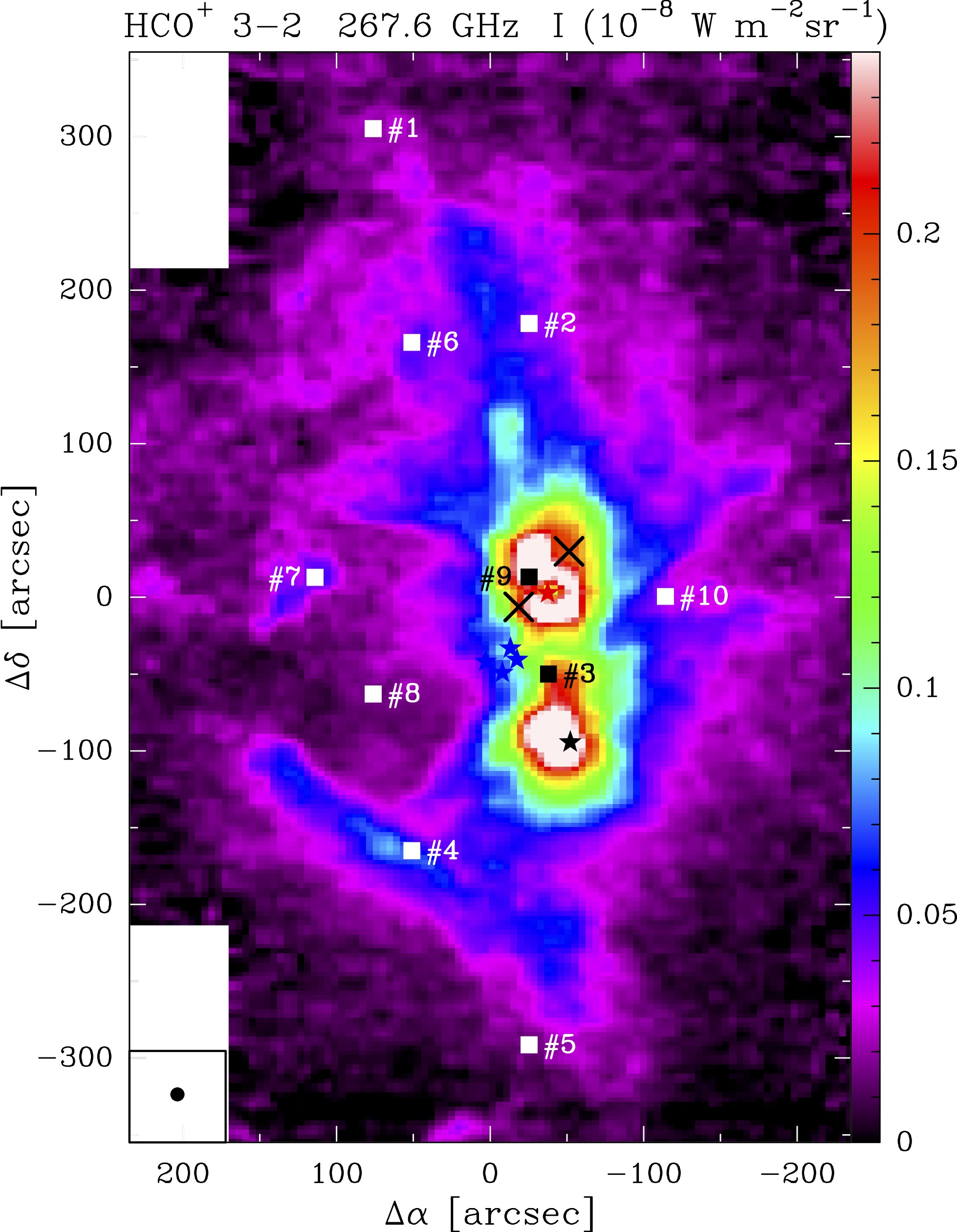}\hspace{0.5cm}
\includegraphics[scale=0.125, angle=0]{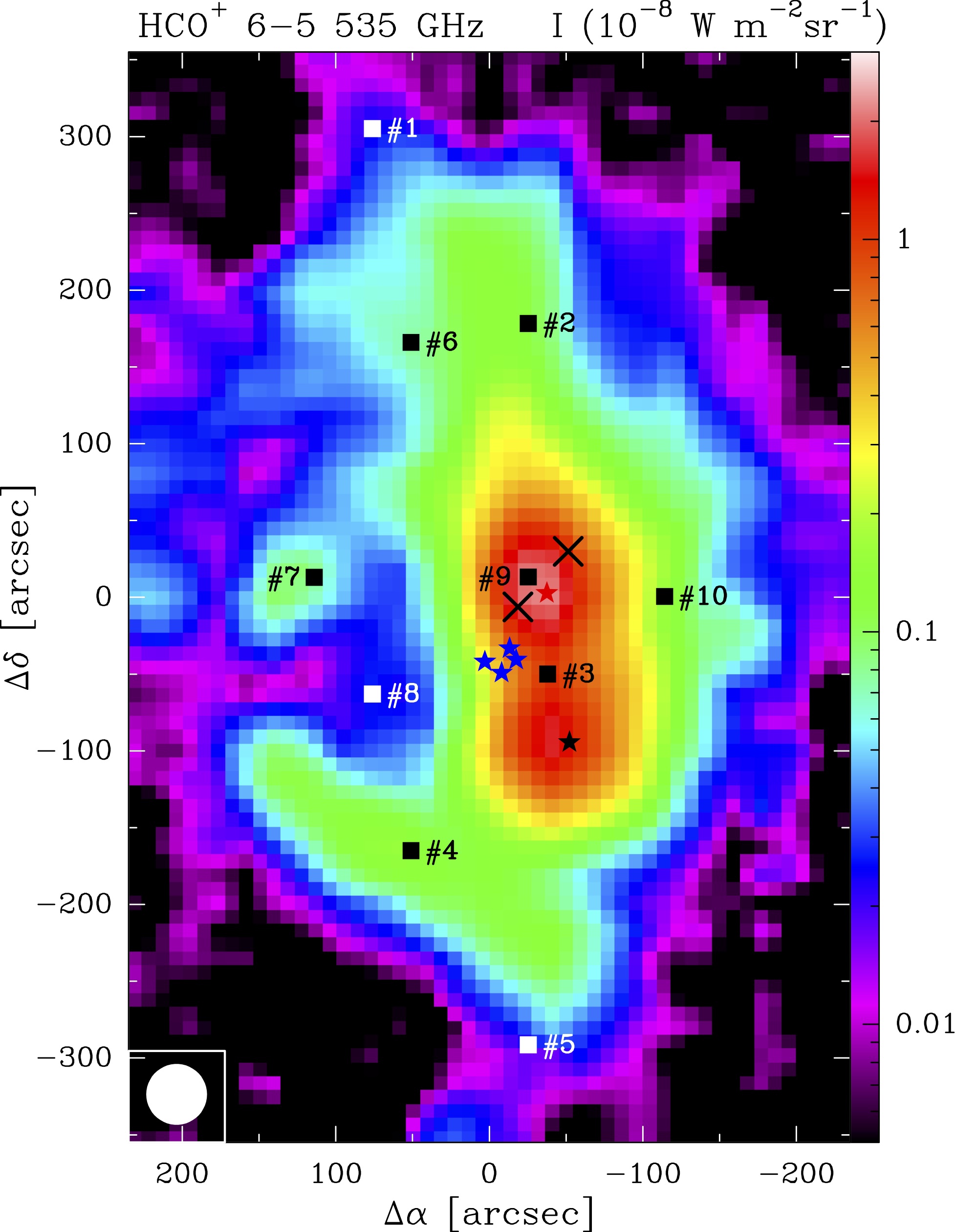}\vspace{0.5cm}
\includegraphics[scale=0.125, angle=0]{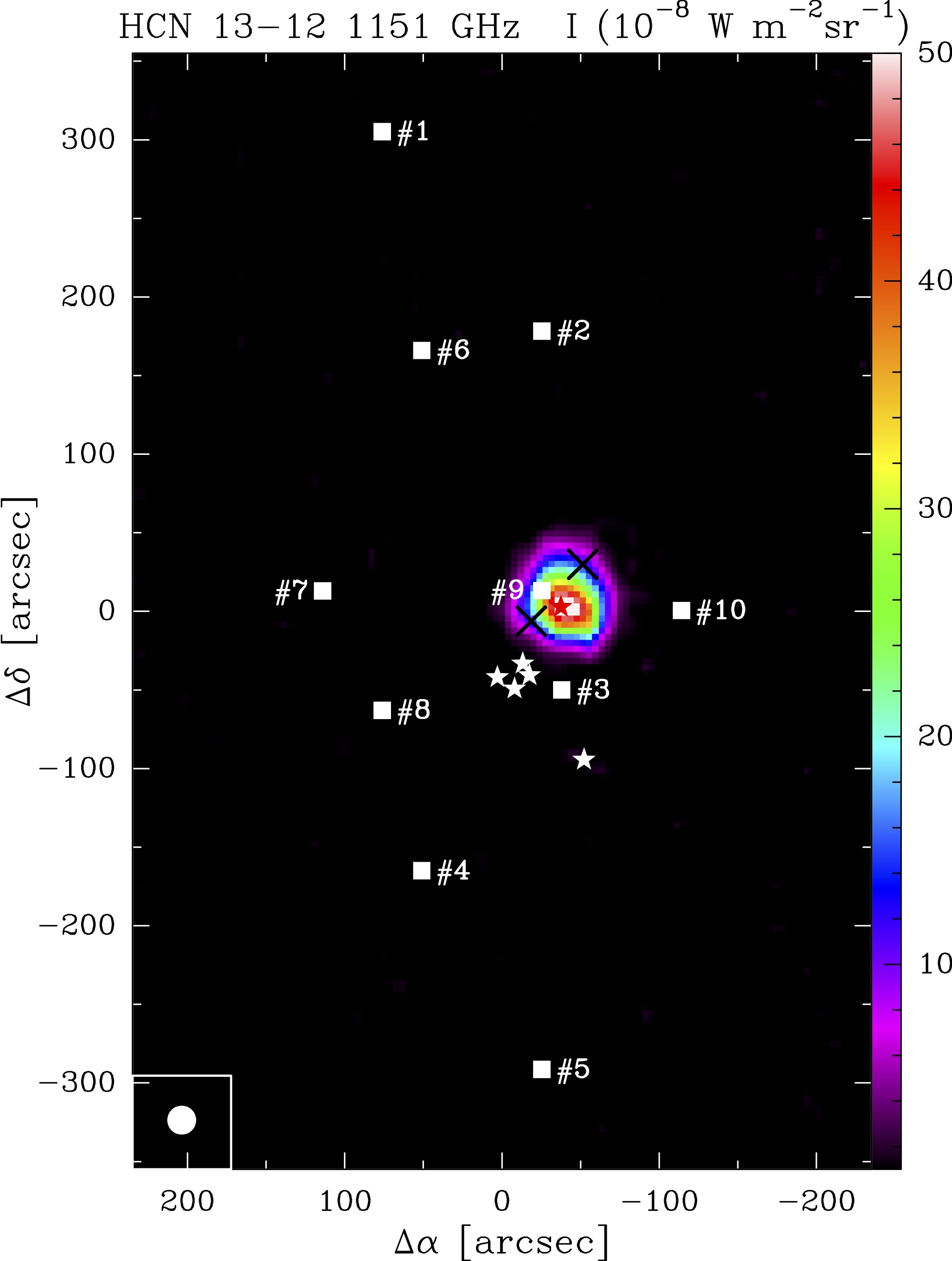}\hspace{0.5cm}
\includegraphics[scale=0.125, angle=0]{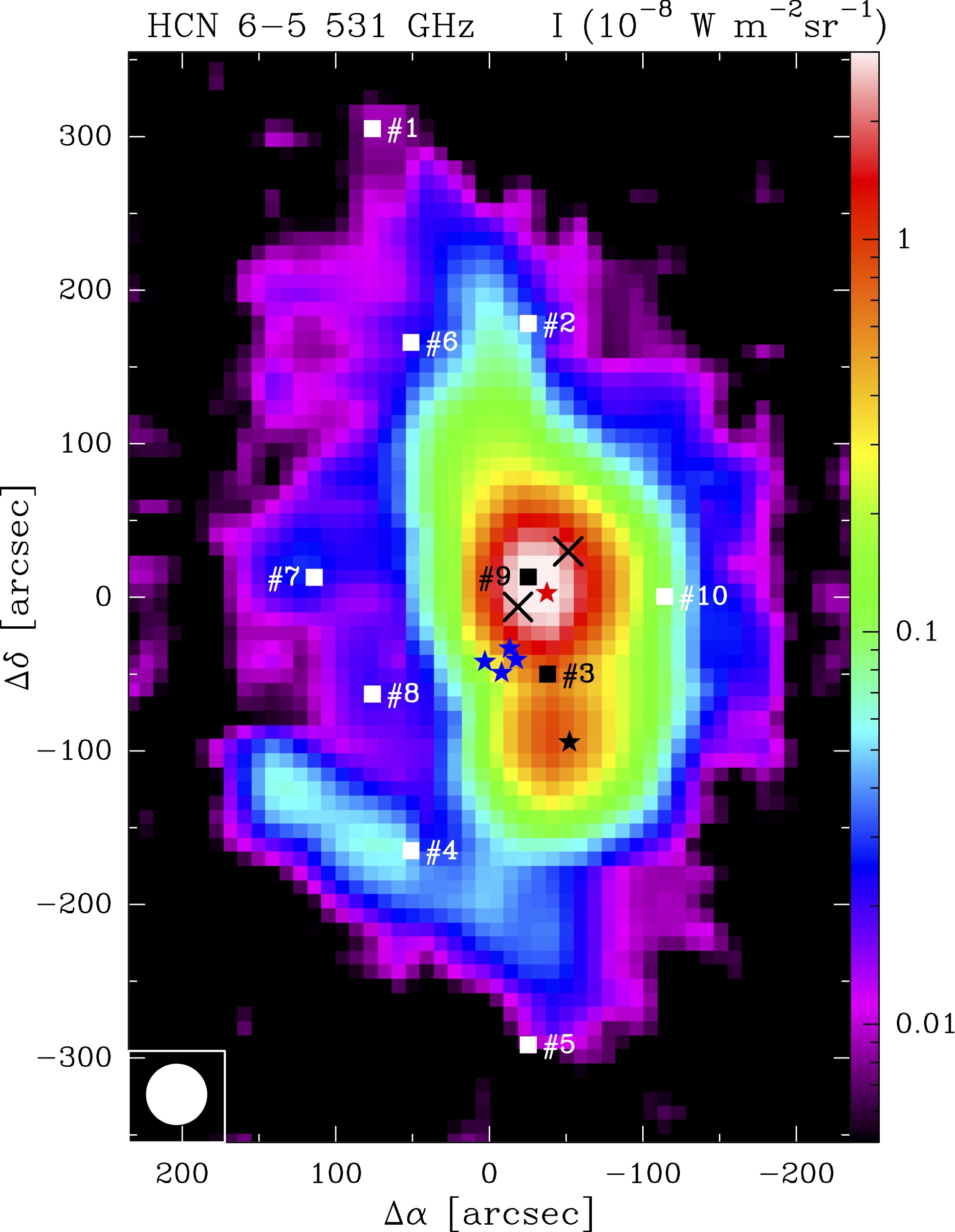}\\
\caption{\textit{Herschel}/HIFI and IRAM-30m maps of different high critical-density 
tracers (mid-$J$ HCN and HCO$^+$ lines). 
The color scale shows the  integrated line intensity in \mbox{W\,m\,$^{-2}$\,sr$^{-1}$}.
The native angular-resolution of each observation, the HPBW, is plotted in the bottom-left corner. 
Representative positions discussed in the text
are indicated with numbers (see Sect.~\ref{sec-profiles}).}\label{fig:Original_maps3}
\end{figure*}

\begin{figure*}[ht]
\centering
\includegraphics[scale=0.115, angle=0]{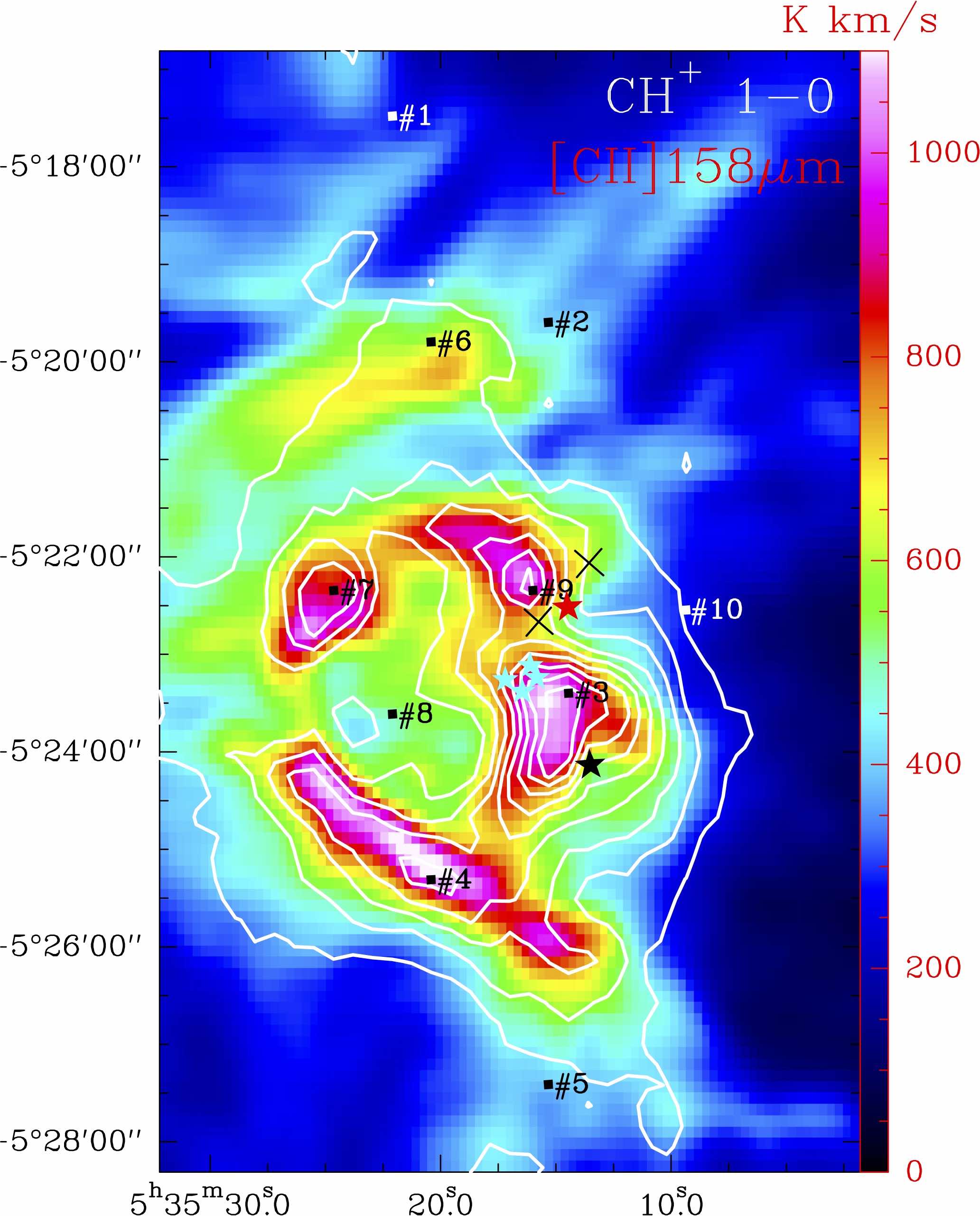}\hspace{1.5cm}
\includegraphics[scale=0.115, angle=0]{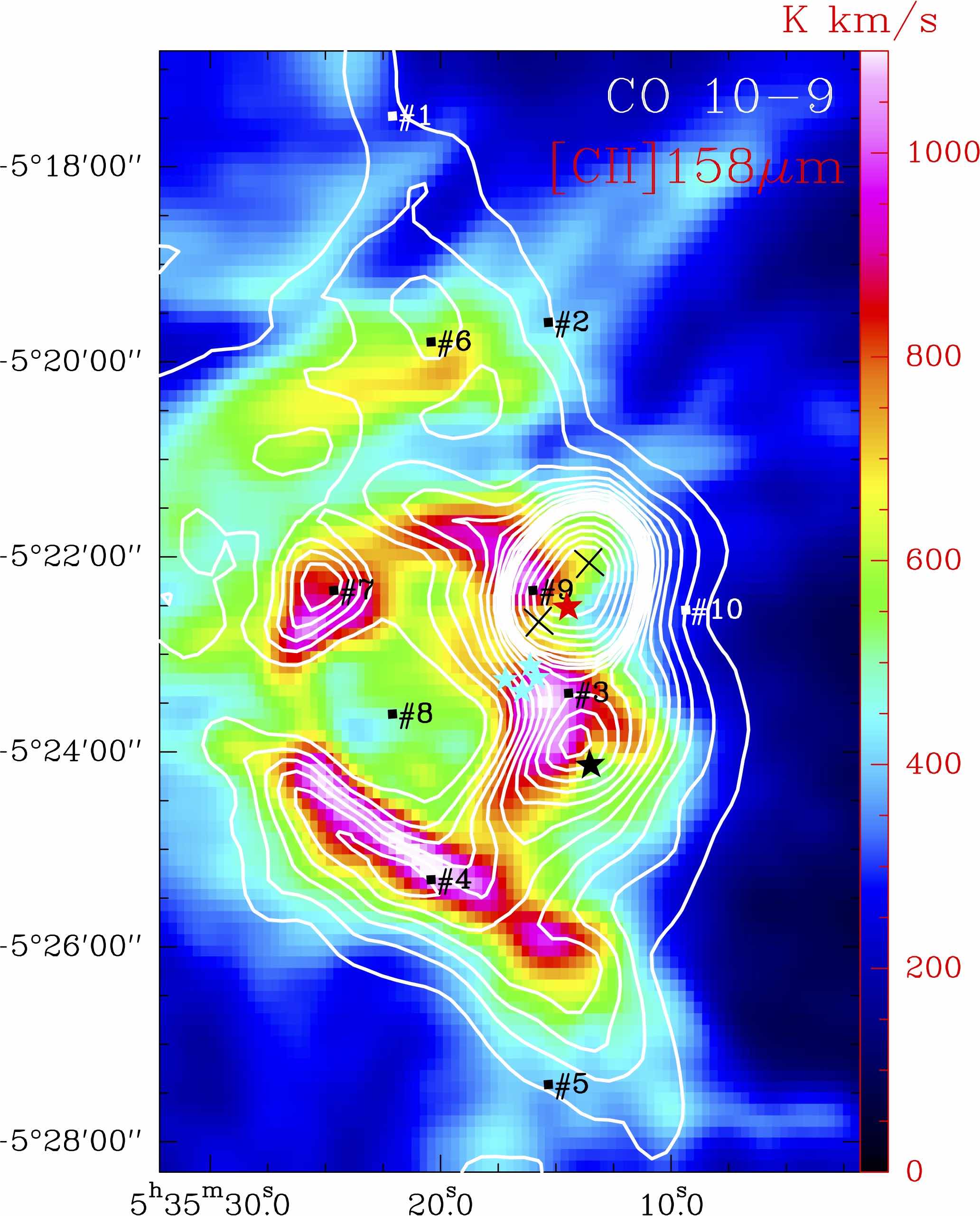}
\caption{Integrated line emission from [\CII]\,158\,$\upmu$m  \citep[in color scale;][]{Goi15},
\mbox{CH$^+$ $J$\,$=$\,1--0}, and \mbox{CO $J$\,$=$\,10--9}  in white contours.
Contours go from 5 to 45~K\,km\,s$^{-1}$ in steps of  5~K\,km\,s$^{-1}$ for CH$^+$,
and from 50 to 600~K\,km\,s$^{-1}$ in steps of  50~K\,km\,s$^{-1}$ and
700 to 1600~K\,km\,s$^{-1}$ in steps of  200~K\,km\,s$^{-1}$ for  CO.
Maps have been convolved to the same angular resolution (27$''$).}\label{fig:maps_Cp_CHp_CO_compa}
\end{figure*}

\begin{figure*}[ht]
\centering
\includegraphics[scale=0.157, angle=0]{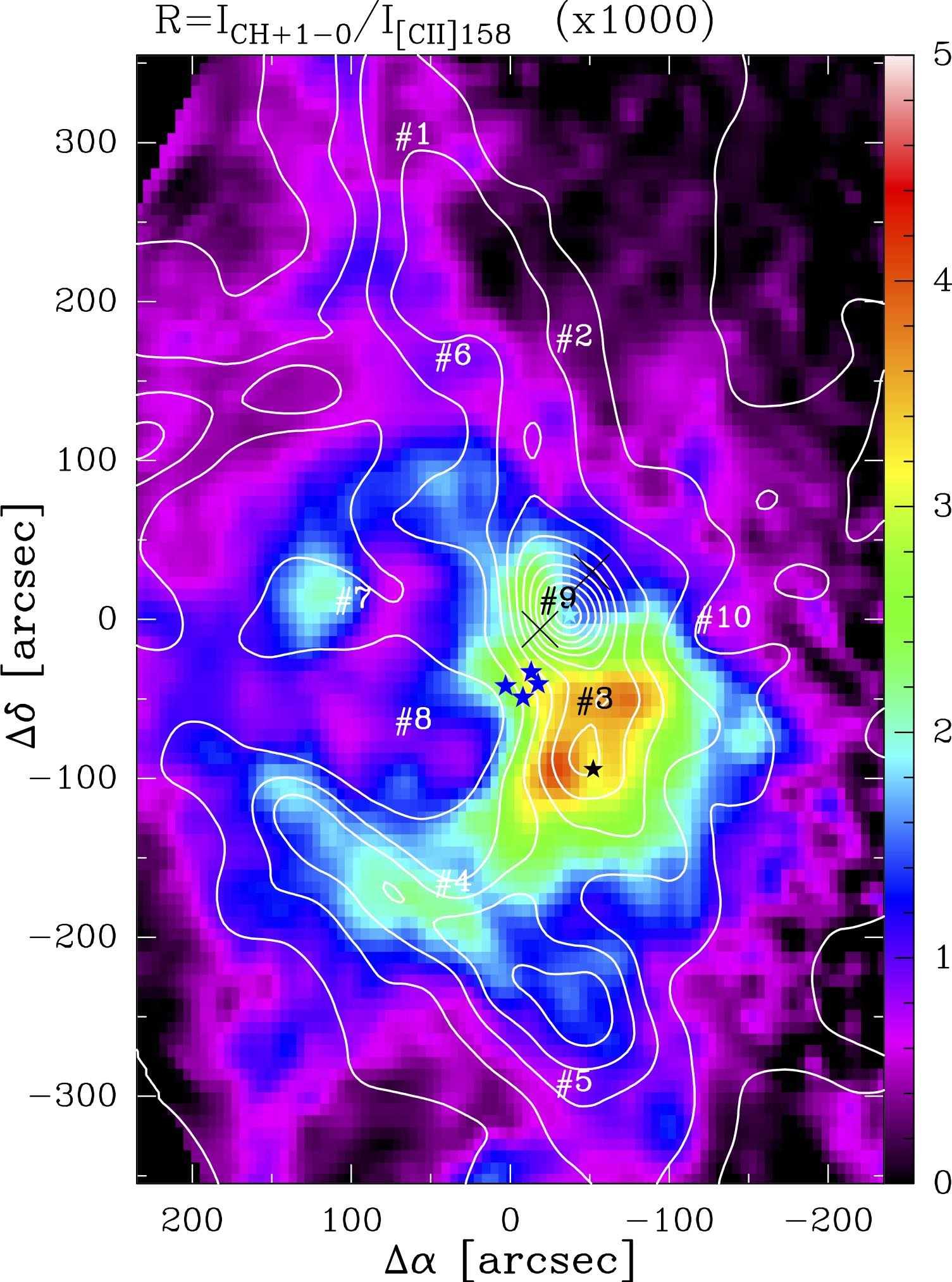}\hspace{1cm}
\includegraphics[scale=0.125, angle=0]{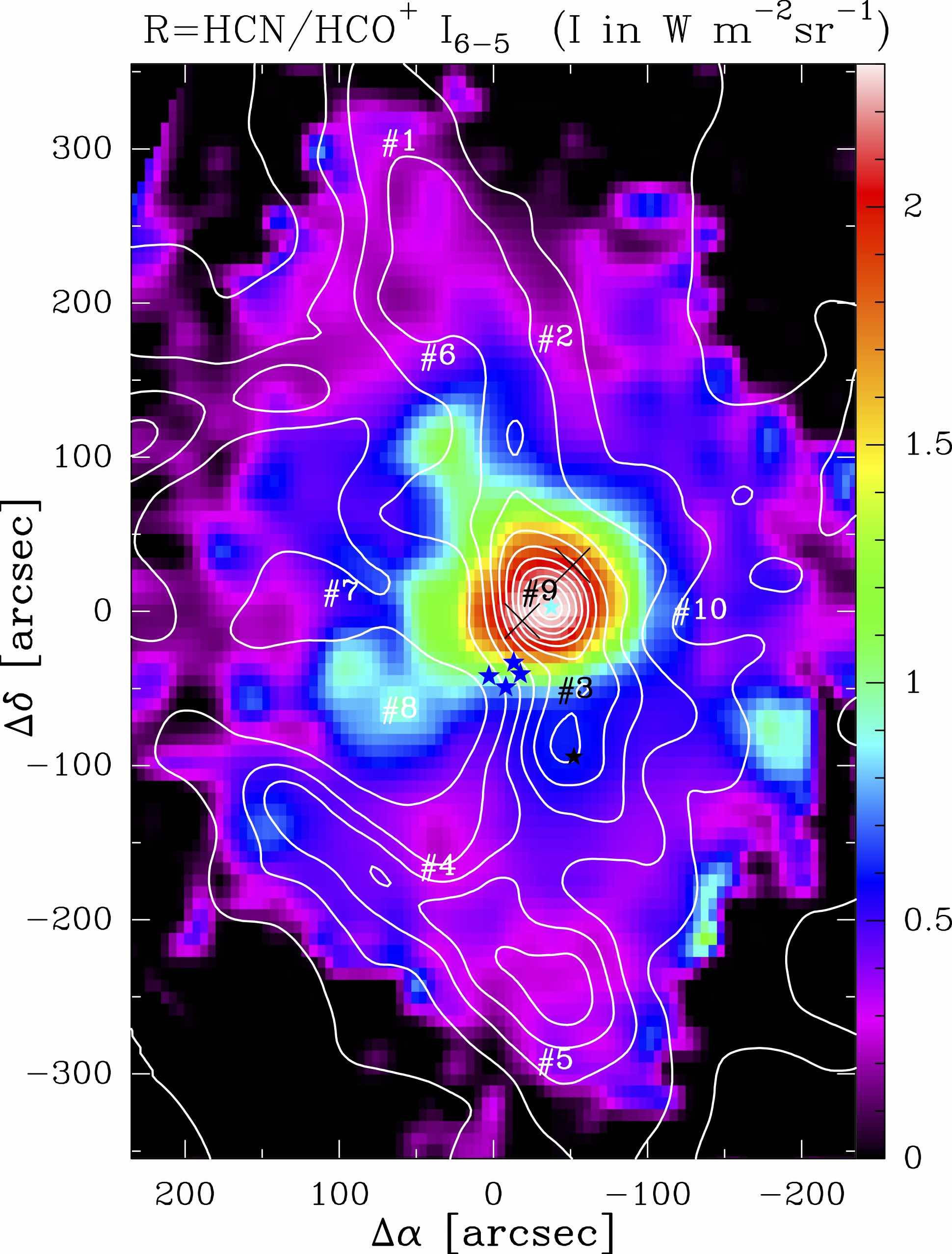}\\
\caption{\textit{Left:} 
Map of the \mbox{CH$^+$~$J$=1--0 to [\CII]\,158\,$\mu$m} line intensity ratio
(in units of W\,m$^{-2}$\,sr$^{-1}$) roughly proportional to gas density variations
at the irradiated cloud surfaces.
\textit{Right:}
Map of the \mbox{HCN~$J$\,$=$\,6--5 to HCO$^+$~$J$\,$=$\,6--5} line intensity ratio.
In both maps, contours represent the $^{13}$CO $J$\,$=$\,2--1 optically thin emission.
}\label{fig:peaks}
\end{figure*}

\begin{figure*}[h]
\centering
\includegraphics[scale=0.47, angle=0]{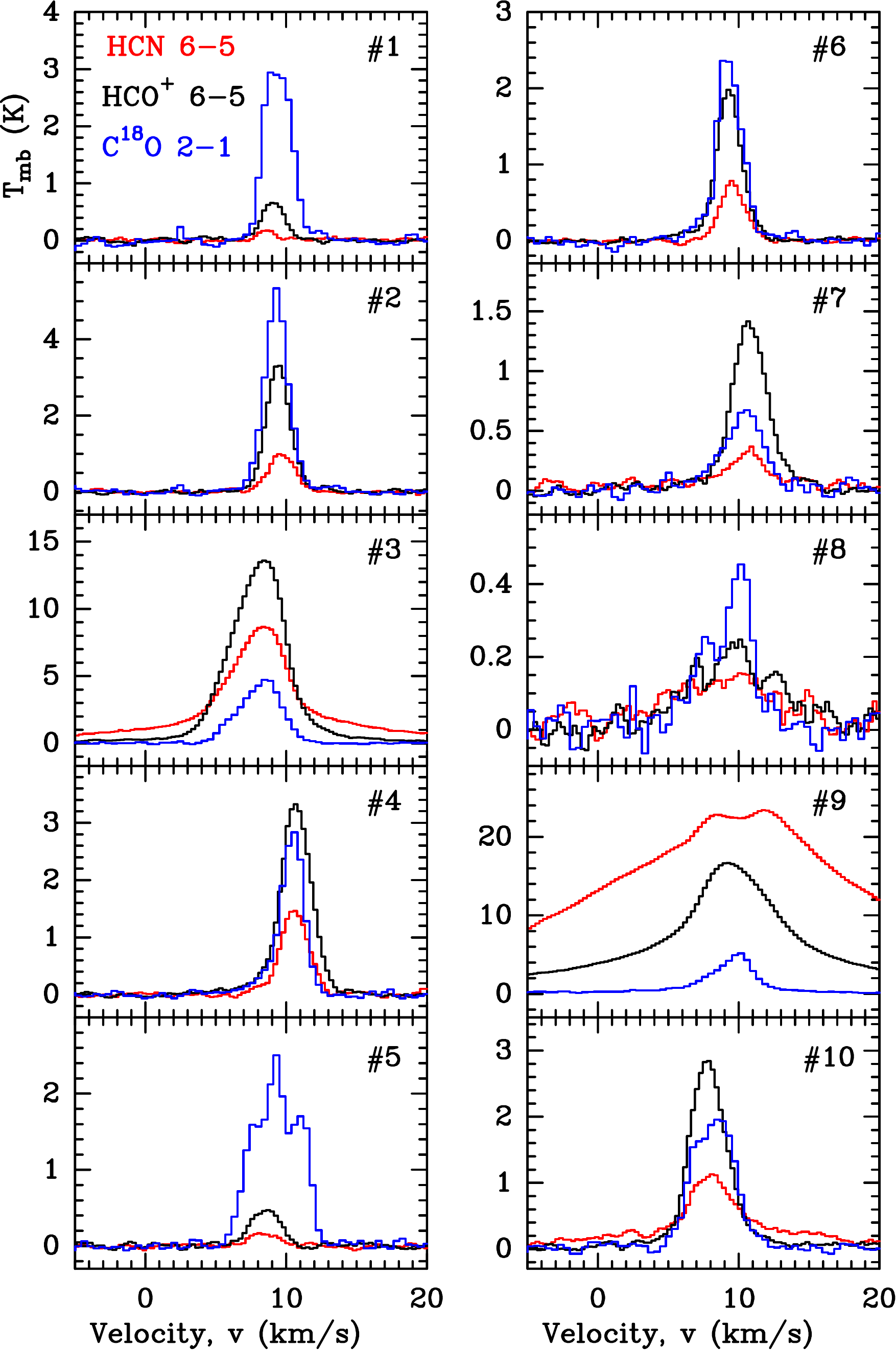}\hspace{1cm}
\caption{Line profiles at representative positions (see Table \ref{table:positions}). Spectra were extracted from maps convolved to a uniform resolution of  43$''$. 
}\label{fig:spectra_43}
\end{figure*}

\begin{figure*}[h]
\centering
\includegraphics[scale=0.33, angle=0]{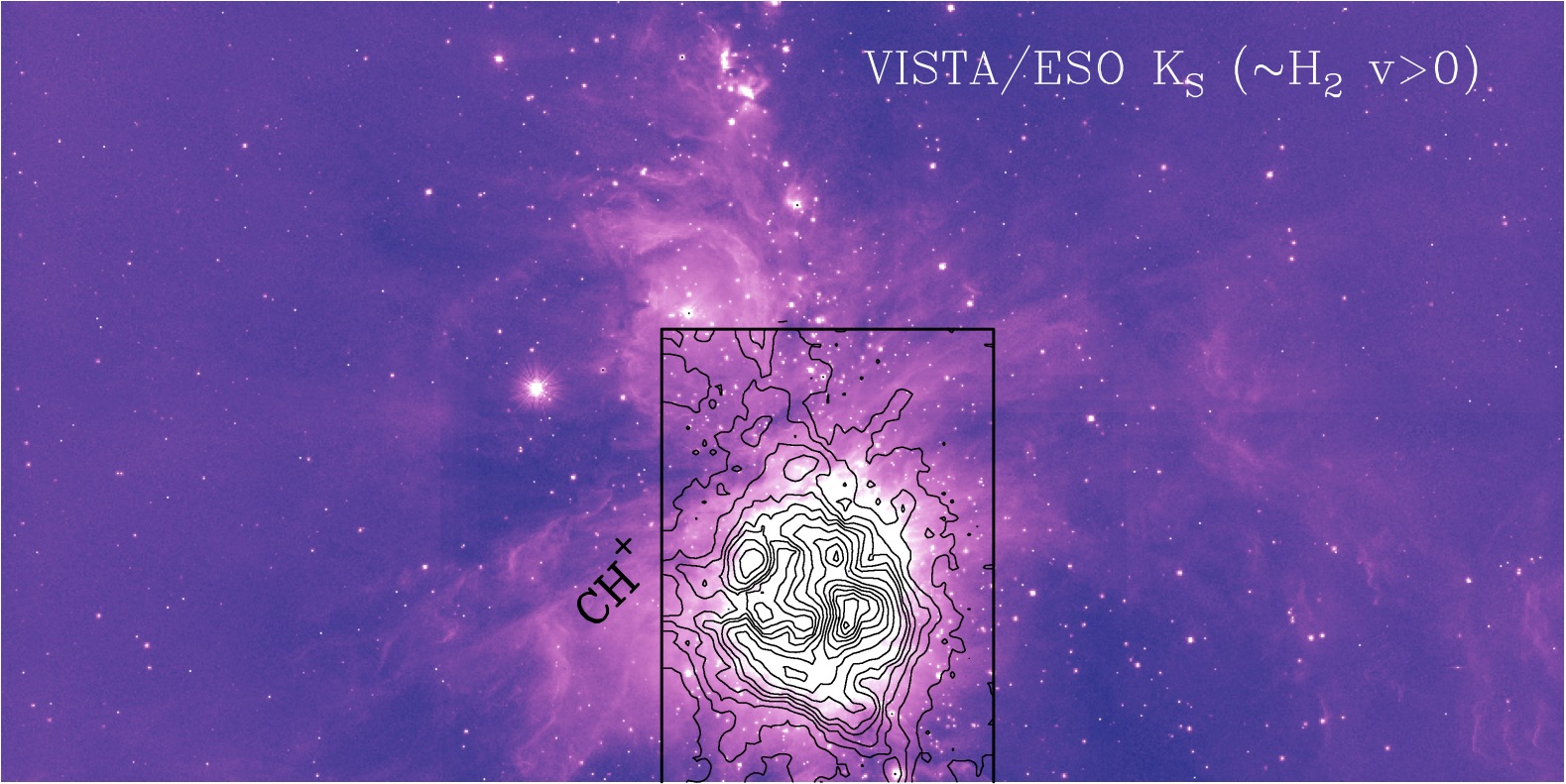}\hspace{1cm}
\caption{Large scale (40$'$$\times$20$'$) VISTA-NIR image of M42 and M43 in the $K_S$ band
around 2.15\,$\upmu$m (Meingast et al. 2016) tracing vibrationally excited H$_2$ emission.
The intensity scale is chosen to enhance the faint and extended NIR emission.
Black contours show the CH$^+$ $J$\,$=$\,1--0 line integrated intensities from 1 to 50\,K\,km\,s$^{-1}$.}\label{fig:Vista_large}
\end{figure*}

\clearpage

\section{Warm gas mass from CO\,$J$\,=\,10--9}\label{App-masses}

\begin{figure}[ht]
\centering
\includegraphics[scale=0.125, angle=0]{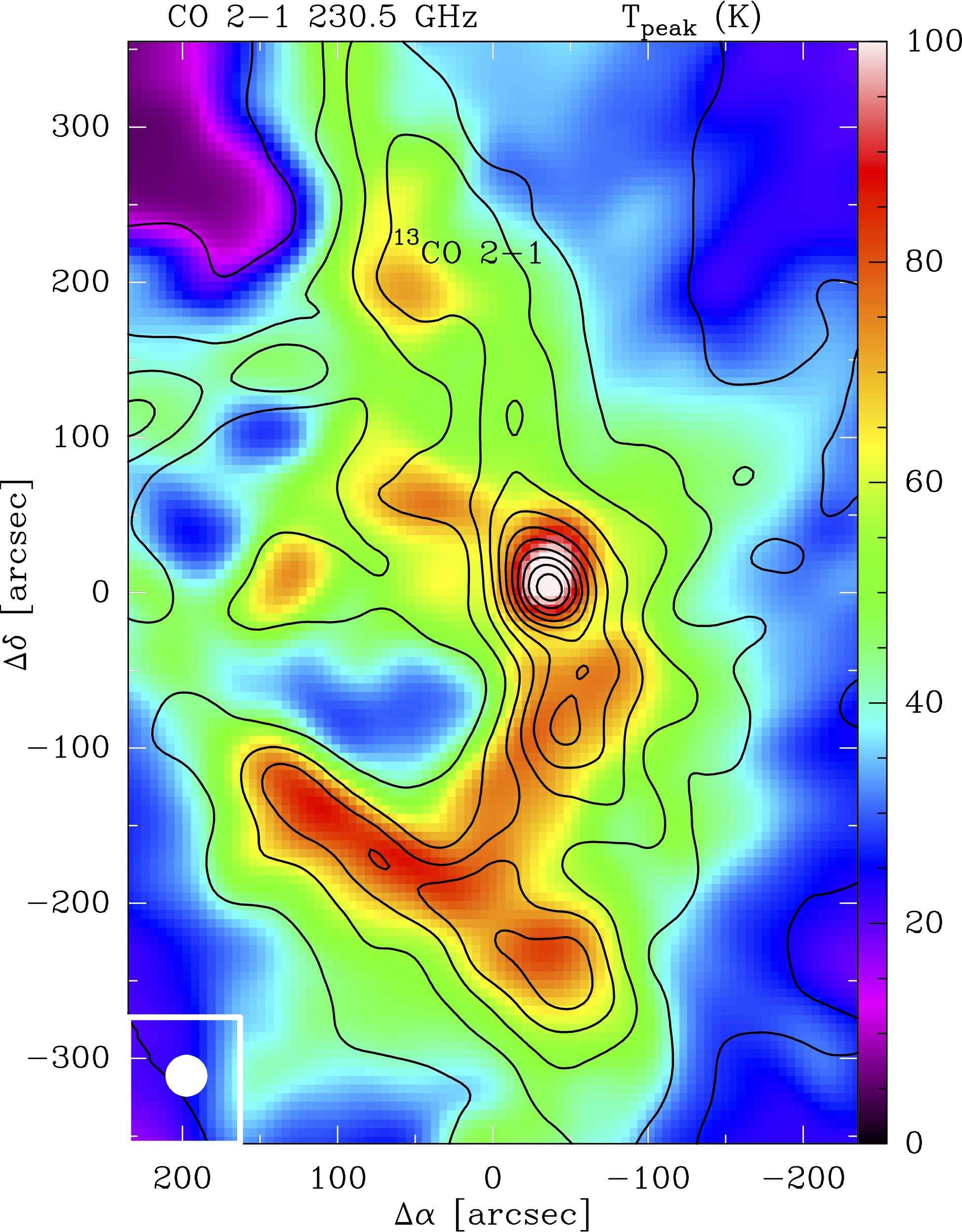}
\caption{Map of the CO~$J$=2-1 line intensity peak ($T_{\rm peak}$ in K; shown in color scale). Because the CO~$J$=2-1 emission is optically thick in most of the field, this map provides a lower limit to the gas temperature in the layers where the line emission arises, that is,   \mbox{$T_{\rm peak}$\,$\simeq$\,$T_{\rm rot}$\,(CO 2-1)\,$\leq$\,$T_{\rm k}$}.
}\label{fig:peak_co}
\end{figure}

In Sect.~\ref{sec-props} we estimated the mass contained in the high pressure molecular PDR layers, 
$M_{\rm mPDR}$(H$_2$). In particular, we integrate the column density equations for optically thin line emission
\begin{equation}\label{eq-N}
N_{\rm thin}({\rm CO})=8\pi\,\frac{\nu^3}{c^3} \frac{Q(T_{\rm rot})}{g_{10} \, A_{10-9}} 
\frac{1}{1 - e^{E_u/kT_{\rm rot}}} 
\frac{W_{\rm 10-9}}{J(T_{\rm rot})-J(T_{\rm bg})} \,\,\, {\rm (cm^{-2})},
\end{equation}over the \mbox{OMC-1} map. $W_{\rm 10-9}$ refers to the
\mbox{CO\,$J$\,$=$\,10--9}  line integrated intensity in K\,km\,s$^{-1}$. We assume that the rotational temperature at each position of the map, $T_{\rm rot}$, is the same as \mbox{$T_{\rm rot}$(2--1)}. 
Because the \mbox{CO\,$J$\,$=$\,2--1} line emission is optically thick through most of the field, we can determine \mbox{$T_{\rm rot}$} at each pixel of the map from the line peak temperature, 
$T_{\rm peak}$(2--1) (see Fig.~\ref{fig:peak_co}), using
\begin{equation}
J(T_{\rm rot})=T_{\rm peak}(2-1)+J(T_{\rm bg})
\end{equation}where
\mbox{$J(T)$\,$=$\,$T^*$/($e^{T^*/\,T} - 1$)} is the equivalent brightness temperature, 
 $T^*$\,$=$\,$E_{\rm u}/k$, and $T_{\rm bg}$ is the temperature of the cosmic microwave background.
If the \mbox{CO\,$J$\,$=$\,10--9} emission is slightly optically thick, line opacity $\tau_{\rm 10-9}$\,$\simeq$\,1,  we add an optical depth correction  \mbox{$N({\rm CO})$= $N_{\rm thin}({\rm CO})\cdot\tau_{\rm 10-9}/(1-e^{-\tau_{\rm 10-9}})$}, where                   
\begin{equation}\label{ec-tau_emi}
\tau_{\rm 10-9}=-{\rm ln}\,\left[1-\frac{T_{\rm P}(10-9)}{J(T_{\rm rot})-J(T_{\rm bg})}\right], 
\end{equation}
is calculated from observations. 
Once we created a map of the warm CO column density by the \mbox{CO\,$J$\,=\,10--9} emission,
 we convert it to a total gas column density map through $N_{\rm H}$\,=\,$x$(CO)$\cdot$$N$(CO)\,$\simeq$10$^{-4}$$\cdot$$N$(CO)  using a representative value for the  CO abundance in the PDR.
The mass of the warm PDR molecular gas  in \mbox{OMC-1} is then \mbox{$M({\rm mPDR}) \simeq \mu\, m_{\rm H}\, \sum_{i} \, (N_{\rm H}^{i}\,A_i)$}, where $A_i$ is the pixel~$i$ area, $m_{\rm H}$ is the H atom mass, and $\mu$ is the mean molecular weight.  For this mass estimation we used the
\mbox{CO\,$J$\,$=$\,2--1} and \mbox{10--9} maps convolved to 27$''$.

\end{appendix}

\end{document}